\documentclass[12pt]{iopart}
\usepackage{iopams}  
\usepackage{epsfig}
\usepackage{graphicx}
\expandafter\let\csname equation*\endcsname\relax
\expandafter\let\csname endequation*\endcsname\relax
\usepackage{amsmath}

\begin{document}

\title[Higher Harmonic Flow]{Initial State Fluctuations and Final State Correlations in Relativistic Heavy-Ion Collisions}

\author{Matthew Luzum${}^{1,2}$, Hannah Petersen${}^{3,4}$\\[.4cm]}

\address{
${}^1$~Department of Physics, McGill University, 3600 University Street, Montreal, Quebec, H3A 2T8, Canada\\
${}^2$~Nuclear Science Division, Lawrence Berkeley National Laboratory, MS70R0319, 1 Cyclotron Road, Berkeley, California 94720, USA\\
${}^3$~Frankfurt Institute for Advanced Studies, 
Ruth-Moufang-Strasse 1, 60438 Frankfurt am Main, Germany\\
${}^4$ Institut f\"ur Theoretische Physik, Goethe Universit\"at, Max-von-Laue-Strasse 1, D-60438 Frankfurt am Main, Germany.
}

\eads{\mailto{mwluzum@lbl.gov}, \mailto{petersen@fias.uni-frankfurt.de}}

\begin{abstract}
We review the phenomenology and theory of bulk observables in
ultra-relativistic heavy-ion collisions, focussing on recent
developments involving event-by-event fluctuations
in the initial stages of a heavy ion collision, and how they manifest in observed correlations.
We first define the relevant observables and show how each measurement is related
to underlying theoretical quantities. 
Then we review the prevailing picture of the various stages of a collision, 
including the state-of-the-art modeling of the initial stages of a collision and 
subsequent hydrodynamic evolution, as well as hadronic scattering and freeze-out in the later stages.
We then discuss the recent results that have shaped our current understanding and identify 
the challenges that remain.  
Finally, we point out open issues and the potential for progress in the field.
\end{abstract}


\pacs{25.75.-q,25.75.Ag,24.10.Lx,24.10.Nz}


\maketitle

\section[Introduction]{Introduction}
\label{intro}

Ultra-relativistic heavy ion collisions offer the opportunity to study strongly interacting matter under extreme conditions. 
In nuclear collisions at the Relativistic Heavy Ion Collider (RHIC) with incident energies up to 200 GeV per nucleon pair 
and at the Large Hadron Collider (LHC) at even higher energies of $\sqrt{s_{\rm NN}}=2.76$ GeV, 
nuclear matter is heated to temperatures up to $\sim$ 200-500 MeV. At these high energies lattice calculations of quantum chromodynamics in equilibrium predict a cross-over transition to a new state of matter, where quarks and gluons are the relevant degrees of freedom, the so called quark-gluon plasma (QGP) \cite{Aoki:2006we}. 

The most important observables supporting the discovery of the QGP are the high values of elliptic flow and the suppression of high $p_T$ hadrons that can be explained within theoretical approaches assuming a strongly coupled quark-gluon liquid to be formed, which exhibits strong partonic collectivity and parton energy loss \cite{Adams:2005dq,Back:2004je,Arsene:2004fa,Adcox:2004mh}. Relativistic viscous hydrodynamics has proven to be a very successful tool to capture the bulk dynamics of heavy ion collisions \cite{Luzum:2008cw,Song:2007ux}. 
One of the major goals of the field is to determine precise quantitative properties of the QGP, such as transport coefficients and their temperature dependence. 

For many years, most of the dynamical descriptions based on hydrodynamics used smooth and symmetric initial density profiles to simulate an ``average'' collision.  
In reality, there are quantum fluctuations in the earliest stages of the collision, that can come from sources as mundane as the nucleonic structure of the nucleus.
This causes the energy density at early times to be lumpy and asymmetric, and to fluctuate from one collision event to the next, even for nuclei colliding at a fixed impact parameter. 
The importance of these fluctuations was realized at least as early as 2005 \cite{Miller:2003kd, Alver:2006wh}, but the full implications were not realized until later \cite{Alver:2010gr}.  Since then, the importance of event-by-event fluctuations for understanding bulk phenomena has been extensively investigated, both experimentally and theoretically (see Ref.~\cite{Adare:2012kf} and references therein). 

This review describes the recent developments to understand initial state fluctuations and final state correlations in a rather comprehensive fashion, though some bias by the authors cannot be avoided in the selection of results and approaches presented. The structure of this manuscript proceeds from the well established to the rather unknown parts of the theoretical description, before results are shown that support the current picture. First, a historical introduction is given from an experimental (Section \ref{sec_histex}) and a theoretical point of view (Section \ref{sec_histtheo}). In Section \ref{sec_obs}  the relevant observables are defined and meaningful comparisons of theory calculations to experimental data are discussed. The state-of-the-art understanding of hydrodynamical modeling is presented in Section \ref{sec_hydro}. The treatment of the final non-equilibrium hadronic stage of the reaction focusing on the relevance for fluctuation and correlation observables is described in Section \ref{sec_foresc}. The physics of the initial stage of a heavy ion reaction is the major unknown and in Section \ref{sec_ic} the different ideas and parametrizations that have been developed are explained. 
Section \ref{sec_results} reviews the results that have been achieved so far and Section \ref{sec_issues} outlines future opportunities and challenges. The manuscript ends with a summary in Section \ref{sec_sum}.

\subsection{A little bit of history -Experimental point of view}
\label{sec_histex}
Anisotropic flow is one of the most prominent bulk observables in heavy ion collisions since the early days of the first heavy ion experiments \cite{Danielewicz:1985hn,Ollitrault:1992bk}. In peripheral collisions the overlap region in the plane transverse to the beam direction has an ellipsoidal shape. This geometric anisotropy between the ``reaction plane'' spanned by the beam direction ($z$) and the impact parameter ($x$), with respect to the perpendicular ($y$) direction can be quantified by the standard eccentricity \cite{Sorge:1998mk}.

\begin{equation}
\label{stdeps2}
\varepsilon_{\rm standard} \equiv \frac{\{ y^2-x^2 \}}{\{ y^2+x^2 \}} \quad ,
\end{equation}
where here and in the following, curly brackets indicate an average over the energy density $\epsilon$ of the system at an early stage of the collision
\begin{equation}
\label{spaceavg}
\{\ldots\} \equiv \frac {\int d^3x\ \epsilon(\vec x) \ldots} {\int d^3x\ \epsilon(\vec x)}\quad .
\end{equation}

This initial coordinate space anisotropy is converted to a final state momentum space anisotropy, if large enough collective pressure gradients drive the evolution of the system. The anisotropy  in momentum space is defined as the second Fourier coefficient of the azimuthal distribution of the produced particles \cite{Ollitrault:1992bk},
\begin{equation}
v_2 \equiv \left\langle \frac{ p_x^2-p_y^2}{ p_x^2+p_y^2 }\right\rangle_p\quad .
\end{equation}
Here the average is performed over the final distribution of particles
\begin{equation}
\langle \ldots \rangle_p \equiv \frac 1 N \int d^3 p \frac {dN} {d^3p} \ldots \quad .
\end{equation}
 The original claim that the quark gluon plasma behaves like a perfect liquid is based on the agreement of the measurements of elliptic flow at RHIC with the predictions of ideal hydrodynamic calculations \cite{Huovinen:2001cy}. The elliptic flow measured for different centrality classes in Au+Au collisions and at different beam energies, scaled by the corresponding eccentricity exhibits a scaling as a function of the charged particle density per unit area. Even though this analysis is highly model dependent since the overlap area and the eccentricity cannot be observed experimentally, this scaling provided substantial evidence for the  hydrodynamic picture. 

In 2005 collisions of smaller nuclei Cu+Cu were carried out at RHIC. 
Using the standard definitions given above for the elliptic flow and the eccentricity, the ratio of $v_2/\varepsilon_2$ 
was surprisingly much higher than previously observed in Au+Au collisions,
and the elliptic flow did not appear to vanish as the collisions become more central. 
The resolution of this puzzle was that fluctuations in the initial state geometry are important, 
and are naturally more pronounced in smaller systems such as the Cu+Cu collision system
and for central collisions, where the average standard eccentricity is small.
Further, the relevant spatial eccentricity that drives elliptic flow is not the standard eccentricity above, 
which is defined with respect to the impact parameter of the incoming nuclei, but
an eccentricity with respect to a plane defined
by the participating nucleons (the ``participant plane''), which can be rotated in a different direction.
 
By using a new generalized definition of the ``participant eccentricity'' given by the expression
\begin{equation}
\label{eps2pp}
\varepsilon_{\rm part}=\frac{\sqrt{(\{ y^2 \} - \{ x^2 \})^2+4 \{ xy \}^2}}{\{ y^2\} +\{ x^2 \}} ,
\end{equation}
defined with respect to a coordinate system where $\{x\} = \{y\} = 0$, it became clear that the elliptic flow results in Cu+Cu collisions were completely consistent with the previous larger systems in a hydrodynamic picture \cite{Alver:2006wh}. This was the first piece of direct evidence of initial state fluctuations, 
and for several years, the study of the effects of these fluctuations was confined to existing elliptic flow analyses. 

The second piece of evidence is related to other details of two-particle correlation measurements
\footnote{The above mentioned measurements of $v_2$ actually represent the second Fourier harmonic of a two-particle correlation, 
though this fact was not obvious when ``event plane'' analyses were first performed, and it was only understood more recently \cite{Alver:2008zza}.  See Sec.~\ref{sec_obs} for more details.}.
Triggered correlations of one particle at higher transverse momentum with other particles at lower transverse momentum, usually plotted as a function of the relative azimuthal angle of the pair $\Delta \phi$ and relative pseudorapidity $\Delta \eta$, contain contributions from different physics sources. Originally, these correlations were supposed to give insights about the interactions of a high energetic parton with the surrounding medium. By losing energy a shock wave in the hydrodynamic medium is formed and the Mach cone may be observable as a double-hump structure opposite in $\phi$ to the high transverse momentum particle \cite{Stoecker:2004qu,Renk:2007rv}. These measurements are affected by the soft background containing flow correlations like elliptic flow. The other striking phenomenon was the discovery of long-range correlations over many units in  pseudorapity on the near-side, the 'ridge' \cite{Abelev:2009af}. 

\begin{figure}[h]
\begin{center}
\includegraphics[width=0.7\textwidth]{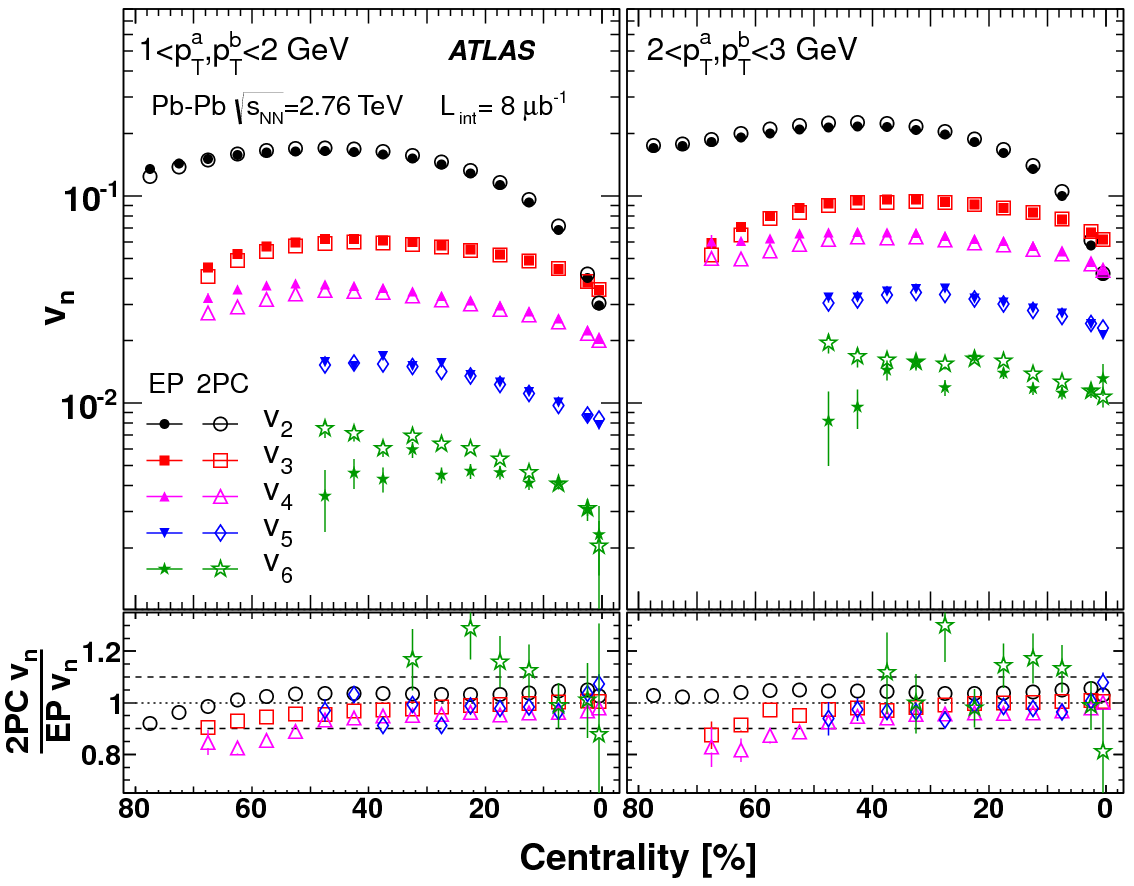}
\end{center}
\caption{
\label{fig_vn_centrality_atlas} 
Two-particle cumulant (2PC) and event-plane (EP) measurements of $v_2$--$v_5$ as a function of centrality for particles in different ranges of transverse momentum at the LHC
(taken from \cite{ATLAS:2012at}).}
\end{figure}

To understand the medium background, untriggered two-particle correlations were measured. Excluding the midrapidity region that is highly affected by back-to-back jet correlations and removing the first and second azimuthal Fourier coefficients $V_{1\Delta}$ and $V_{2\Delta}$ revealed a striking $\langle cos(3\phi) \rangle$ structure. A further analysis by Alver and Roland using the AMPT model demonstrates that this can be understood by introducing the so called triangular flow \cite{Alver:2010gr}. In analogy to the participant eccentricity a triangularity that reflects the geometric initial state fluctuations  can be defined. Since 2010, the whole series of higher Fourier coefficients has been explored (see, e.g., Fig.~\ref{fig_vn_centrality_atlas}) and a similar hydrodynamic response for odd moments as for the even ones has been established that we will describe in the later sections of this review. The main feature of the odd components is that at mid-rapidity, they are only existent because of event-by-event fluctuations and have little or no geometry component, in contrast to the even moments. Based on the discovery of long-range correlations in pseudorapidity in the untriggered two-particle correlations, the dynamical fluctuations of elliptic flow have been quantified as well \cite{Alver:2010rt}.

\subsection{A little bit of history - theory} 
\label{sec_histtheo}
In the beginning of heavy ion research the quark gluon plasma was expected to form at energy densities higher than some critical value of 1--10 GeV/fm$^3$. Since 
lattice QCD calculations showed only small deviations from the Stefan-Boltzmann limit, it was suggested to behave like an ideal gas of quarks and gluons. In contrast to this expectation, the first experimental data 
for elliptic flow in gold-gold collisions at RHIC 
were in good agreement with ideal hydrodynamic calculations. Since then the quark gluon plasma at temperatures up to 2--3 $T_c$  seems to be a rather strongly coupled liquid instead of the anticipated dilute gas of partons. The realization of the nearly perfect fluid dynamic behaviour of the QGP has triggered connections to other systems with similar properties in condensed matter physics \cite{Gelman:2004fj} and was the basis for exploiting the AdS/CFT correspondance \cite{Schafer:2009dj}. 

Most of the hydrodynamic calculations have been carried out using smooth initial state distributions and making use of the symmetries of the heavy ion collision. Still the idea of lumpy initial conditions or `hot spots' is not new and has been mentioned from early on \cite{Gyulassy:1996br}. Especially the Brasilian group has been investigating the effect of initial state fluctuations on observables within the NEXspheRIO approach since the early 2000s \cite{Aguiar:2001ac,Andrade:2006yh,Andrade:2008xh}.

Bulk observables in heavy ion collisions like, e.g., particle yields, spectra and collective flow are well described by hybrid approaches based on hydrodynamics for the hot and dense stage and hadron transport theory for the late stages where the matter is diluted and runs out of equilibrium \cite{Hirano:2012kj} (and references therein). The initial conditions are the major unknown for the hydrodynamic evolution. A first principle treatment of non-equilibrium QCD to calculate the initial stage of the collision of two nuclei is the ultimate goal which is not yet achieved. The pragmatic approach is therefore to employ the well-constrained dynamical evolution in a hybrid approach and constrain the needed structures in the initial state profile from experimental data. 

Apart from complex numerical modeling to achieve a realistic description of the dynamics of heavy ion collisions, the motivation to study initial state fluctuations is based on an intriguing analogy to the cosmic microwave background measurements. In cosmology, there is only one event, the evolution of the universe, but from the detailed measurement of thousands of multipole components of the temperature fluctuations present today, one is able to deduct detailed knowledge about the matter content in the universe. In heavy ion collisions the time and length scales of the system impose a restriction on the number of measurable multipole components, but in principle there is an almost unlimited number of events to be analyzed \cite{Mocsy:2010um}. To be able to extract transport properties of the quark gluon plasma like the shear viscosity over entropy ratio on a quantitative level a better understanding of the initial conditions and the importance of fluctuations is needed. The above described experimental development triggered very high interest of the whole heavy ion community in initial state fluctuations and resulting final state correlations and the most recent developments are the content of this review article.

\section[Observables]{Observables}
\label{sec_obs}
In each collision event, experiments detect a large number of particles (up to several thousand) of various types exiting the collision region.  The task is to analyze their properties and devise observables that can give insight into the dynamics of the collision, and properties of the medium that is created.  We begin by defining these observables, and how they relate to underlying theoretical concepts.  See \cite{Voloshin:2008dg} for an earlier review.

In theory, one can describe the particles finally emitted from a collision with an underlying probability distribution.  The particles that are detected represent a finite sample of this distribution.  The azimuthal dependence is of great interest, and can be written as a Fourier series in azimuthal angle $\phi$:
\begin{equation}
\label{vn}
  E\frac {dN} {d^3p} \equiv \frac 1 {2\pi} \frac {\sqrt{m^2+p_T^2\cosh(\eta)^2}} {p_T \cosh(\eta)}  \frac {dN} {p_T dp_T d\eta}\left[1+ \sum_{n=1}^\infty v_n \cos n(\phi - \Psi_n)\right],
\end{equation}
where $p_T$ is the transverse momentum of an emitted particle.   The pseudorapidity $\eta\equiv \ln[(|p|+p_z)/(|p|-p_z)]/2$ is a convenient variable to use for the remaining momentum degree of freedom.
The coefficients $v_n(p_T,\eta)$ represent azimuthal anisotropy in the distribution, and the orientation angles $\Psi_n(p_T, \eta)$ are defined such that all sine terms vanish.  

Explicitly, the entire flow vector can be written as a number in the complex plane,
\begin{equation}
\label{flowvector}
v_n e^{in\Psi_n}(p_T,\eta) \equiv \frac {\int_0^{2\pi} d\phi \frac {dN} {d\phi dp_T d\eta} e^{in\phi}} {\int_0^{2\pi} d\phi \frac {dN} {d\phi dp_T d\eta}}.
\end{equation}
For a general distribution, therefore, both $v_n$ and $\Psi_n$ can depend on transverse and longitudinal momentum.  One can also define an average flow vector with magnitude $\bar v_n$ and direction $\bar \Psi_n$ by generalizing the integrals to include an integral over a range of transverse momentum and (pseudo-)rapidity  
\begin{equation}
\label{intflowvector}
\bar v_n e^{in\bar\Psi_n} \equiv \frac {\int d\phi d\eta dp_T \frac {dN} {d\phi dp_T d\eta} e^{in\phi}} {\int d\phi d\eta dp_T \frac {dN} {d\phi dp_T d\eta}}.
\end{equation}

If the collision is symmetric in the transverse plane, all angles $\Psi_n$ are equal to the direction of the impact parameter and all odd harmonics vanish.  In the past this was often assumed to be a good approximation for the distribution at mid-rapidity in a collision between identical ions.  However, event-by-event fluctuations in the early stages of the collision will break this symmetry in the final particle distribution, and are now known to be important.   Further, these fluctuations ensure that there is a different underlying distribution for every collision event (even for events at the same centrality) and every particle species.

A reconstruction of an anisotropy coefficient $v_n$ from $N$ samples of the distribution has a statistical uncertainty $\simeq1/\sqrt{2N}$.  Typically these anisotropies have a value of a few percent, while 
the number of particles detected in a single event is at most a few thousand, resulting in a typical relative uncertainty of 50\% or more in any single-event measurement (even with the large multiplicities and detector coverage at the LHC). Therefore, one must look at correlations that can be averaged over an ensemble of events, usually selected according to a centrality criterion~\cite{Abelev:2013qoq}, to determine moments of the event-by-event distribution of $v_n$ and $\Psi_n$. 

In general,  particles are not emitted independently, and can be correlated with other particles.  For example, the underlying distribution of pairs can be written as
\begin{equation}
\label{pair}
\frac {dN_{\rm pairs}} {d^3p^a d^3p^b} 
= \frac {dN} {d^3p^a} \frac {dN} {d^3p^b} + \delta_2(p^a,p^b).
\end{equation}
There is a contribution from a product of one-particle distributions \eqref{vn}, and in addition there can be an intrinsic correlation between a pair of particles $\delta_2$, defined as the remainder.  The first term only depends on the single-particle probability and represents a correlation of each particle with the global event.  Such contributions represent collective behavior of the system and can be labelled ``flow'', while the second term represents contributions from intrinsic correlations between particles.  These can be  termed ``non-flow'' correlations, since they are absent in purely hydrodynamic calculations (i.e., when particles are emitted independently from a thermal fluid,  neglecting resonance decays, Bose-Einstein correlations, jet correlations, etc).  Many experimental analyses are designed in the hopes of removing or minimizing non-flow correlations, in order to directly measure properties of the underlying one-particle distribution \eqref{vn}, but in principle they can be present in any measurement.

The orientation of each collision is random, so only rotationally invariant quantities can be measured.  The simplest example is the average yield of a particular particle.  However, to get any information about azimuthal dependence, one must look at correlations between more than one particle.   The simplest of these is the pair distribution, whose azimuthal dependence can only be measured as a function of the relative angle between the pair $\Delta\phi$.  The dependence on $\Delta\phi$ can be captured by the set of Fourier coefficients, with each coefficient depending on the transverse and longitudinal momentum (or equivalently pseudorapidity $\eta$) of each of the particles
\begin{align}
\label{Vndelta}
V_{n\Delta} (p_T^a,\eta^a, p_T^b, \eta^b) &\equiv \left\langle\langle \cos n(\phi^a-\phi^b) \rangle_{\rm pairs}\right\rangle\\
\label{twoparticle}
&= \langle v_n^a v_n^b \cos n(\Psi_n^a - \Psi_n^b) \rangle + \langle \delta_{2,n}^{a,b} \rangle 
\end{align}
where the inner brackets represent an average over pairs of particles detected in a collision event and ($a,b$) represent a particular bin in transverse momentum and pseudorapidity to which each particle is restricted, respectively, and over which the underlying probability distribution is integrated to define, e.g., $v_n^a$ and $\Psi_n^a$.  

In this manuscript, unadorned angular brackets always indicate an average over events $\langle \ldots \rangle \equiv \sum_{\rm events} \ldots / N_{\rm events}$. 

The notation $\delta_{2,n}^{a,b}$ refers to the appropriate angular projection of the non-flow correlation $\delta_2$,
\begin{equation}
\delta_{2,n}^{a,b} \equiv \frac 1 {N_{\rm pairs}} \int d^3p^a d^3p^b \delta_2(p^a,p^b)  \cos n(\phi^a-\phi^b),
\end{equation}
where the integrals go over the appropriate phase space of the chosen bins ($a$,$b$).

The second line shows how the measured observable \eqref{Vndelta} is related to the underlying probability distribution~\eqref{vn} and non-flow correlations, using Eq.~\eqref{pair} and assuming each term in Eq.~\eqref{twoparticle} is small.  By calculating Eq.~\eqref{twoparticle} in a given theoretical model, one can  directly compare to the measurement, Eq.~\eqref{Vndelta}.  I.e., as long as one can calculate the underlying single-particle and pair distributions, Eqs.~(\ref{vn}, \ref{pair}), there is no need to simulate details of a detector response and an experimental analysis, or even generate discrete particles. 

The remaining 4 dimensions of the 5-dimensional phase space of pair correlations ($\Delta\phi$, $p_T^a$, $\eta^a$, $p_T^b$, $\eta^b$), plus correlations between different particle species, can be investigated in various ways.  One of the most common is the two-particle cumulant flow measurement $v_n\{2\} $, which restricts only one particle of the pair to a narrow region in transverse momentum and/or pseudorapidity (and often a particular identified particle species), while the second particle in each pair is an unidentified hadron that is allowed to be taken from a wide range in phase space.  This correlation is then divided by the square root of a fully integrated correlation to obtain~\cite{Borghini:2000sa, Luzum:2011mm,Gardim:2012im}
\begin{align}
\label{v22}
v_n\{2\} (p_T, \eta)  &\equiv \frac{\langle V_{n\Delta} (p_T,\eta,p_T^b,\eta^b) \rangle_{p_T^b,\eta^b}} {\sqrt{\langle V_{n\Delta} (p_T^a,\eta^a,p_T^b,\eta^b) \rangle_{p_T^a,\eta^a,p_T^b,\eta^b}}}\\
\label{vn2theory}
&= \frac {\langle v_n(p_T,\eta) \bar{v}_n \cos n(\Psi_n(p_T,\eta) - \bar{\Psi}_n)  \rangle} {\sqrt{\langle \bar{v}_n^2 \rangle}} + \langle \delta_{2,n} \rangle\\
\label{vn2approx}
&\simeq \sqrt{\langle v_n(p_T,\eta)^2 \rangle}.
\end{align}
Equation~\eqref{vn2theory} shows the general sensitivity to the underlying probability distribution, with the integrated flow
$\bar{v}_n$ and $\bar\Psi_n$ defined as in Eq.~\eqref{flowvector}, but integrated over $p_T$ and $\eta$.  The fully momentum-integrated measurement reduces to $\bar v_n\{2\} = \sqrt{\langle \bar v_n^2 \rangle + \langle \bar\delta_{2,n} \rangle}$.  

This is the correct quantity to calculate in a theoretical model, while
Eq.~\eqref{vn2approx} is the result if one makes the following approximations:  non-flow correlations are negligible ($\delta_2\simeq 0$), the event plane depends little on transverse momentum and pseudorapidity ($\Psi_n(p_T,\eta) \simeq \bar\Psi_n$), and event-by-event fluctuations are such that $v_n(p_T,\eta)/\bar v_n$ is approximately constant (e.g., if $v_n$ at all momenta scale only with an eccentricity $\varepsilon_n$).  None of these assumptions are expected to be exactly true, but can be reasonable approximations depending on the situation and desired accuracy.  Lastly, one can assume no event-by-event fluctuations are present, in which case $ v_n\{2\} = \langle v_n \rangle$, but this typically introduces at least a $\sim10$\% error in any model with realistic fluctuations.

\begin{figure}
\begin{center}
\includegraphics[width=0.7\textwidth]{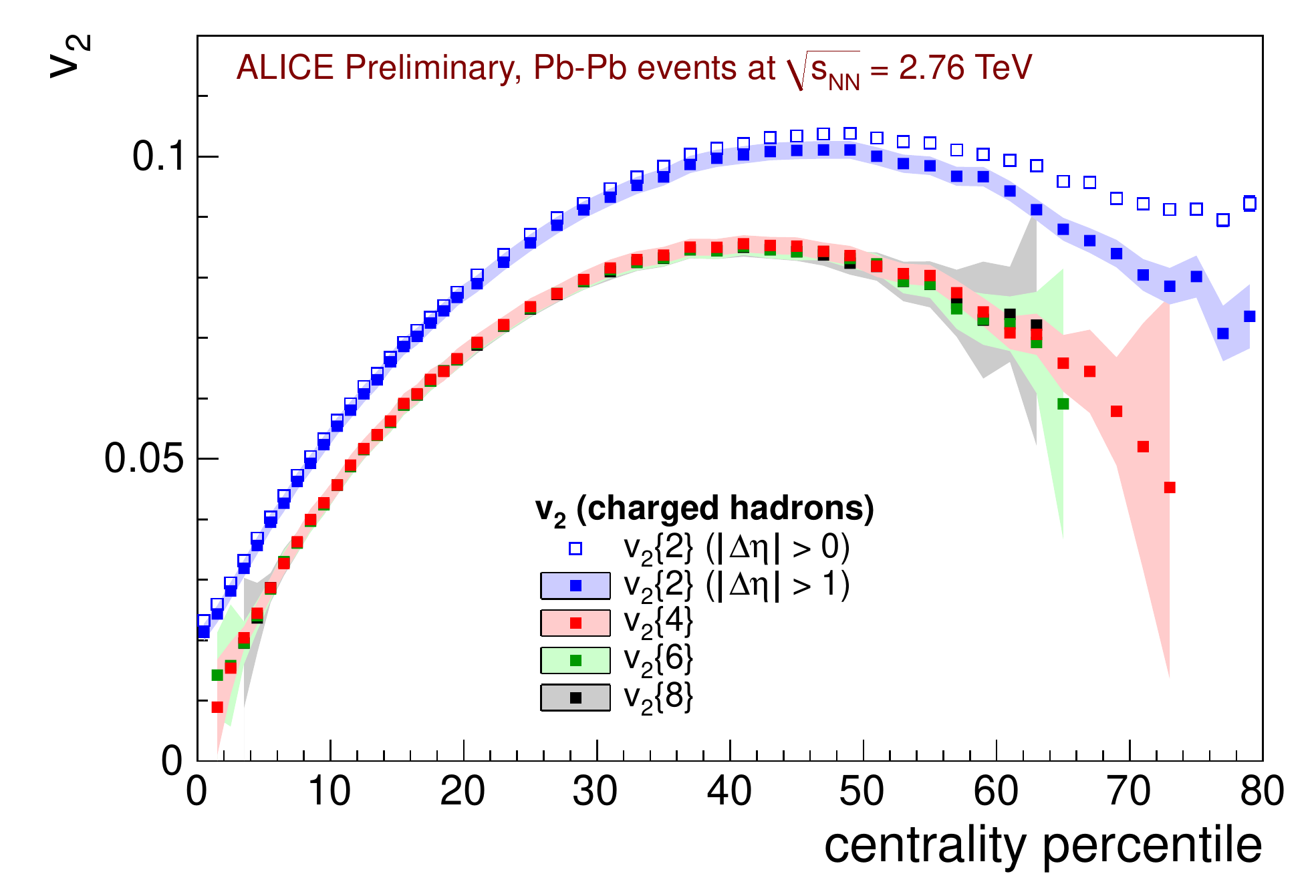}
\end{center}
\caption{Momentum integrated elliptic flow $\bar v_n\{m\}$ measured as a function of centrality percentile, from~\cite{Bilandzic:2011ww}.  Each measurement has a different sensitivity to event-by-event flow fluctuations and to non-flow correlations [see, e.g., Eqs.~(\ref{vn2theory},~\ref{4ptheory})].   Typically, the difference between $v_2\{2\}$ measurements with and without an enforced gap in pseudorapidity is interpreted as a contribution from short-range non-flow correlations, while the difference between $v_2\{2\}$ and higher particle cumulants, and the equality of higher particle cumulants, can be explained by flow fluctuations.}
\label{fig_cumulants} 
\end{figure}

Often the unidentified particles in the second set are given non-equal weights in the average --- one can, e.g., give each particle a weight proportional to its transverse momentum.   In this case, Eq.~\eqref{vn2theory} is the same, except with $\bar v_n$ and $\bar{\Psi}_n$ now derived from a weighted average distribution.  Another common case is when one correlates with the energy deposited in a calorimeter instead of individual particles. This is the same as giving particles a weight proportional to their energy.  

Note that occasionally, this same notation for a two-particle cumulant is used when both particles in the pair are chosen from a particular momentum bin (see, e.g., Fig.~\ref{fig_vn_centrality_atlas}), in which case Eq.~\eqref{vn2approx} is an exact result.  In addition, while in this manuscript we denote all momentum-integrated quantities with a bar (e.g., $\bar v_n\{2\}$), a distinction in notation is not always made (see, e.g., Fig.~\ref{fig_cumulants}, which represents an integrated measurement).
Finally, depending on the details of how the analysis is done, it is also often referred to as a scalar product, and the observable labelled $v_2\{{\rm SP}\}$~\cite{Adler:2002pu}, but it is equivalent to a two-particle correlation.

Many known sources of non-flow correlations are suppressed when the relative longitudinal momentum of the pair is large, and so one of the most common methods to isolate the effect of the underlying anisotropies $v_n$ is to only consider pairs with a large relative pseudorapidity~\cite{Ackermann:2000tr, Adler:2003kt}, in the hopes that any non-flow correlation $\delta_2$ is then negligible.  In Fig.~\ref{fig_cumulants}, one can see the effect of enforcing a gap in pseudorapidity between particle pairs on a measurement of $\bar v_2\{2\}$.  The difference between measurements with and without a rapidity gap is believed to be mostly due to non-flow correlations, which become increasingly important as one moves to the right in the plot toward more peripheral collisions.

A larger space of information opens up with correlations between more than two particles.  
In general, one can measure any $m-$particle correlation of the form~\cite{Bhalerao:2011yg,Bilandzic:2013kga}
\begin{align}
\label{corr}
\langle m \rangle_{n_1, n_2, \ldots, n_m}
&\equiv \left\langle \langle \cos(n_1 \phi^{a_1} + n_2 \phi^{a_2}\ldots + n_m \phi^{a_m}) \rangle_{m\ \rm particles}\right\rangle\\
\label{corrflow}
&\simeq \left\langle v_{n_1}^{a_1} v_{n_2}^{a_2} \ldots v_{n_m}^{a_m} \cos(n_1\Psi_{n_1}^{a_1} + n_2\Psi_{n_2}^{a_2}\ldots n_m \Psi_{n_m}^{a_m}) \right\rangle 
\end{align}
with $\sum n_i = 0$ (the angular structure is again restricted because one can only measure rotationally-invariant quantities).  The labels $a_i$ again represent bins in phase space and/or particle species.

In the absence of non-flow correlations, the observable \eqref{corr} relates to the underlying single-particle distributions by Eq.~\eqref{corrflow}.

However, in general the underlying $m$-particle distribution can again depend on both flow and non-flow contributions.  E.g.,
\begin{align}
\label{3p}
\frac {dN_{\rm triplets}} {d^3p^a d^3p^b d^3p^c} 
=& \frac {dN} {d^3p^a} \frac {dN} {d^3p^b} \frac {dN} {d^3p^c}\nonumber\\
 &
 +  \frac {dN} {d^3p^a} \delta_2(p^b,p^c)
 +  \frac {dN} {d^3p^b} \delta_2(p^a,p^b)
 +  \frac {dN} {d^3p^c} \delta_2(p^a,p^b)\nonumber\\
&+ \delta_3(p^a,p^b,p^c),\\
\label{4p}
\frac {dN_{\rm quad.}} {d^3p^a d^3p^b d^3p^c d^3p^d}
=&  \frac {dN} {d^3p^a} \frac {dN} {d^3p^b} \frac {dN} {d^3p^c} \frac {dN} {d^3p^d}
+ \frac {dN} {d^3p^a} \delta_3(p^b,p^c, p^d)
+ \frac {dN} {d^3p^b} \delta_3(p^a,p^c, p^d)\nonumber\\
&+ \frac {dN} {d^3p^c} \delta_3(p^a,p^b, p^d)
+ \frac {dN} {d^3p^d} \delta_3(p^a,p^b, p^c)\nonumber\\
&+ \frac {dN} {d^3p^a} \frac {dN} {d^3p^b}  \delta_2(p^c,p^d)
+ \frac {dN} {d^3p^a} \frac {dN} {d^3p^c}  \delta_2(p^b,p^d)
+ \frac {dN} {d^3p^a} \frac {dN} {d^3p^d}  \delta_2(p^b,p^c)\nonumber\\
&+ \frac {dN} {d^3p^b} \frac {dN} {d^3p^c}  \delta_2(p^a,p^d)
+ \frac {dN} {d^3p^b} \frac {dN} {d^3p^d}  \delta_2(p^a,p^c)
+ \frac {dN} {d^3p^c} \frac {dN} {d^3p^d}  \delta_2(p^a,p^b)\nonumber\\
&+ \delta_2(p^a,p^b) \delta_2(p^c,p^d)
+ \delta_2(p^a,p^c) \delta_2(p^b,p^d)
+ \delta_2(p^a,p^d) \delta_2(p^b,p^c)\nonumber\\
&+ \delta_4(p^a,p^b,p^c,p^d)
,
\end{align}
etc.
So the correlations~\eqref{corr} depend on potentially complicated combinations of flow and non-flow.  On can again suppress the non-flow correlations with strategically-placed gaps in pseudorapidity between two or more of the particles in the muliplet~\cite{Bhalerao:2011yg,Bhalerao:2013ina}.
However, one can also suppress non-flow correlations by considering particular combinations of correlations.
Most common are $m$-particle cumulants, which are designed to suppress non-flow correlations of order less than $m$~\cite{Borghini:2000sa}.

For example, a momentum-integrated four-particle cumulant is:
\begin{align}
\label{4pcum}
\bar v_n\{4\}^4 &\equiv 2\bar v_n\{2\}^4 - \langle \langle \cos n(\phi_1 + \phi_2 - \phi_3 - \phi_4) \rangle_{\rm quadruplets} \rangle\\
\label{4ptheory}
&=2\langle \bar v_n^2 + \delta_{2,n} \rangle^2 - \langle \bar v_n^4+2\delta_{2,n}^2+4\bar v_n^2\delta_{2,n} + 2 \delta^v_{2,n} + 4 \delta^v_{3,n} +\delta_{4,n} \rangle\\
\label{4pflow}
&\overset{\rm{(flow)}}{\simeq} 2\langle \bar v_n^2 \rangle^2 - \langle \bar v_n^4 \rangle\\
\label{4pnonflow}
&\overset{\rm{(non-flow)}}{\simeq} 2\left(\langle \delta_{2,n} \rangle^2 - \langle \delta_{2,n}^2 \rangle\right) 
- \langle \delta_{4,n} \rangle
,
\end{align}
where Eq.~\eqref{4ptheory} is obtained from Eqs.~\eqref{3p} and \eqref{4p} and represents the general dependence on the underlying probability distribution.  

Equation \eqref{4pflow} is the result when there are only flow correlations.  When event-by-event fluctuations of the flow vector are small or are distributed as a 2D Gaussian~\cite{Voloshin:2007pc}, this is the average flow vector projected onto the direction of the impact parameter (the ``reaction plane''), $\langle v_n \cos n (\phi- \Psi_{RP}) \rangle$.  For non-central collisions, therefore, it can be large (though in principle Eq.~\eqref{4pflow} can be negative, even in the absence of non-flow).  

Equation \eqref{4pnonflow} is the result when there are only non-flow correlations.  This quantity is typically expected to be small, and so cumulants can be a powerful test of the nature of azimuthal correlations.

When non-flow and flow are both present, there can be a contribution from three-particle non-flow correlations and an additional angular component of a two-particle correlation, coupled to flow,
\begin{align}
\delta^v_{2,n} &\equiv \frac {\bar v_n^2} {N_{\rm pairs}} \int d^3p^a d^3p^b\ \delta_2(p^a,p^b)  \cos n(\phi^a+\phi^b - 2 \bar\Psi_n)\\
\delta^v_{3,n} &\equiv \frac {\bar v_n} {N_{\rm triplets}} \int d^3p^a d^3p^b d^3p^c\ \delta_3(p^a,p^b,p^c)  \cos n(\phi^a+\phi^b - \phi^c - \bar\Psi_n),
\end{align}
in addition to the term ($\langle \bar v_n^2 \delta_{2,n}\rangle - \langle \bar v_n^2 \rangle \langle \delta_{2,n} \rangle$).  

It should be noted, therefore, that in most cases, non-flow correlations are not completely removed by constructing cumulants, even though they are typically suppressed.

Similarly, higher-order cumulants $v_n\{m\}$ have been measured, 
in addition to $p_T$-differential measurements \cite{Borghini:2000sa}. 
E.g., for reference, a differential fourth cumulant is
\begin{align}
\label{diff4pcum}
v_n\{4\}(p_T,\eta)  &\equiv \frac {2 v_n\{2\}(p_T,\eta) \bar v_n\{2\}^3 - \left\langle \langle \cos n(\phi + \phi_1 - \phi_2 - \phi_3) \rangle_{\rm quadruplets} \right\rangle} {\bar v_n\{4\}^3}\\
\label{diff4cumth}
&\simeq \frac {2\left\langle v_n(p_T,\eta) \bar v_n \cos n(\Psi_n - \bar \Psi_n)\right\rangle \left\langle \bar v_n^2  \right\rangle -  \left\langle v_n(p_T,\eta) \bar v_n^3 \cos n (\Psi_n - \bar \Psi_n) \right\rangle} 
 {\left(2\langle \bar v_n^2 \rangle^2 - \langle \bar v_n^4 \rangle\right)^{3/4}},
\end{align}
where only one of the particles in each quadruplet of the second term of Eq.~\eqref{diff4pcum} is restricted to be an identified or unidentified particle in the desired phase space bin.   In the absence of non-flow, this measures Eq.~\eqref{diff4cumth}, where $v_n(p_T,\eta)$ and $\Psi_n(p_T,\eta)$ are the underlying differential flow coefficient and event plane of the particle of interest, while $\bar v_n$ and $\bar \Psi_n$ are again the momentum-integrated flow and event plane of unidentified charged hadrons.

Note that, although different cumulants have different sensitivity to non-flow,  when there are event-by-event fluctuations in $v_n$, they also measure different moments of the event-by-event distribution, and so one can not isolate non-flow correlations by simply comparing, e.g., $v_n\{2\}$ and $v_n\{4\}$.  In fact, when there are fluctuations, there is no rigorous way to separate flow and non-flow (as was hoped to be the case for cumulants when fluctuations were thought to be negligible).  On the other hand, if non-flow is negligible or can be controlled in other ways, these cumulants each give an independent handle on the nature of flow and event-by-event flow fluctuations.   For example, as seen in Fig.~\ref{fig_cumulants}, measured cumulants $v_2\{m\}$ of order $m\geq 4$ are all equal to each other within the reported error bars~\cite{Bilandzic:2011ww}.  This is exactly what is expected if non-flow is negligible and if flow fluctuations are Gaussian~\cite{Voloshin:2007pc} --- i.e., if the vector defined by $v_n$ and $\Psi_n$ fluctuates event-by-event according to a two-dimensional Gaussian within a given centrality class.  The relationship between $v_2\{2\}$ and $v_2\{4\}$ then indicates the magnitude of the fluctuations, which are directly related to properties of the initial stage of the collision~\cite{Bhalerao:2011yg}.

A number of other $m-$particle correlations have now been measured, with many more to come \cite{Bilandzic:2013kga}, giving a large number of independent constraints on theory.  Especially notable are various mixed harmonic correlations, which contain information about flow angles $\Psi_n$~\cite{Bilandzic:2012an}.

The finite multiplicity in each event limits the number of particles $m$ in Eq.~\eqref{corr}, as well as the harmonics $n_i$, leaving a finite (though large) number of possible observables.  In principle, all information about correlations between particles is contained in the complete set of measurements of this form (suitably generalized to include momentum-dependent particle weights and multiplicity-dependent event weights).  However, it is often useful to cast this information in different forms.

As an example, if one considers momentum-integrated correlations of the form \eqref{corr} and assumes non-flow correlations are negligible, one can directly measure all even moments of the event-by-event distribution of a flow coefficient,  
$\langle \bar v_n^{2k} \rangle$.  A quantity such as $\langle \bar v_n \rangle$ or a plot of the entire flow-angle-averaged distribution, however, can only be obtained indirectly.

In principle, it can be reconstructed from this full set of (even) moments.  
Alternatively, it can be obtained by an unfolding analysis, as follows.
First one defines an observed single-event anisotropy $v_n^{\rm obs}$, and plots its event-by-event distribution.  In the absence of non-flow correlations, this distribution represents the true distribution of $\bar v_n$ folded with a statistical smearing due to the finite number of detected particles.  If the form of the statistical fluctuations is known, the true distribution can be obtained by an unfolding procedure.  This can be done by assuming a particular form for the response function (from, e.g., a Monte Carlo calculation)~\cite{Alver:2006zz, Timmins:2013hq} or by extracting it from other data, such as the correlation between the observed anisotropy at forward and backward rapidities~\cite{Aad:2013xma}, since the difference is expected to be mostly due to statistical fluctuations rather than the (weak) rapidity dependence of the underlying distribution $v_n(\eta)$.

\begin{figure}
\begin{center}
\includegraphics[width=0.36\textwidth]{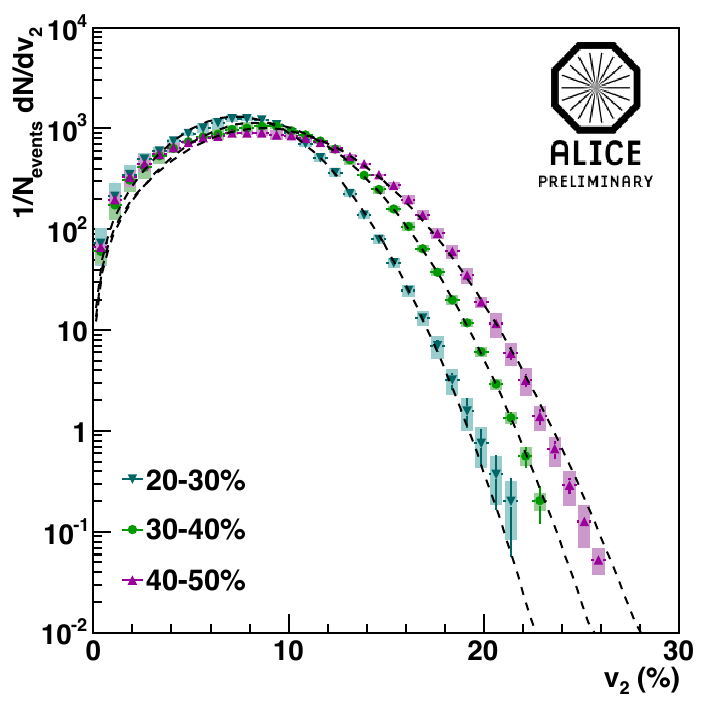}
\includegraphics[width=0.63\textwidth]{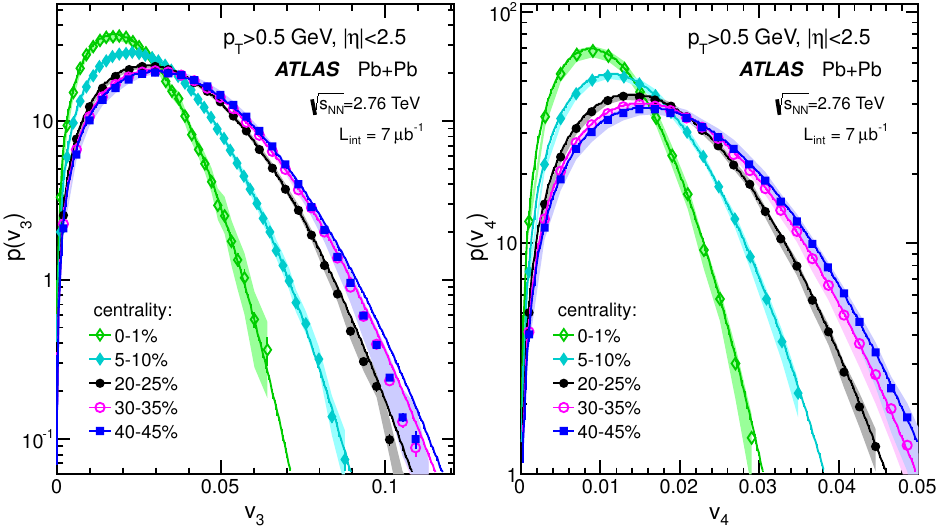}
\end{center}
\caption{\label{fig_unfolded} Unfolded event-by-event distributions of $v_2$ \cite{Timmins:2013hq}, $v_3$, and $v_4$ \cite{Aad:2013xma}.} 
\end{figure}

After the unfolding, one has an entire event-by-event distribution of $\bar v_n$ \cite{Alver:2006zz, Alver:2007qw, Timmins:2013hq, Aad:2013xma}, as shown in Fig. \ref{fig_unfolded}.  Interestingly, although the directly-measured cumulants of the distribution appear to be consistent with Gaussian fluctuations as mentioned above, the unfolded distribution of $v_2$ shows a deviation at peripheral centralities~\cite{Aad:2013xma}.   It is not yet known whether this information was actually contained in the previously-measured cumulants (errors are correlated, so higher cumulants may not actually be consistent with each other despite overlapping error bars), or if there is a systematic difference in the two analyses.%

Historically, one of the most common measurements has been the event-plane anisotropy coefficient $v_n\{\rm EP\}$.  In these analyses, one identifies an observed event plane $\Psi_n^{\rm obs}$ from each of two or more subsets of particles (``subevents'', labeled, e.g., $A$ and $B$) in each event:

\begin{equation}
\Psi_n^{A,B} \equiv \frac 1 n \tan^{-1}\left(\frac {\sum_{i\in A,B}  w_i\sin n \phi_i} {\sum_{i\in A,B}  w_i\cos n \phi_i} \right),
\end{equation}
where the sum goes over the chosen set of particles (usually all charged hadrons without cuts on momentum) and $w_i$ is a possible momentum- or energy-dependent weight.

Particles of interest are then correlated with this azimuthal angle.  In the simplest case of two subevents, the final observable is:

\begin{align}
\label{v2EP}
v_n\{\rm EP\} &\equiv \frac {\left\langle \cos n (\phi_i - \Psi_n^A) \right\rangle_{\rm particles, events}} { \sqrt{\left\langle\cos n (\Psi_n^A - \Psi_n^B)\right\rangle}},\\
\label{lowres}
&\xrightarrow[\text{low res.}]{} v_n\{2\},\\
\label{highres}
&\xrightarrow[\text{high res.}]{} \langle v_n \rangle,
\end{align}
where the average in the numerator is over all particles of interest in a class of events, and the average in the denominator is the usual average over events.  Unlike the above observables, there is no general formula for what this observable measures in terms of the underlying flow distribution and non-flow correlations.  It depends on what is called the event-plane resolution.  In the low resolution limit, as the number of particles $N$ used to estimate the event plane goes to one, it coincides exactly with a two-particle correlation (including the same non-flow contribution), as indicated in Eq.~\eqref{lowres}.  
As the resolution increases, the measurement begins to have a contribution from correlations between more than two particles, and  
in the high resolution limit, when the square root of the number of particles is much larger than the inverse of the anisotropy, $\sqrt{N} v_n\gg 1$,  one measures the mean anisotropy, \eqref{highres}~\cite{Alver:2008zza}.  
This limit typically differs from the low-resolution limit by $\sim$10\% or more.

In most cases, the resolution is low enough that the result is essentially a two-particle cumulant.  However, in some cases the resolution is high enough to impart a non-negligible deviation from the low-resolution result (roughly when the reported resolution parameter --- the square of the denominator in Eq.~\eqref{v2EP} --- $\mathcal R\gtrsim$ 0.8--0.9).  
See, e.g., Fig.~\ref{fig_vn_centrality_atlas}.  The high multiplicity of high-energy collisions at the LHC combined with the large acceptance of the ATLAS detector allows the resolution for $v_2$ to exceed 0.9 in mid-peripheral collisions, resulting in a small difference between the event-plane result and the two-particle cumulant (note the log scale).
One can estimate the deviation from the low resolution limit based on this experimental resolution factor~\cite{Ollitrault:2009ie}, but one must make some general assumptions.

It should therefore be explicitly noted that, although the denominator of Eq.~\eqref{v2EP} is often referred to as a `resolution correction', the final result still has a systematic dependance on the resolution, i.e.~the number of particles used in a particular analysis to obtain the reconstructed event planes $\Psi_n^{A,B}$, whenever there are event-by-event fluctuations.  In principle, this adds an ambiguity when comparing different measurements.

\begin{figure}
\begin{center}
\includegraphics[width=1.0\textwidth]{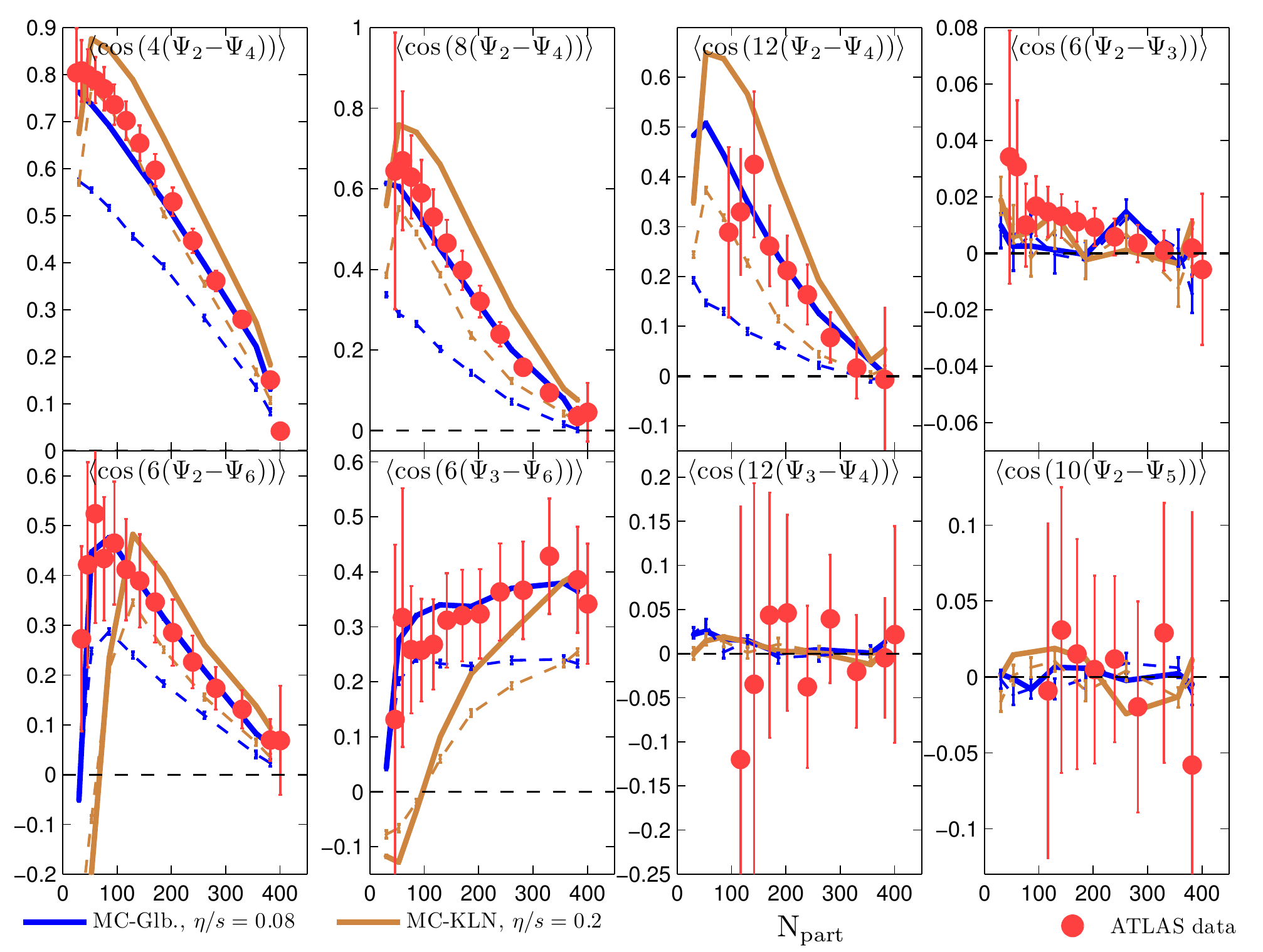}
\end{center}
\caption{\label{fig_mixed} Mixed harmonic correlators measured with the event-plane method by the ATLAS Collaboration \cite{Jia:2012sa} compared with event-by-event viscous hydrodynamic calculations using two models for initial conditions.  The labels and dashed lines~\cite{Qiu:2012uy} represent a calculation of the quantity that would be measured in the high-resolution limit, while the solid lines~\cite{HeinzPrivate} are calculations of the low-resolution limit, Eq.~\eqref{mixedlowres}.  In general it is expected that the measurements represent a quantity that is in between these two limits, but closer to the low-resolution limit~\cite{Bhalerao:2013ina}.} 
\end{figure}

This same method has been used to measure various mixed-harmonic observables.  The projection of both $v_1$ and $v_4$ have been measured with respect to the second event plane $\Psi_2$ at RHIC~\cite{Back:2005pc, Abelev:2007qg, Adare:2010ux}, while more recently a large set of mixed harmonic event-plane correlations were measured at the LHC~\cite{Jia:2012sa}, providing a large number of independent constraints to theory.  Unfortunately, the resolution dependence for mixed-harmonic correlations is much larger than for single-harmonic event plane measurements, and so it is more difficult to properly compare theoretical calculations to these results~\cite{Luzum:2012da}, compared to the analogous mixed-harmonic $m$-particle correlations~\eqref{corr}.  See, e.g., Fig.~\ref{fig_mixed}.  For this reason, the use of traditional event-plane methods is declining, in favor of multiparticle correlations of the form~\eqref{corr} (e.g., $v_n\{2\}$ and $v_n\{\rm SP\}$).

For reference, the low-resolution limit of the mixed-harmonic event-plane correlations as analyzed by ATLAS~\cite{Jia:2012sa}, in the absence of non-flow correlations, is:
\begin{equation}
\label{mixedlowres}
\cos (k_1 \Psi_{1} + 2 k_2 \Psi_2 \ldots +  n k_n \Psi_{n})\{{\rm EP}\}
\xrightarrow[\text{low res.}]{} 
\frac{
\left\langle v_1^{k_1} \ldots v_n^{k_n}\cos(k_1\Psi_{1} + \ldots  + n k_n \Psi_{n}) \right\rangle
}
{
\sqrt{
\langle v_1^{2 k_1}\rangle \langle v_2^{2 k_2}\rangle \ldots \langle v_n^{2 k_n} \rangle
}
},
\end{equation}
where the coefficients $k_i$ are always integers satisfying $\sum i k_i = 0$.  This will also be the correct quantity with which to compare
when the analysis is redone with a scalar product method or $m$-particle correlation.

Finally, it should be mentioned that there are often slight differences in experimental analysis compared to the definitions presented here.  In particular, if multiplicity fluctuations are important within a given centrality bin, one may need to include in a theoretical calculation factors of the yield inside the event averages, depending on the details of the experimental analyses.
\section{Hydrodynamics}
\label{sec_hydro}

In the following Sections we will describe the ingredients for a theoretical description of the dynamical evolution of heavy ion reactions starting with the basics of relativistic fluid dynamics.   
For recent reviews see Refs.~\cite{Heinz:2013th,Gale:2013da}.

If the constituents of the collision fireball interact strongly enough, 
the system will behave as a fluid.  Indeed, calculations based
on viscous hydrodynamics have been very successful at describing measured 
observables, and this has become the standard picture of the bulk evolution of a relativistic 
heavy-ion collision.

The relativistic theory of viscous fluid dynamics has been a significant topic of study
in high-energy physics in recent years.  Here, we include only a brief sketch of the main ideas.
For a more thorough discussion see, e.g.,  Refs.~\cite{Romatschke:2009im,Denicol:2012cn}.  

Fluid dynamics is closely connected to the idea of local thermal equilibrium.  If a system is in 
thermal equilibrium, its macroscopically-averaged energy and momentum are distributed isotropically
it its rest frame.  In this frame, there is no net momentum, and so its energy-momentum tensor has the form
\begin{equation}
\label{idealrest}
T^{\mu\nu}_{\rm ideal, rest} = \left( \begin{array}{cccc}
          \epsilon & 0 & 0 & 0 \\
                      0 & p & 0 & 0 \\
                      0 & 0 & p & 0 \\
                      0 & 0 & 0 & p 
                     \end{array} \right).
\end{equation}
The only two remaining degrees of freedom can be identified as the energy density $\epsilon$ and pressure $p$.

\textit{Local} thermal equilibrium is a generalization of thermal equilibrium, when the rest frame and thermodynamic properties of a 
system are allowed to vary in space and time, though slowly enough that every local region of the system  can be 
said to be arbitrarily close to thermal equilibrium.  In a general lab frame related to the local rest frame by a four-velocity
$u^\mu$, Eq.~\eqref{idealrest} becomes
\begin{equation}
\label{idealTmunu}
T_{\rm ideal}^{\mu\nu} =  (\epsilon+p)\ u^{\mu} u^{\nu} - p\ g^{\mu\nu}, 
\end{equation}
where $u^\mu$, $\epsilon$, and $p$ are now functions of spacetime, and we have chosen a mostly negative Minkowski metric $g^{\mu\nu} = diag(1,-1,-1,-1)$.  The equations of ideal relativistic hydrodynamics
are obtained by demanding local conservation of energy and momentum 
\begin{equation}
\partial_\mu T^{\mu\nu} = 0.
\end{equation}
Any other conserved quantity is governed by its own conservation equation
\begin{equation}
\partial_\mu j_i^\mu = 0,
\end{equation}
which also has a simple form when a system is in local thermal equilibrium
\begin{equation}
j^\mu_{i,\rm ideal} = \rho_i u^\mu,
\end{equation}
where $\rho_i$, is the conserved charge density in the local rest frame.

These equations describe universal behavior of any system with sufficient separation of scales between
 macroscopic scales and the microscopic dynamics that bring the system toward equilibrium.  The only
connection to the particular physical system is the relationship between $\epsilon$, $p$, and $\rho_i$, which for 
a system in local thermal equilibrium is the equation of state $p(\epsilon, \rho_i)$.  This relation closes the above
set of equations, and the evolution of the system is completely determined.

This framework can be generalized to systems with a small departure from local equilibrium, which exhibit
dissipative effects. 
In this case, the ideal stress-energy tensor and conserved current include a small additive correction
\begin{align}
T^{\mu\nu} & = T_{\rm ideal}^{\mu\nu}  + \Pi^{\mu\nu}\\
j_i^\mu &= j^\mu_{i, \rm ideal} + \delta j_i^\mu.
\end{align}
In general the independent terms in these corrections 
each contain an associated transport coefficient, such as shear viscosity $\eta$, bulk viscosity $\zeta$, or 
charge diffusion coefficients $\kappa_i$:
\begin{align}
\label{NS}
\Pi^{\mu\nu}\ &= \eta \nabla^{\langle \mu} u^{\nu\rangle} + \zeta \partial_\alpha u^\alpha \Delta^{\mu \nu} + \ldots \\
\label{NSj}
\delta j_i^\mu &=  \kappa_i \nabla^\mu \frac {\mu_i} T  + \ldots
\end{align}
Here we use convenient notation for a projector orthogonal to the four velocity $\Delta^{\mu \nu}  \equiv g^{\mu \nu}-u^\mu u^\nu$, which can be used to define a spatial gradient in the local rest frame $\nabla_\mu = \Delta_\mu^\alpha \partial_\alpha$, and the brackets indicate a symmetrized, traceless tensor $\nabla^{\langle\mu} u^{\nu\rangle}\equiv \nabla^{\mu} u^{\nu} + \nabla^{\nu} u^{\mu} -\frac{2}{3} \Delta^{\mu \nu} \nabla_\alpha u^\alpha$.    So Eq.~\eqref{NS} contains the two possible independent terms containing a single derivative of the ideal hydrodynamic quantities in Eq.~\eqref{idealTmunu}, which is usually separated as the traceless (shear) and scalar (bulk) part.
Each conserved current has an associated thermodynamic chemical potential $\mu_i$, related to the charged density $\rho_i$ and temperature $T$ by a thermodynamic relation, and all of these coefficients can depend on, e.g., the local energy and charge densities. 

It turns out that these first-order equations are unstable, so to perform numerical calculations one typically uses a second order formulation, represented by ellipses in Eqs.~\eqref{NS} and \eqref{NSj}, of which slightly different variants exist~\cite{Romatschke:2009im, PeraltaRamos:2009kg}.  
The extra terms typically involve time derivatives that transform the equations into relaxation equations that are then stable and solvable with numerical algorithms.
These equations also contain additional second-order transport coefficients.  Typically their effect is not large~\cite{Luzum:2008cw} (but also not negligible~\cite{Luzum:2012wu}).

More recently, there has been a development of ``anisotropic hydrodynamics'', formulated around an anisotropic momentum distribution, which may be a better description of a collision system,  in particular at its earliest stages when the system may be far from an isotropic energy-momentum distribution \eqref{idealTmunu} \cite{Martinez:2012tu}.  
In addition, a thermal system contains intrinsic fluctuations, which can contribute to the event-by-event fluctuations generated from the initial conditions~\cite{Kapusta:2011gt}.

Like the equation of state, the transport coefficients are determined by the underlying dynamics of the medium in question (e.g., a high-temperature quark-gluon plasma).  The rule of thumb is that stronger interactions result in smaller transport coefficients (e.g., viscosity).  The formalism is then valid if the length or time scale set by these transport coefficients is much smaller than the macroscopic length or time scale set by the associated gradients.  

Hence, these transport coefficients are of significant interest in a heavy-ion collision, to learn
about the dynamics of QCD matter near the deconfinement phase transition, and the
success of hydrodynamic descriptions has lead to the standard picture of a strongly-coupled,
low-viscosity quark-gluon plasma having been created.
 
The equation of state at zero net baryochemical potential and finite temperature can be reliably calculated using Lattice QCD methods~\cite{Borsanyi:2010bp,Bazavov:2011nk}.  However, 
transport coefficients are much more difficult to calculate.  Typically, they are treated as
free parameters in calculations, and experimental data can then provide constraints. So a system in a hydrodynamic regime has universal behavior, with only a few parameters
that represent the physics of the particular system [$p(\epsilon, \rho_i)$, $\eta(\epsilon, \rho_i)$, $\zeta(\epsilon, \rho_i)$, $\kappa_i(\epsilon, \rho_i)$].  

However, these equations require
an initial condition --- the values of the hydrodynamic variables 
($\epsilon$, $u^\mu$, $\Pi^{\mu\nu}$,  $\rho_i$, $\delta j_i^\mu$)  after the system has sufficiently 
thermalized at the beginning of the collision.   The hydrodynamic equations then describe how the system evolves forward in time. Likewise, the end result of a collision is a set of particles that hit the detectors, whose distribution, Eq.~\eqref{vn}, we would like to calculate.
At some point, then, one must switch from a fluid description to a description of the system in terms of particles.  

We discuss these stages of the collision in the following sections.

\section{Hadronic Rescattering and Freeze-Out}
\label{sec_foresc}
At later stages in the heavy ion reaction, a fluid description will eventually break down.  However, if there is some
region where a fluid and a kinetic description are both valid, we can calculate the local distribution of particles $f(x,p)$ in the kinetic theory
that corresponds to the values of the hydrodynamic variables in the fluid description.  This distribution can then be modified by
further collisions and decays, with the evolution described using the same transport theory, to obtain the final momentum space distribution~\eqref{vn}. The basic idea to match fluid dynamical calculations to hadron transport models has been developed around the year 2000 \cite{Bass:2000ib,Teaney:2001av}. 

Since then, so called hybrid approaches that combine the advantages of hydrodynamics and kinetic transport have become the primary choice for developing dynamical models that are able to capture the whole evolution from initial to the final stages of the heavy ion collisions \cite{Nonaka:2006yn,Hirano:2005xf,Hirano:2007ei,Petersen:2008dd,Werner:2010aa,Song:2010mg,Ryu:2012at}. A recent review article about hybrid approaches can be found here \cite{Hirano:2012kj}. The typical effects of the final state hadronic rescatterings are: 

\begin{itemize}
\item
an increase of the mean transverse momentum/radial flow by up to 30\% (for protons),
\item
the mass splitting effect for anisotropic flow is more pronounced,
\item
the separation of chemical and kinetic freeze-out is dynamically taken into account.
\end{itemize}

In general, the switching procedure between hydrodynamics and hadron transport should happen in a regime where both descriptions are equally applicable and yield similar results for the bulk evolution. In practice, it is important to ensure that the degrees of freedom do not change during the change of theoretical description. Hadronization is taken care of by a change of the equation of state during the hydrodynamic evolution. Then, some switching criterion has to be defined: In most cases, a constant temperature on the order of 150-170 MeV is chosen, but an energy density criterion or an isochronous transition might also be good approximations. 

In practice, the hydrodynamic evolution is performed completely, and afterward a hypersurface finder is employed to extract a smooth transition surface from hydrodynamics to the transport description. The finding of the hypersurface is rather straight forward in lower dimensions and for smooth initial conditions, but becomes more complex for 3+1 dynamical calculations including event-by-event fluctuations. It is important to employ an algorithm that avoids double-counting and does not leave holes in the surface. One example for a sophisticated implementation is Cornelius, that is publicly available on the OSCAR code archive (\emph{https://karman.physics.purdue.edu/OSCAR}) as a Fortran90 and C++ routine \cite{Huovinen:2012is}. 

On this hypersurface, the distribution of particles is given by the Cooper-Frye formula that describes the flux of particles through the surface \cite{Cooper:1974mv}
\begin{equation}
E \frac{dN}{dp^3}= \frac {g} {(2\pi)^3}  \int_\sigma d\sigma_\mu p^\mu f(x,p)
\end{equation} 
where $f(x,p)$ is the grand-canonical boosted particle distribution function in each hypersurface element, with normal $d\sigma_\mu$, and the total momentum distribution is an integral over the entire surface $\sigma$.  Here, $g$ is the degeneracy of the particle in question.

In any equilibrium system with a kinetic theory description, the distribution function is a Fermi-Dirac or Bose-Einstein distribution for a particular species of fermion or boson, respectively, corresponding to the temperature $T$ and chemical potential $\mu$,
\begin{equation}
f_{eq}(x,p) = \frac {1} {e^{(p_\mu u^\mu(x)-\mu(x))/T(x)} \pm 1} ,
\end{equation}
where $u^\mu$ is the fluid four-velocity.

In the limit of ideal hydrodynamics, then, this process of switching from a fluid to a particle description is well understood.  However, a system with a small departure from equilibrium --- i.e., a system described by viscous hydrodynamics --- differs from the equilibrium distribution: $f(p^\mu) = f_{eq}(p\cdot u) + \delta f(p^\mu)$.    The viscous correction $\delta f$ is not universal, and depends on the particle interactions in a particular system.  In principle it could even depend on the history of the evolution, rather than only the value of local hydrodynamic variables.   What is used most often in practice for the case of shear viscosity is the ``quadratic'' and ``democratic'' ansatz, referring to the assumption for the momentum dependence and that the form is the same for every particle species:
\begin{equation}
\delta f (p^\mu) = f_{eq}(1\pm f_{eq}) \frac {p_\mu p_\nu \Pi^{\mu\nu}} {T^2(\epsilon + p)}\quad .
\end{equation}
However, there is no reason to believe it is correct for a hadronic system at the later stages of a heavy-ion collision, and it is a current topic of research what is the correct form of $\delta f$.  Typically, altering the form of the viscous correction gives a small change to momentum-integrated quantities, but can have a significant effect on differential measurements.

The Cooper-Frye formula can in principle have both positive and negative contributions, representing particles being emitted from the fluid, and flowing into the fluid, respectively.  However, in a typical simulation of a heavy-ion collision, the expansion is so fast, that the negative contributions are on the order of 5\% or even smaller and can be neglected.

The resulting distribution function can then be evolved with the Boltzmann equation, which is valid in principle for the remainder of the system lifetime.  The most common way to do this is to sample a finite number of particles from the local distribution function on the Cooper-Frye surface, and evolve them forward in time with a cascade algorithm.  This method has the advantage of allowing for the addition of (non-flow) correlations, which are not present in a purely fluid description, in a straightforward way.
 
For the sampling procedure itself, there are many different algorithms developed by different groups \cite{Huovinen:2012is} (and references therein). It would be desirable to standardize the procedure, clarify the crucial steps and create an efficient routine that is available as open source code. Numerical efficiency is especially important for event-by-event calculations as they are needed for a quantitative understanding of many observables in heavy ion collisions.

\begin{figure}[h]
\includegraphics[width=0.5\textwidth]{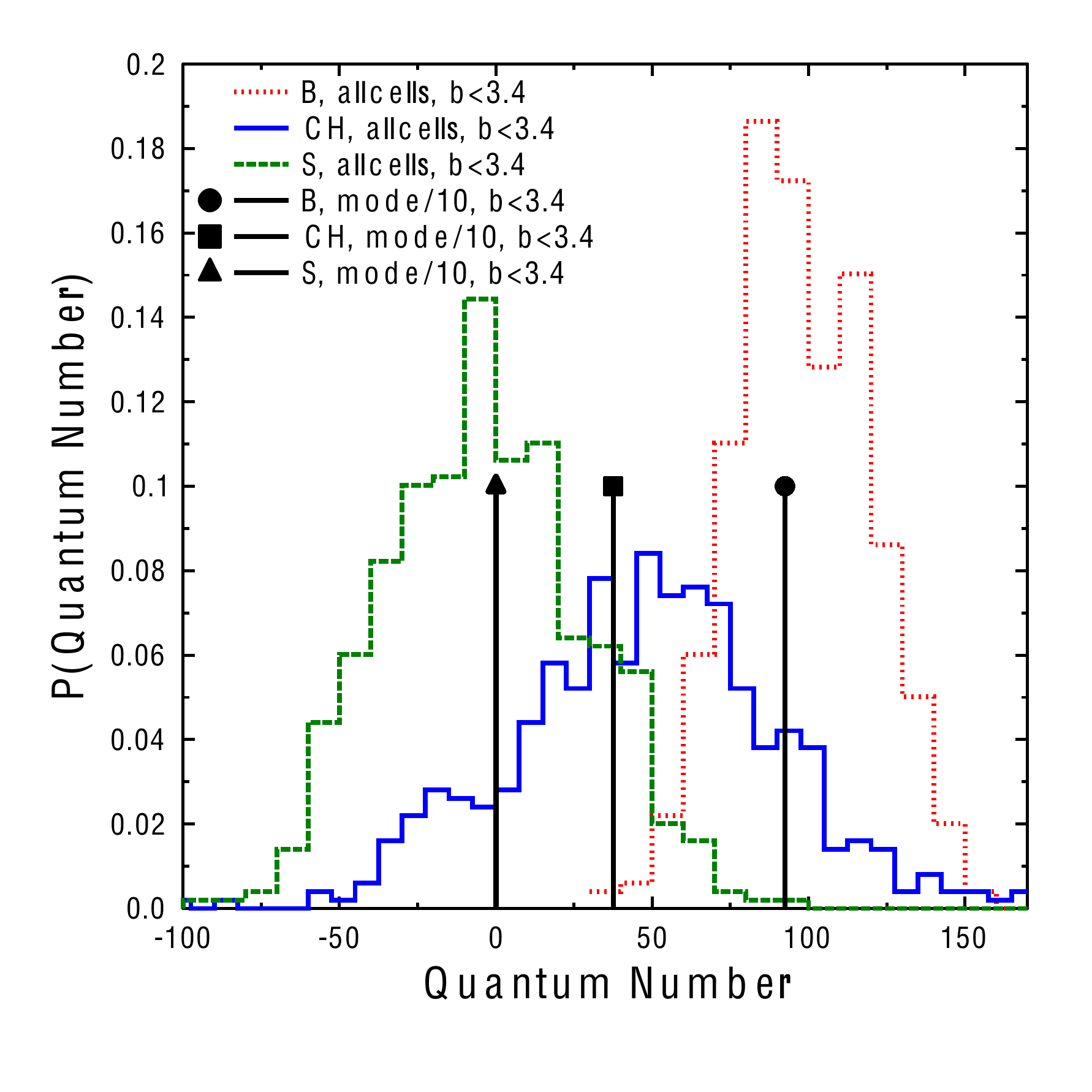}
\includegraphics[width=0.5\textwidth]{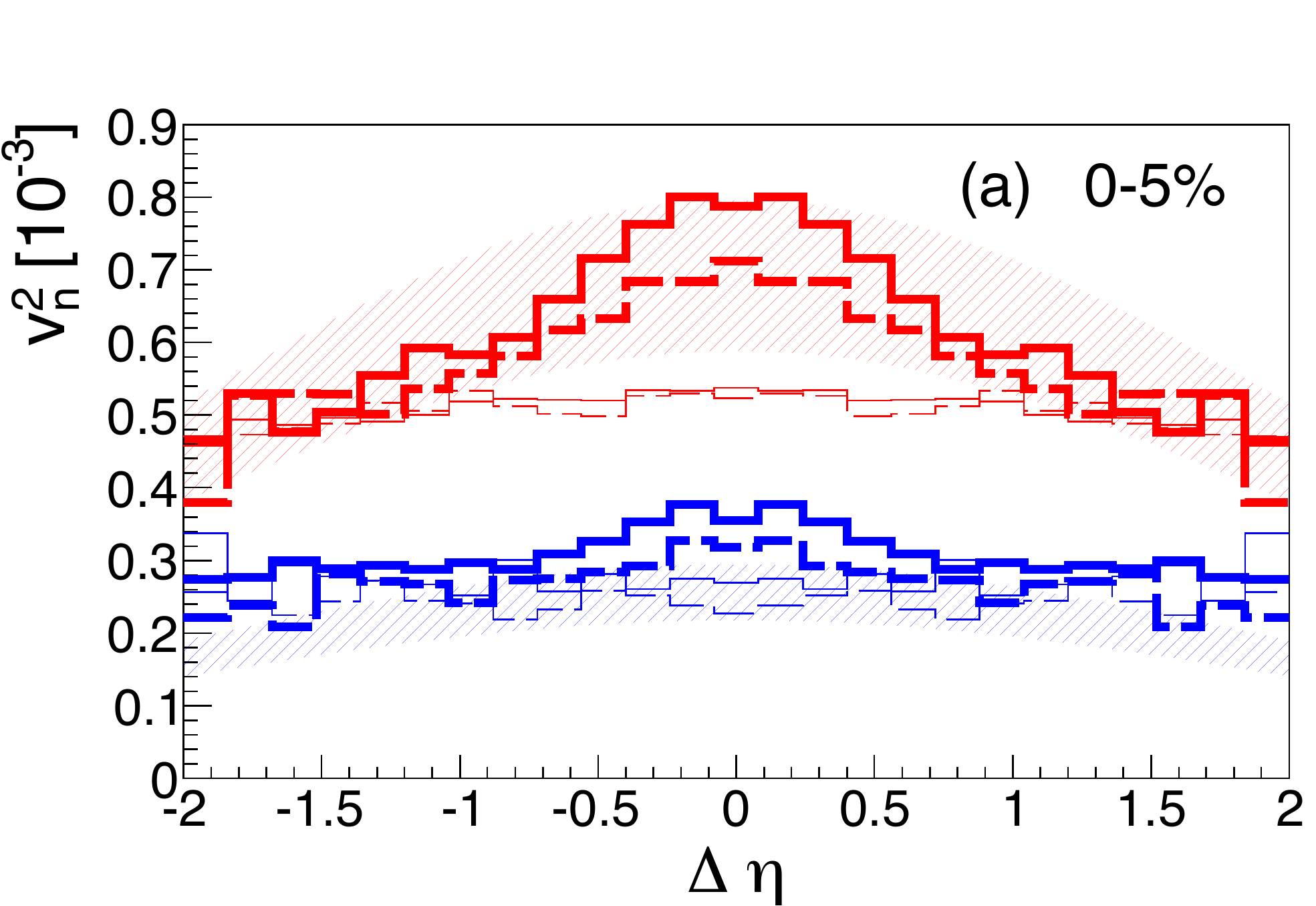}
\caption[Hypersurface Sampling]{Left: Distribution of quantum numbers compared to exact conservation laws (taken from \cite{Huovinen:2012is}). Right: Influence of local charge conservation on $v_n^2 (\Delta \eta)$ (taken from \cite{Bozek:2012en}).}
\label{fig_sampling} 
\end{figure}

One obvious source of non-flow correlations comes from conservation laws --- if one knows that a finite set of particles has a particular total
momentum, charge, etc., this implies a correlation between the particles.
One of the open questions is what effect various conservation laws have on correlation observables, and what is the most realistic way of implementing them in the sampling process. 
In \cite{Huovinen:2012is} the effect of global conservation of quantum numbers in each event has been demonstrated. The fluctuations of the energy, net baryon number, electric charge and net strangeness can be very large, if their conservation is not enforced in the sampling procedure. Those fluctuations will have a non-negligible effect on correlation observables even though it is important to emphasize that ensuring global conservation laws of quantum numbers does not include any local conservation laws for the same quantum numbers or for momentum. These local conservation laws are even more crucial for the detailed investigation of correlation observables as shown in the next example.   

The Krakow group \cite{Bozek:2012en} has investigated the effect of local charge conservation on correlation observables. This implementation reproduces the measured charge balance functions and results in a significant dependence of pair correlations on relative pseudorapidity, as well as explaining the difference in correlations for same charge versus opposite charge pairs. More recently, Becattini \textit{et al.}~\cite{Becattini:2013vja} have pointed out that the angular momentum conservation on the Cooper-Frye transition surface has an influence on the spin correlations of outgoing particles and might be important for polarization measurements. 

In the future, there should be more studies of locally conserved transverse momentum or strangeness. In principle, it makes only sense to apply such conservation laws in the sampling procedure if the associated currents are also locally conserved in the hydrodynamic evolution. If net baryon number current conservation is implemented in the hydrodynamic algorithm, this requires only the addition of one more additional differential equation for each conserved current  like isospin or net strangeness in the framework. 
\section[Initial State]{Physics of the Initial State}
\label{sec_ic}
The physics of the initial stages of a heavy-ion collision is currently one of the biggest open questions,
and is one of the largest contributions of uncertainty for many observables. The fundamental question is how the initial state of two ultrarelativistic nuclei before a collision 
evolves into a system that can eventually be described by hydrodynamics.
Reliable first-principles calculations from non-equilibrium quantum chromodynamics are not yet possible, so the current state-of-the-art
is to construct models, based on what we know about the relevant physics, and constrained by agreement with data after subsequent
fluid / transport evolution. In this section we describe how the initial state is modeled in general and then describe commonly used parametrizations and dynamic approaches, highlighting the improvements during the recent years and with a focus on fluctuations in the initial state. 

Much progress has been made in identifying the physical processes that contribute to the shape of the initial state profile and the amount of fluctuations. 
For heavy-ions, the most important source of azimuthal fluctuations is the nucleonic structure of the incoming nuclei --- at the moment of impact, each nucleon 
acquires a definite transverse position in the nucleus --- followed by local fluctuations in energy deposition / entropy production.  
However, many other aspects can have a non-negligible effect, such as local sub-nucleonic structure and correlations between nucleons in the nucleus, especially in smaller systems.

To specify initial conditions for a hydrodynamic evolution the energy momentum tensor is needed as a function of coordinate space at a certain time $\tau_0$ or $t_0$. In terms of hydrodynamic variables, this translates to an energy density, flow velocity, and values for the viscous tensor.  
Some of the more commonly used assumptions for these initial conditions include:
\begin{itemize}
\item
Longitudinal profile: boost-invariance or Gaussian parametrization as a function of spatial rapidity
\item 
Transverse profile: Glauber model or color glass inspired approaches
\item
Total energy/entropy: chosen to reproduce final particle multiplicity at one centrality
\item
Initial velocity distributions: Bjorken scaling in longitudinal direction and zero initial flow in the transverse direction
\item 
Fluctuations: Monte Carlo versions of above mentioned models, flux tube/hot spot toy models
\end{itemize}

It is common to characterize initial transverse density profiles with generalizations of the participant eccentricity \eqref{eps2pp}, with
strength $\varepsilon_n$ and orientation $\Phi_n$ that can be compactly written as the magnitude and phase of a complex number:
\begin{align}
\label{eqn_epsilon}
\varepsilon_n e^{in\Phi_n} &\equiv - \frac {\{r^n e^{in\phi}\}} {\{r^n\}};\quad  n\geq 2 \\
\varepsilon_1 e^{in\Phi_1} &\equiv - \frac {\{r^3 e^{i\phi}\}} {\{r^3\}},
\end{align}
where $r$ and $\phi$ are the polar coordinates in the transverse plane and the curly brackets are again an average over the initial transverse density, Eq.~\eqref{spaceavg}.   For $n$=2, this corresponds to Eq.~\eqref{eps2pp}.

In the literature, sometimes a different radial weight has been used.
For example, the original proposal kept $r^2$ weights for the generalization to the third harmonic.
However, weighting the coefficients with the $n^{\rm th}$ power of $r$ generally are better correlated to final $v_n$ values \cite{Gardim:2011xv}. 
We will see in Section~\ref{sec_results} that this choice has a natural interpretation as the lowest momentum mode of a Fourier transform of the initial transverse density;  although this set of eccentricity parameters lacks information about small-scale structures, they contain most of the information relevant to the hydrodynamic response.

Systematic studies of using either the energy density or the entropy density for the weighted average have demonstrated that this distinction does not make a difference to the results \cite{Gardim:2011xv,Qiu:2011iv}. Still, it is important to verify that the same definition for initial state eccentricity is used when comparing results from different theoretical calculations. It would be very helpful to define a standard for the characterization of initial state profiles and their features that allows for an apples-to-apples comparison. In \cite{ColemanSmith:2012ka,Floerchinger:2013rya} new methods based on a two dimensional Fourier expansion in polar coordinate space have been proposed, that allows for such comparisons and can be extended to three dimensions and other properties (e.g. initial velocity fields and viscous tensors).   

\subsection{Glauber-based Approaches}

Parametrizations based on Monte Carlo Glauber approaches with different degrees of sophistication concerning finite size of the nucleons, nucleon-nucleon correlations and fluctuations in the energy deposition have been widely applied. The basic version of the optical and Monte Carlo Glauber approach has been nicely reviewed in \cite{Miller:2007ri} and \cite{Alver:2008aq} contains a description of the PHOBOS Glauber Monte Carlo. 
The main idea in a Glauber approach is independent straight-line scattering of nucleons.  One starts with a spatial probability distribution for nucleons usually obtained from the measured Woods-Saxon charge density distribution.  In the Monte Carlo version,  nucleon positions are sampled accordingly and their position in the transverse plane in conjunction with the energy-dependent total nucleon-nucleon cross-section is used to decide if the nucleons interact with each other. In this way, the number of participant nucleons $N_{\rm part}$ and the number of binary collisions $N_{\rm coll}$ can be calculated.   In the optical version, densities of participants and collisions are calculated directly from the probability distribution. One then typically sets the initial energy density or entropy density in the transverse plane to be proportional to the density of participants or collisions, or a linear combination of both. 

\begin{figure}[h]
\includegraphics[width=0.5\textwidth]{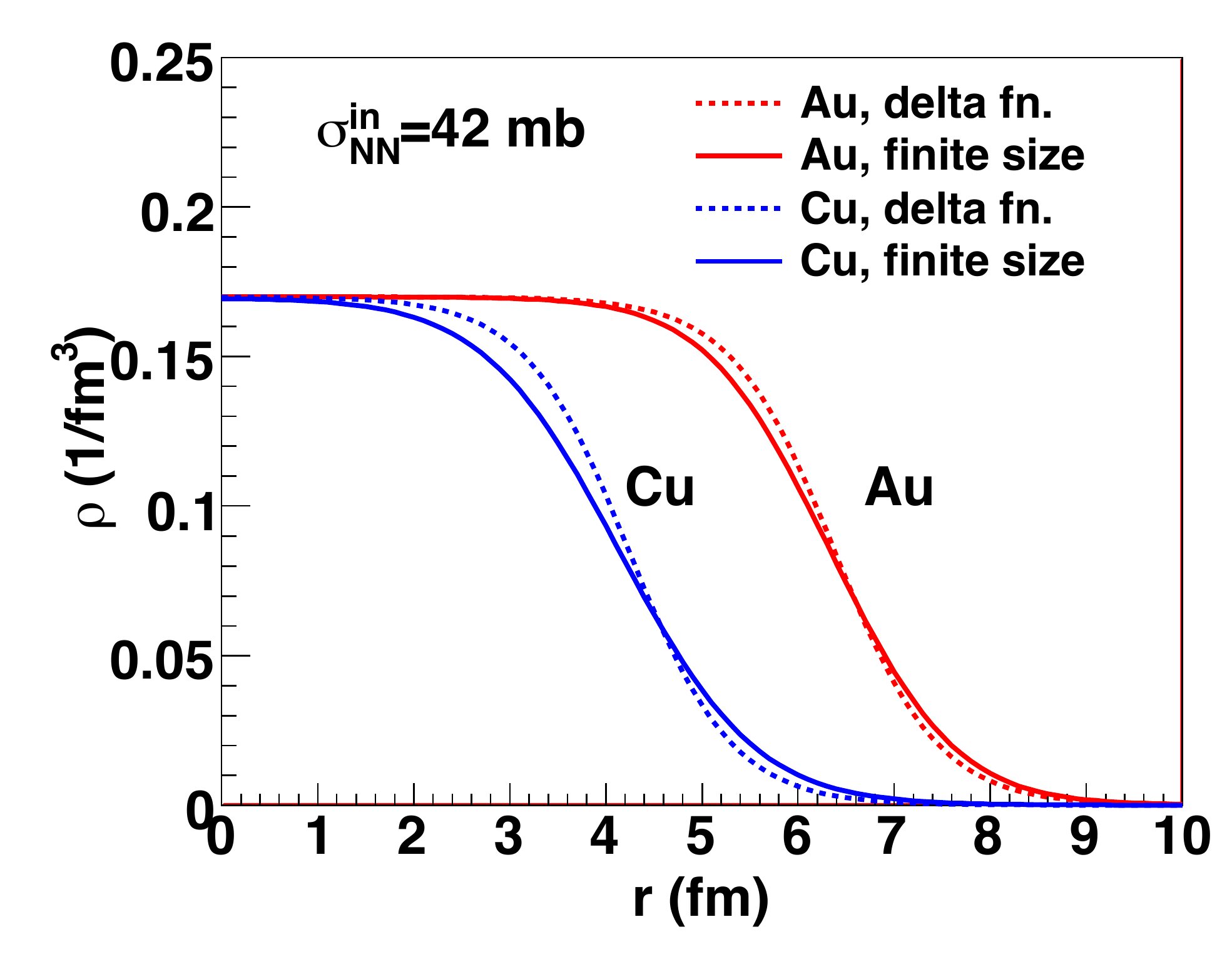}
\includegraphics[width=0.5\textwidth]{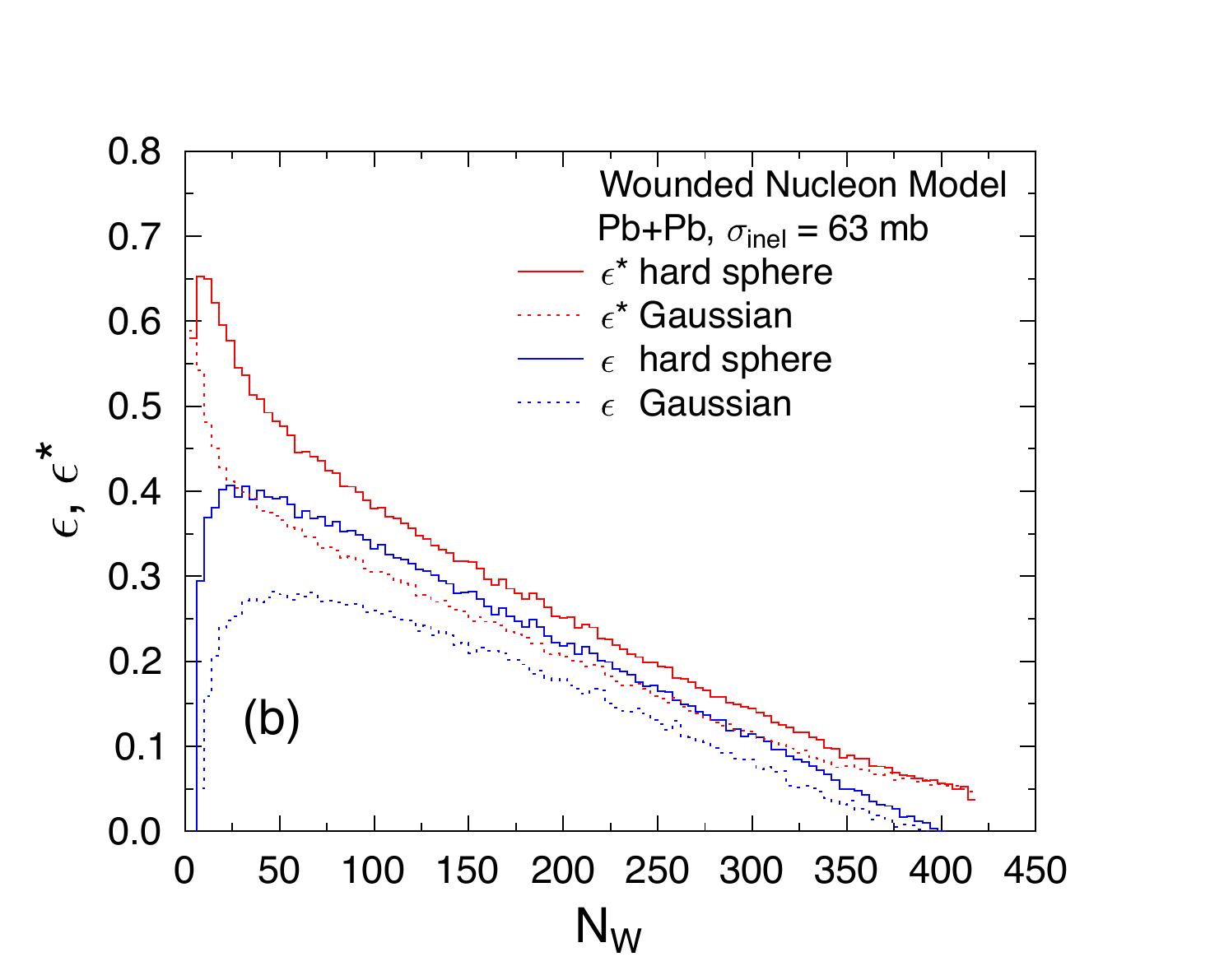}
\caption[finite_size_glauber]{Left: Comparison of Wood-Saxon distribution resulting from point-like nucleons or nucleons with a finite size (taken from \cite{Hirano:2009ah}). Right: Comparison of the standard $\epsilon$ and participant eccentricity $\epsilon^*$ as a function of wounded nucleon number with black disk and a Gaussian profile for the nucleons (taken from \cite{Rybczynski:2011wv}).}
\label{fig_glauber_finite} 
\end{figure}

The finite extent of the nucleons is taken into account either with the black-disk approximation or by assuming a Gaussian profile for the individual nucleons. The GLISSANDO package offers different possibilities to model the nucleons and their distribution \cite{Broniowski:2007nz,Broniowski:2007ft}. The parameters of the Woods-Saxon distributions need to be readjusted such that the average profile of the nucleus still matches the wanted distribution (see Fig. 1) \cite{Hirano:2009ah}. Different implementations of the finite size effect have an influence on eccentricity already for smooth initial conditions \cite{Rybczynski:2011wv}. 

The correlations of the nucleons within the wave-function of the nuclei are explored in \cite{Broniowski:2010jd} up to the two-body level. The conclusion is that the correlations lead to a 10-20 \% change in the initial eccentricities. This change can be accounted for by adjusting the hard-core expulsion distance between the center of the nucleons to $d=0.9$ fm. In \cite{Alvioli:2011sk} the influence of nucleon-nucleon correlations on the centrality dependence of the eccentricities has been calculated. The authors find agreement of the central correlations to the previous publication \cite{Broniowski:2010jd}, but the inclusion of full three-body correlations cancels the effect and the eccentricities appear very similar to the ones from the standard Monte Carlo Glauber approach. 

In addition to more sophisticated descriptions of the initial geometry of the nucleons, the Glauber approach has been extended to include a more realistic mechanism for particle production. In \cite{Qin:2010pf} the fluctuations of the number of charged particles produced in elementary nucleon-nucleon collisions has been taken into account. From measurements in p-p collisions it has been found that the multiplicity distribution can be well described by a negative binomial distribution. Once the number of particles has been determined, momenta are sampled according to a power law distribution. This approach can then be used to include initial non-equilibrium evolution at least on a qualitative level, since the whole phase-space distribution of the initially produced particles is known from the extended Glauber Monte Carlo. It is shown that the change in the eccentricities resulting from initial free-streaming is similar in magnitude to the decrease attributed to larger smearing factors in the conversion from particles to initial energy density distributions.  

Overall, the Monte Carlo Glauber approach is suitable to generate initial conditions for a hydrodynamic expansion. It delivers a realistic overlap profile as a function of impact parameter and in its Monte Carlo version, the observed fluctuations are matched rather well. Still, it is important to specify which incarnation of Glauber model implementation has been used in the calculation, when such an approach is used to describe the initial state of a heavy ion collision. 

\subsection{CGC-based Approaches}

The other common class of initial conditions is based on the gluon saturation picture. Examples are models based on kt factorization such as the Monte Carlo KLN model \cite{Kharzeev:2000ph,Kharzeev:2001gp,Drescher:2006ca} and MCrcBK model incorporating unintegrated gluon densities from the running-coupling BK equation and negative binomial fluctuations,  or the IP-Glasma model including dynamical evolution of SU(3) classical Yang-Mills equations for the gluon fields \cite{Schenke:2012wb}. In the high energy limit, the parton distribution function of the nucleons are dominated by gluons. The gluon contribution raises rapidly towards smaller x values and exceeds the one of valence and sea quarks by more than an order of magnitude. Therefore, the intuitive picture is that the gluons within the nuclei start to overlap and build a coherent condensate state. For high occupation numbers the Bose distribution coincides with the Boltzmann distribution and a description of the gluon dynamics in terms of classical fields can be applied. The initial state dynamics based on saturation ideas are summarized in a recent review article \cite{Albacete:2013tpa}.

On the level of smooth initial conditions, the KLN model leads to sharper edges and therefore higher eccentricities which require a higher shear viscosity during the hydrodynamic evolution to reproduce the same elliptic flow values as Glauber initial conditions \cite{Song:2010mg}. An obvious difference between Glauber Monte Carlo and CGC based Monte Carlo approaches are the degrees of freedom. Concerning fluctuations that implies that there are structures on much smaller scales in gluon based approaches compared to hadron based approaches. The hope is that by incorporating fluctuations and looking at higher harmonics in the flow anisotropies, there are more constraints to understand the initial state. 

\begin{figure}[h]
\includegraphics[width=0.5\textwidth]{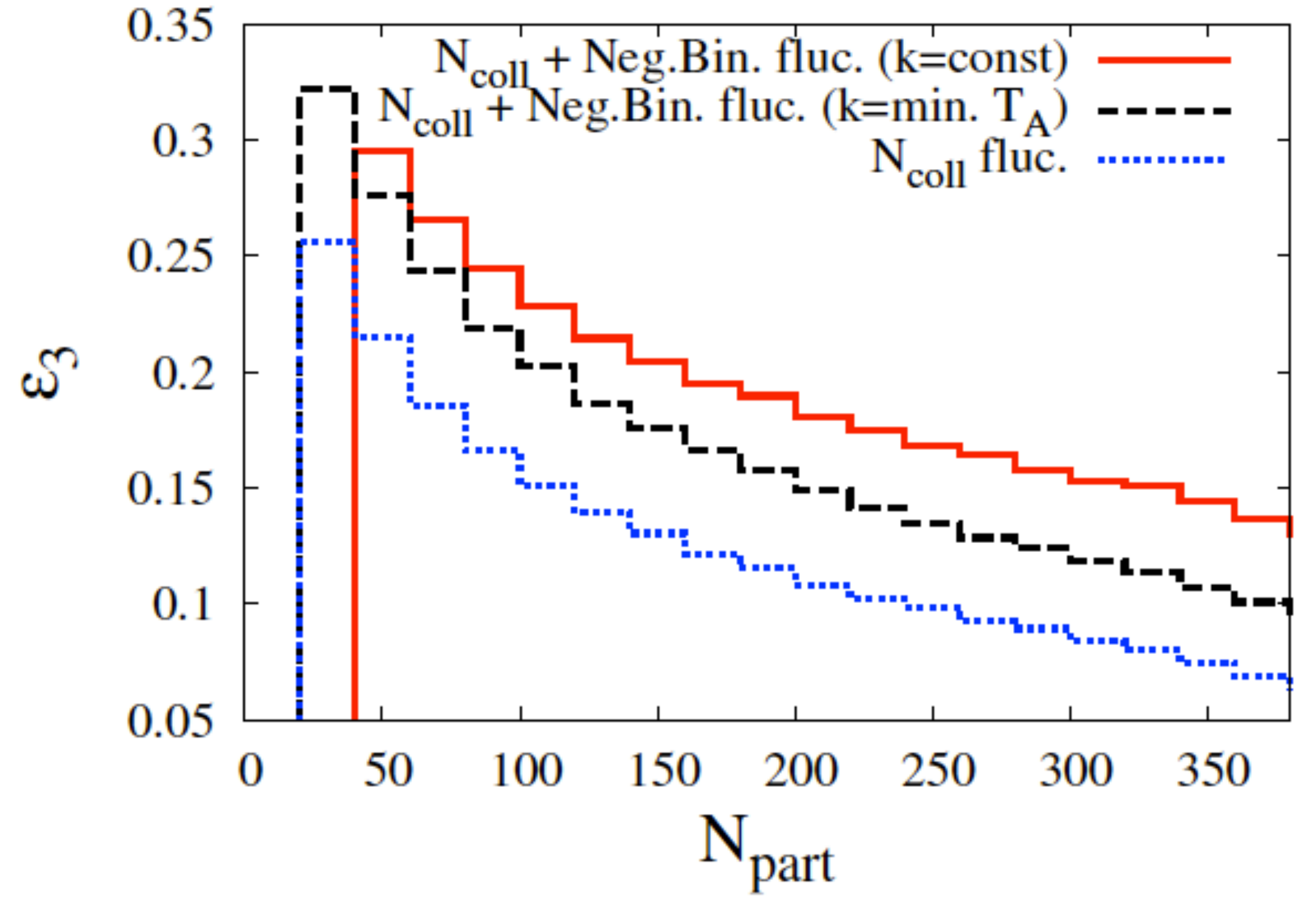}
\includegraphics[width=0.5\textwidth]{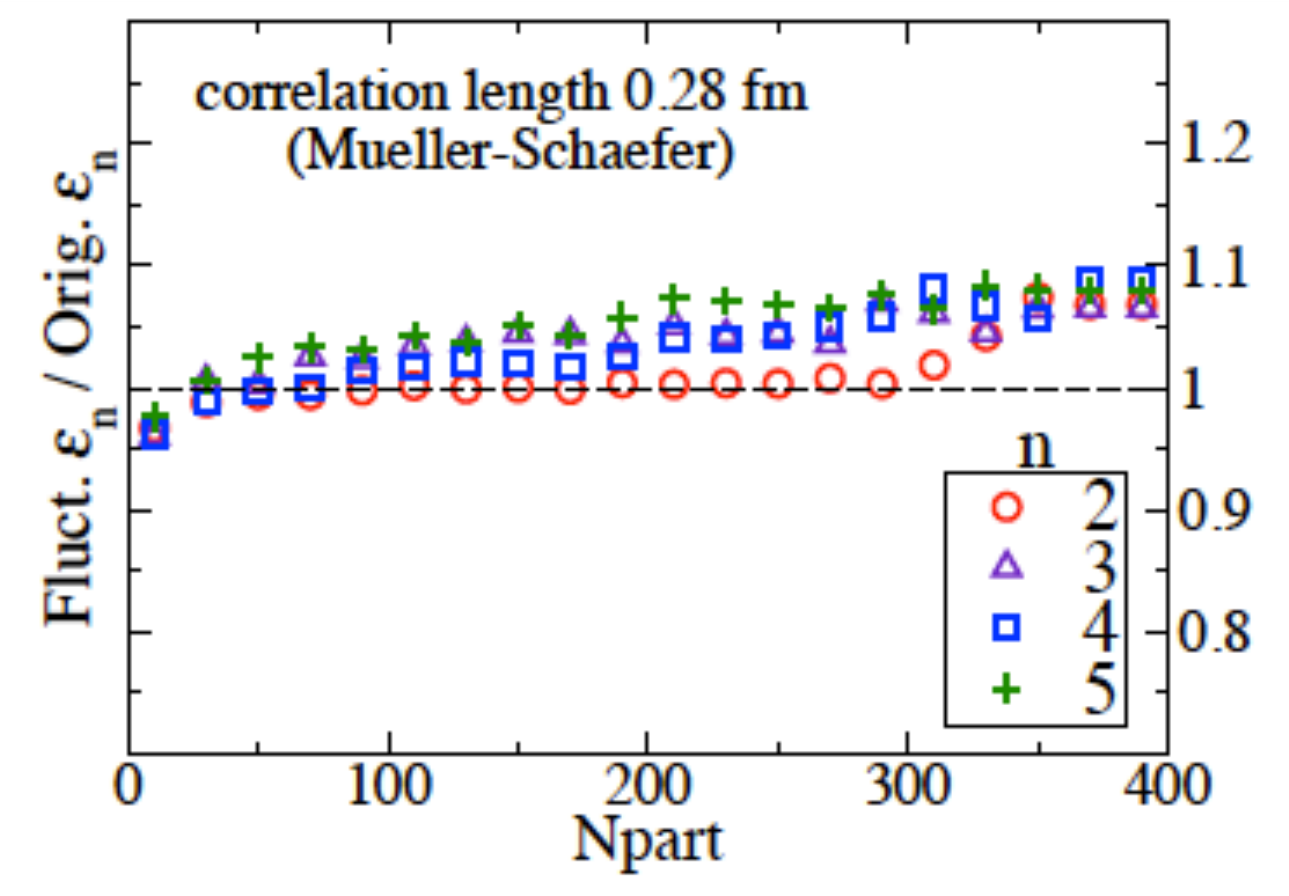}
\caption[cgc_flucs]{Left: Triangularity as a function of centrality in Au+Au collisions at $\sqrt{s_{\rm NN}}=200$ GeV increases with inclusion of multiplicity fluctuations (taken from \cite{Dumitru:2012yr}). Right: Ratio of eccentricity of coefficients with fluctuations imprinted over the original values (taken from \cite{Moreland:2012qw}).}
\label{fig_kln_mulfluc} 
\end{figure}

In the standard MC-KLN \cite{Drescher:2006ca} the initial sources for the gluon fields fluctuate only according to the positions of participating nucleons in each nucleus, as in a Glauber model. However, even an elementary nucleon-nucleon collision, the multiplicity of charged particles fluctuates from one event to the next.   This distribution has been measured and can be well described by a negative binomial distribution (NBD)

\begin{equation}
P(n)=\frac{\Gamma(k+n)}{\Gamma(k)\Gamma(n+1)}\frac{\bar{n}^n k^k}{(\bar{n}+k)^{n+k}}
\end{equation}
where $\bar{n}$ is the mean multiplicity and $k$ is the fluctuation parameter with smaller $k$ corresponding to larger fluctuations around the mean value. Koba-Nielsen-Olesen (KNO) scaling expresses the finding that the multiplicity distributions exploit a universal behaviour when normalized by the mean $\bar{n}$ also for larger systems. Dumitru and Nara have explored the effect of fluctuations in the particle production with the MC-KLN framework and found rather large effects on the initial eccentricity and triangularity \cite{Dumitru:2012yr}, see Fig. \ref{fig_kln_mulfluc} (left).

In \cite{Muller:2011bb} the correlations between the transverse energy density deposition has been calculated in the Gaussian Color Glass Condensate model. Even though there might be some remaining uncertainties related to assumptions made in that calculation, the fluctuations in the energy density are rather substantial and the correlation length is on the order of $0.3$ fm, which correspond to typical partonic scales. An algorithm to imprint fluctuations according to an arbitrary covariance function onto smooth profiles has been developed and applied to this calculation in \cite{Moreland:2012qw}. The effect on the eccentricity coefficients shown in Fig. \ref{fig_kln_mulfluc} (right) is between 10 \% and 25 \% depending on the scale of the subnucleonic structures that is assumed. 

The IP-Glasma approach that has been introduced by the BNL group in \cite{Schenke:2012wb} combines the impact parameter dependent saturation model (IP-Sat) with the initial evolution of gluon fields in a SU(3) Yang-Mills simulation. It includes fluctuations of the color charges on length scales of the inverse saturation momentum in addition to the fluctuations of the nucleon positions. Therefore, it intrinsically reproduces the NBD for the multiplicity distributions. 

Initial state models based on saturation ideas and therefore sub-nucleonic degrees of freedom are conceptually more suitable for higher beam energies at RHIC and LHC. There is rapid progress being made in developing these approaches to higher sophistication and more realistic descriptions of the initial non-equilibrium evolution. Even though any statements referring to a binary choice between `Glauber' and `CGC' initial conditions are certainly too simplistic, it is clear that there are qualitative differences in the scale of fluctuations in both types of approaches. 

\subsection{Analytic Approaches}
To study the effect of initial state fluctuations without specific assumptions about the underlying physical processes, it makes sense to introduce parametric initial conditions. A smooth background profile for the energy density distribution is augmented by certain numbers of flux tubes or hot spots to explore the qualitative differences concerning eccentricity coefficients and final state momentum space anisotropy observables. 

In \cite{Bhalerao:2011bp} analytic formulas are derived for various moments of eccentricities in a simple model of random independent sources (e.g., flux tubes or hot spots), characterized only the number of sources, the shape of each individual source and their probability distribution across the transverse plane.  This simple picture accurately reproduces reproduce the behavior of eccentricity coefficients and their correlations as a function of centrality from MC Glauber and MC KLN models, and therefore allows one to gain an understanding of what determines various features.  
For example, changing a smearing parameter, e.g.~in a MC Glauber model, only changes the denominator of the eccentricities \eqref{eqn_epsilon}, and so the effect can be easily calculated and understood as simply a change in the overall size of the system.  In addition, the observation that $\varepsilon_1 < \varepsilon_3$ in known models can be readily understood as a generic result.

Reference \cite{Qin:2011uw} contains a different way of parametrizing the number of hot (or cold) spots in the initial state of a heavy ion reaction. The main conclusion is that correlations between different final state Fourier coefficients contain information about the amount of inhomogeneity of initial conditions. 

The Brasilians have studied the influence of a coarse-graining scale on the hydrodynamic evolution. A flux tube model has been employed where the number of tubes, their individual size, the energy per tube and their position are inspired by NEXUS initial conditions. In \cite{Mota:2012qv} a first attempt at looking at the dependence of event plane angles and their correlations on transverse momentum is made. 

\subsection{Dynamic Approaches}

Transport approaches like NEXUS \cite{Gardim:2012yp,Gardim:2011xv}, EPOS \cite{Werner:2010aa,Werner:2012xh,Werner:2012ca}, UrQMD \cite{Steinheimer:2007iy,Petersen:2008dd,Petersen:2009vx,Petersen:2010md,Petersen:2010zt,Petersen:2011sb} and AMPT \cite{Pang:2012he} have been used to provide initial conditions for hybrid models. Employing a dynamical approach to generate an initial condition for the hydrodynamic evolution allows for a treatment of the non-equilibrium dynamics. All the different pieces to specify the initial state, namely the longitudinal and transverse profile, the maximum value, the velocity distributions and the fluctuations are in principle available in a self-consistent way.  
Fluctuations arising from the positions of the produced particles and fluctuations in the energy deposition per binary collision are naturally included. Since transport approaches are often implemented in three dimensions, they allow for more realistic (and fluctuating) initial conditions in the longitudinal direction.
Once the full energy-momentum tensor is calculated at some early time, this can be uniquely converted into viscous hydrodynamic variables.  However, 
the system in these calculations usually do not thermalize well, and so instant thermalization is still an assumption that had to be enforced in all of these approaches,
until very recently \cite{vanderSchee:2013pia}. 

As one example the procedure to match UrQMD to a hydrodynamic initial state will be described. The nucleons are sampled initially according to Wood-Saxon profiles and the interactions are calculated until a certain thermalization time is reached. For heavy ion reactions at RHIC and LHC usually values for this switching time around $0.5$ fm are assumed. At this time, all the particles are mapped using three-dimensional Gaussian distributions that are lorentz-contracted in the beam direction to the hydrodynamical grid. The energy, momentum and net baryon densities are conserved in this process. Whereas UrQMD represents a purely hadronic transport approach including PYTHIA for hard scatterings, NEXUS and its successor EPOS include a more sophisticated treatment of multi-parton interactions based on Gribov-Regge theory. The initial state configurations in AMPT are based on HIJING and the ZPC parton cascade.

\begin{figure}[h]
\includegraphics[width=0.6\textwidth]{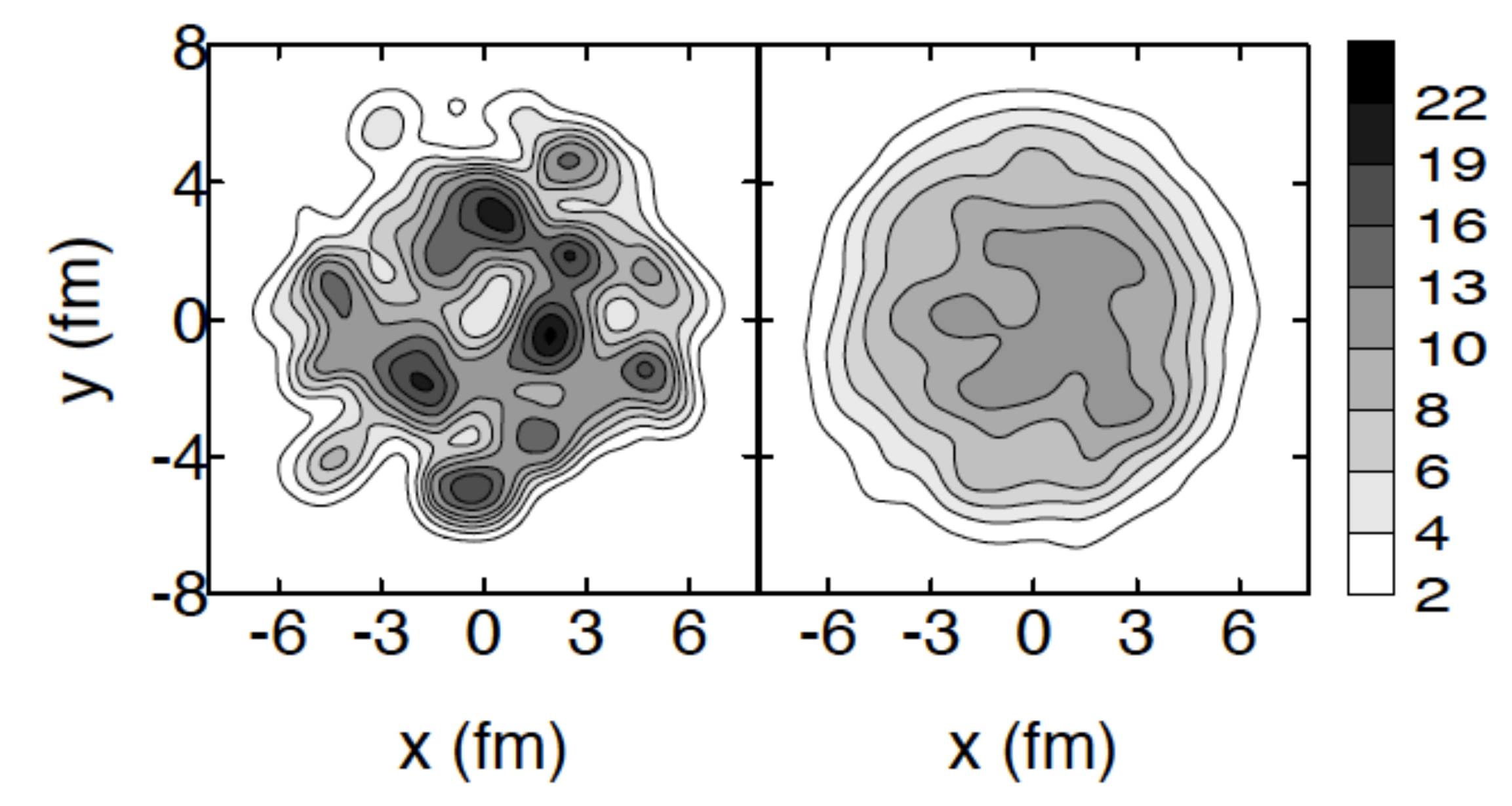}
\includegraphics[width=0.4\textwidth]{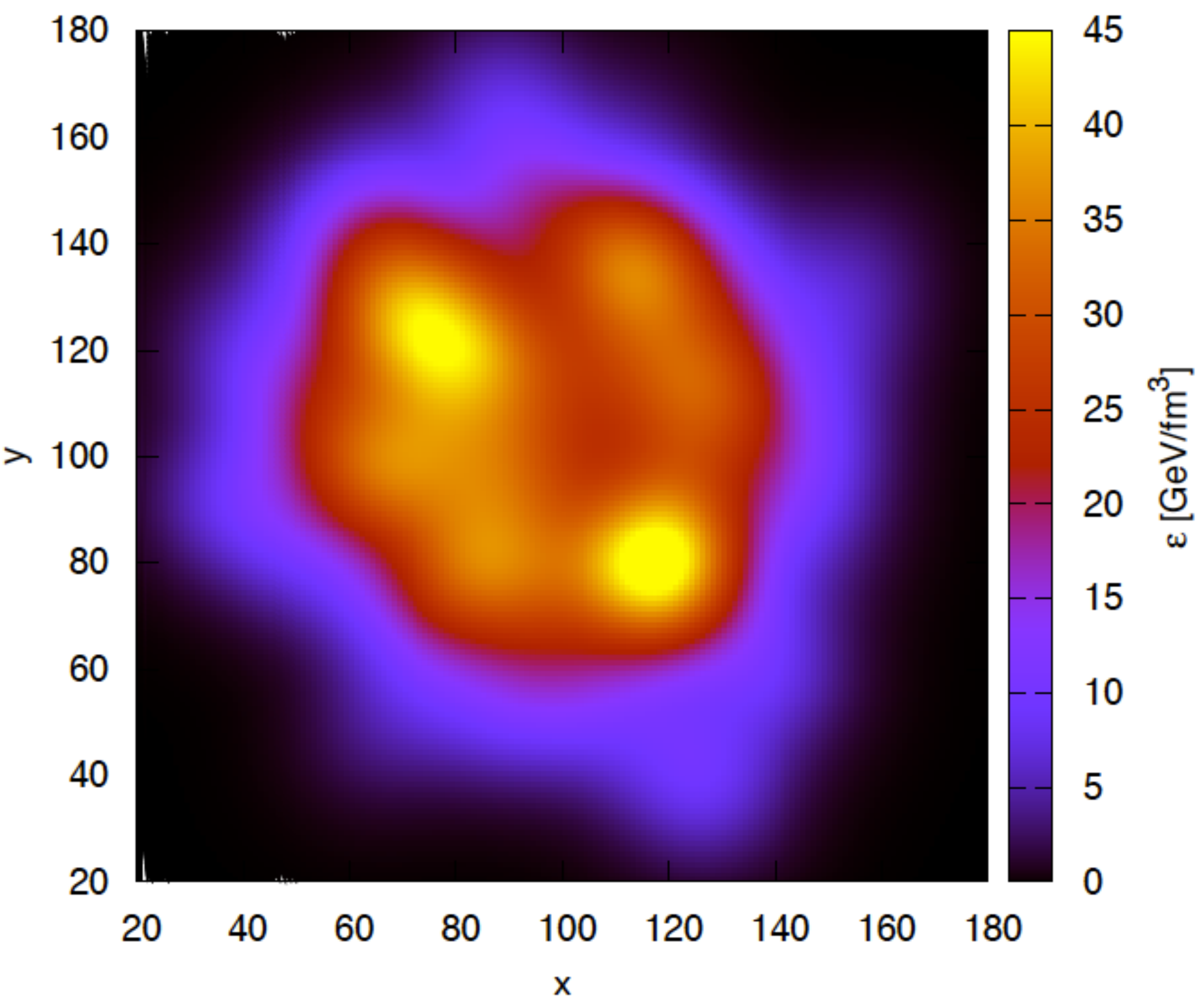}
\caption[ini_dyn]{Left: Energy density distribution from one NEXUS event compared to the distribution after averaging over 30 events (taken from \cite{Andrade:2008xh}). Right: Typical initial energy density distribution in the transverse plane from UrQMD (taken from \cite{Petersen:2012qc}).}
\label{fig_inidyn} 
\end{figure}

An important feature of fluctuating initial conditions is, that the created fireball is not at a constant temperature or density and even not at a fixed entropy per particle. Each cell in the beginning of the hydrodynamic evolution has its own thermodynamic properties as can be seen in the phase diagram evolutions shown in \cite{Bass:2012gy}. Typical initial state profiles in the transverse plane from dynamic approaches are shown in Fig. \ref{fig_inidyn}.

\subsection{Open Issues}
Despite the large effort that is made understanding the initial conditions in terms of parametrizations and dynamical models, it is important to concentrate on the goal to quantify the amount of initial state fluctuations that is associated with the early stage non-equilibrium QCD evolution of two colliding nuclei. Even though the hydrodynamic description of the hot and dense stage of heavy ion collisions seems to work very well, one of the big open questions in the field is how the system thermalizes rapidly. Whatever the process for thermalization actually is, it is likely that it will have an impact on initial state fluctuations. 

Initial non-trivial flow velocity fields have not yet been studied in a comprehensive fashion. There are certainly non-zero initial angular momentum and vorticity in the system, but how they should be treated and what impact they have on observables is not obvious. Matching the full  $T^{\mu\nu}(x^\mu)$ from initial state dynamics including off-diagonal elements to (3+1)d viscous hydrodynamics has not been done so far.

\section{Results}
\label{sec_results}
\subsection{Generic properties of collective expansion}

\begin{figure}
\begin{center}
\includegraphics[width=0.4\textwidth]{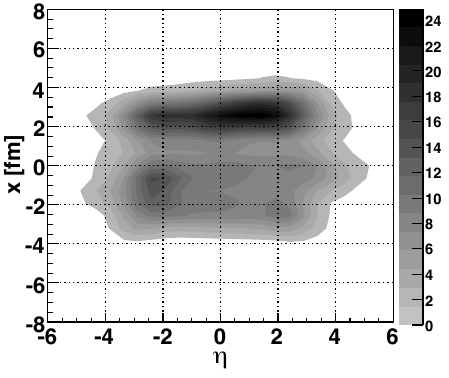}
\includegraphics[width=0.45\textwidth]{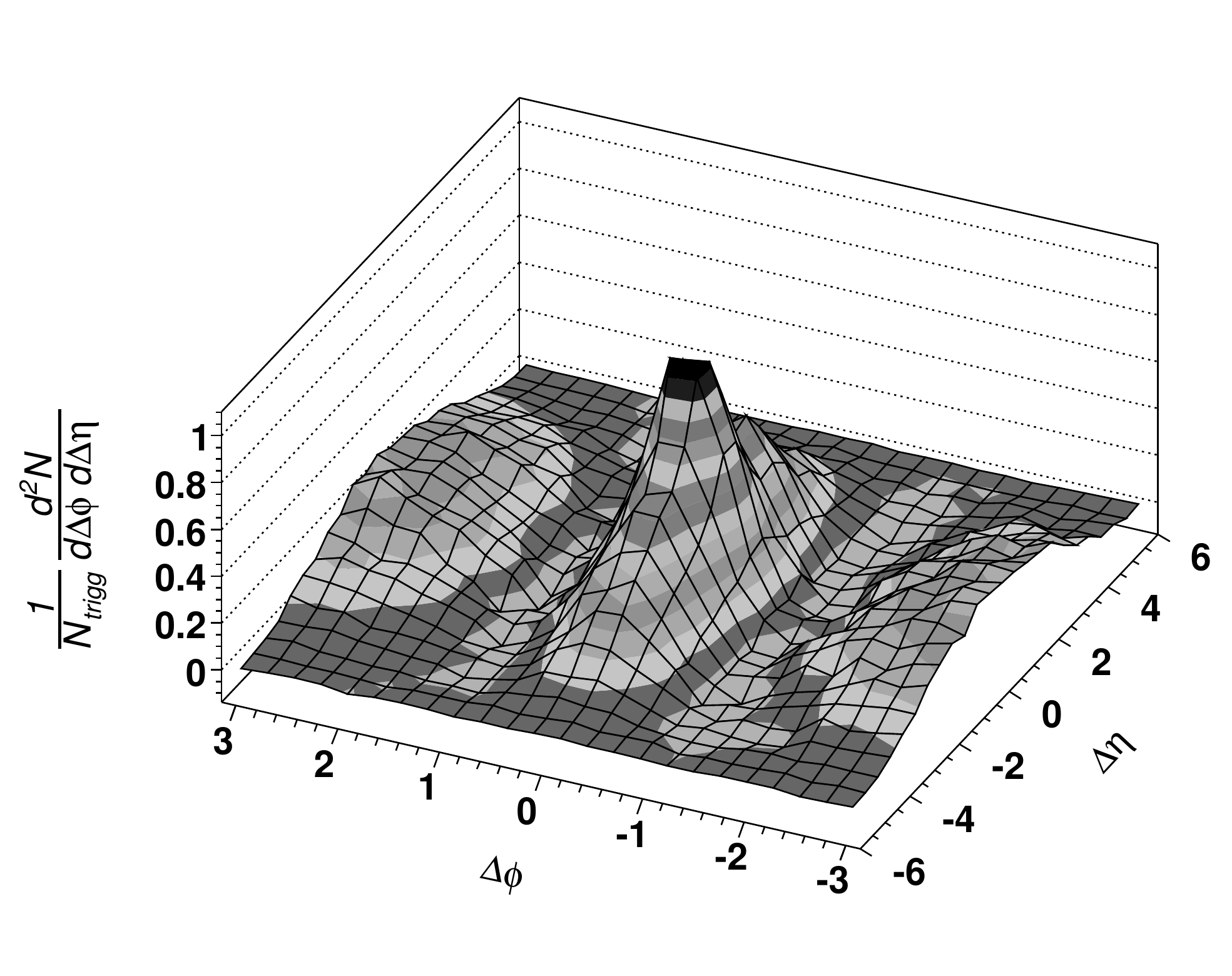}
\end{center}
\caption{Extended longitudinal structures in the initial condition (left) are transferred by hydrodynamics to long-range correlations among the final momenta of particles (right).  
The elliptic anisotropy in the transverse direction leads to a strong $\cos (2\Delta\phi)$ variation in the resulting correlation,  clearly seen in the two pronounced ridges at $\Delta\phi=0$ and $\pi$.
Taken from \cite{Takahashi:2009na}}
\label{fig_LRcorr} 
\end{figure}

If the initial energy density distribution is asymmetric in the transverse plane, 
the resulting anisotropic pressure gradients will generate an anisotropy in the 
final momentum-space distribution of particles.  Due to the ``almond-shape'' 
overlap area between a pair of nuclei colliding at non-zero impact parameter,
it is expected to find an anisotropy in the second Fourier coefficient of Eq.~\eqref{vn}, $v_2$.
This was one of the original proposals to look for collective behavior~\cite{Ollitrault:1992bk}.

Most models for initial conditions contain extended longitudinal structures such as 
``flux tubes'' or ``strings''.  So the transverse density distribution typically depends only weakly on rapidity, 
which results in a relatively weak dependence of $v_n$ and $\Psi_n$ on (pseudo)rapidity.
Thus, particles widely separated in rapidity can be correlated
with each other by their mutual correlation with the same global azimuthal structure (see Fig.~\ref{fig_LRcorr}).
Indeed, a large azimuthal asymmetry is observed in two-particle correlations (i.e., Eqs.~\eqref{v22} and \eqref{v2EP}),
even among pairs separated by a large gap in rapidity~\cite{Ackermann:2000tr, Aamodt:2010pa}.

As we have seen, event-by-event fluctuations will break the apparent symmetry of the naive ``almond-shaped'' 
overlap area of two spherical nuclei, and other Fourier harmonics $v_n$ should also be non-zero --- 
in particular odd ones at midrapidity.  Since they are generated entirely by fluctuations, they are typically smaller than
$v_2$, and have a weaker dependence on centrality --- they depend on centrality only because the magnitude
of fluctuations and the hydrodynamic response depend on, e.g., the size of the system~\cite{Alver:2010dn}. 

\begin{figure}
\begin{center}
\includegraphics[width=0.7\textwidth]{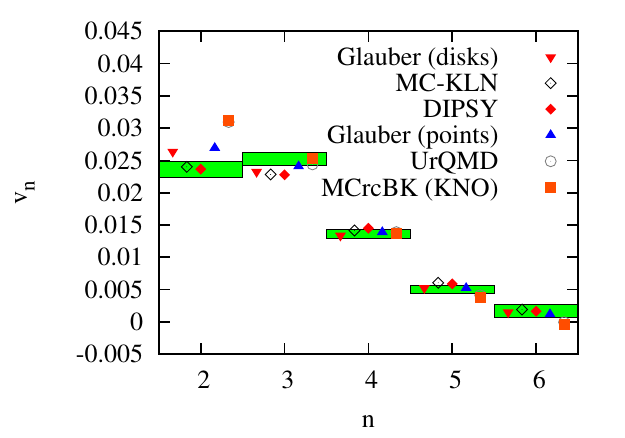}
\end{center}
\caption{$v_n$ vs. $n$ in 0--1\% central Pb+Pb collisions~\cite{ATLAS:2012at} compared to hydro calculations using a number of initial condition models (taken from \cite{Luzum:2012wu}).}
\label{fig_vnvsn} 
\end{figure}

In hydrodynamic models, another generic feature is that higher harmonics are typically suppressed in comparison
to lower harmonics (even in ideal hydrodynamics, although the effect is especially strong when viscosity is present)~\cite{Alver:2010dn}.
See, e.g., Fig.~\ref{fig_vnvsn}.

One more generic property is found by inspecting the form of the pair correlation, Eq.~\eqref{twoparticle}, in the absence of 
non-flow $\delta_2$.  This formula implies a set of inequalities~\cite{Ollitrault:2012cm}:
\begin{align}
V_{n\Delta}^{a,a} &\geq 0 ,\\
\label{CS}
(V_{n\Delta}^{a,b})^2 &\leq V_{n\Delta}^{a,a} V_{n\Delta}^{b,b},
\end{align}
where $V_{n\Delta}^{a,b}\equiv V_{n\Delta}(p_T^a,\eta^a,p_T^b,\eta^b)$.  I.e., diagonal components of the
correlation matrix (where both particles are of the same type and come from the same momentum bin) 
are positive (semi)definite, while off-diagonal components are related to the diagonal components by
a triangle inequality.

Finally, the near-equilibrium distribution of particles at freeze-out generically has a mass ordering, whereby higher
mass particles have a smaller anisotropy at low transverse momentum than lower mass particles~\cite{Huovinen:2001cy, Borghini:2005kd}.

The observed data are consistent with all of these generic expectations, with the exception of the Cauchy-Schwarz 
inequality Eq.~\eqref{CS} in the first harmonic $V_{1\Delta}$~\cite{Ollitrault:2012cm}, indicating the presence of a non-flow correlation
due to momentum conservation that was previously known to exist~\cite{Borghini:2000cm, Retinskaya:2012ky}.
\subsection{Characterizing the initial conditions and hydrodynamic response}
In a typical hydrodynamical calculation, the particle distribution Eq.~\eqref{vn} is given deterministically from a given initial condition. The exceptions are statistical fluctuations if discrete particles are generated, and intrinsic thermal fluctuations.  Experimental observables
are designed to remove the former, while preliminary investigation of the latter indicates it may not be a large effect.

If one can understand in detail which properties of the initial state are most important for determining each coefficient in Eq.~\eqref{vn},
and which are largely irrelevant, one can more easily use experimental data to place constraints on models of the initial conditions.
One could also potentially save a considerable amount of computational resources by avoiding brute force event-by-event calculations
for every possible input parameter or model of the initial conditions.

So far, the most effort has been put into a characterization of the initial transverse density distribution. 
Initial flow $u^\mu$, shear stress $\Pi^{\mu\nu}$, and rapidity dependence appear to be less
important for most mid-rapidity hadronic observables in calculations that have been done so far, though
studies are currently ongoing.

\begin{figure}[hb]
\includegraphics[width=0.5\textwidth]{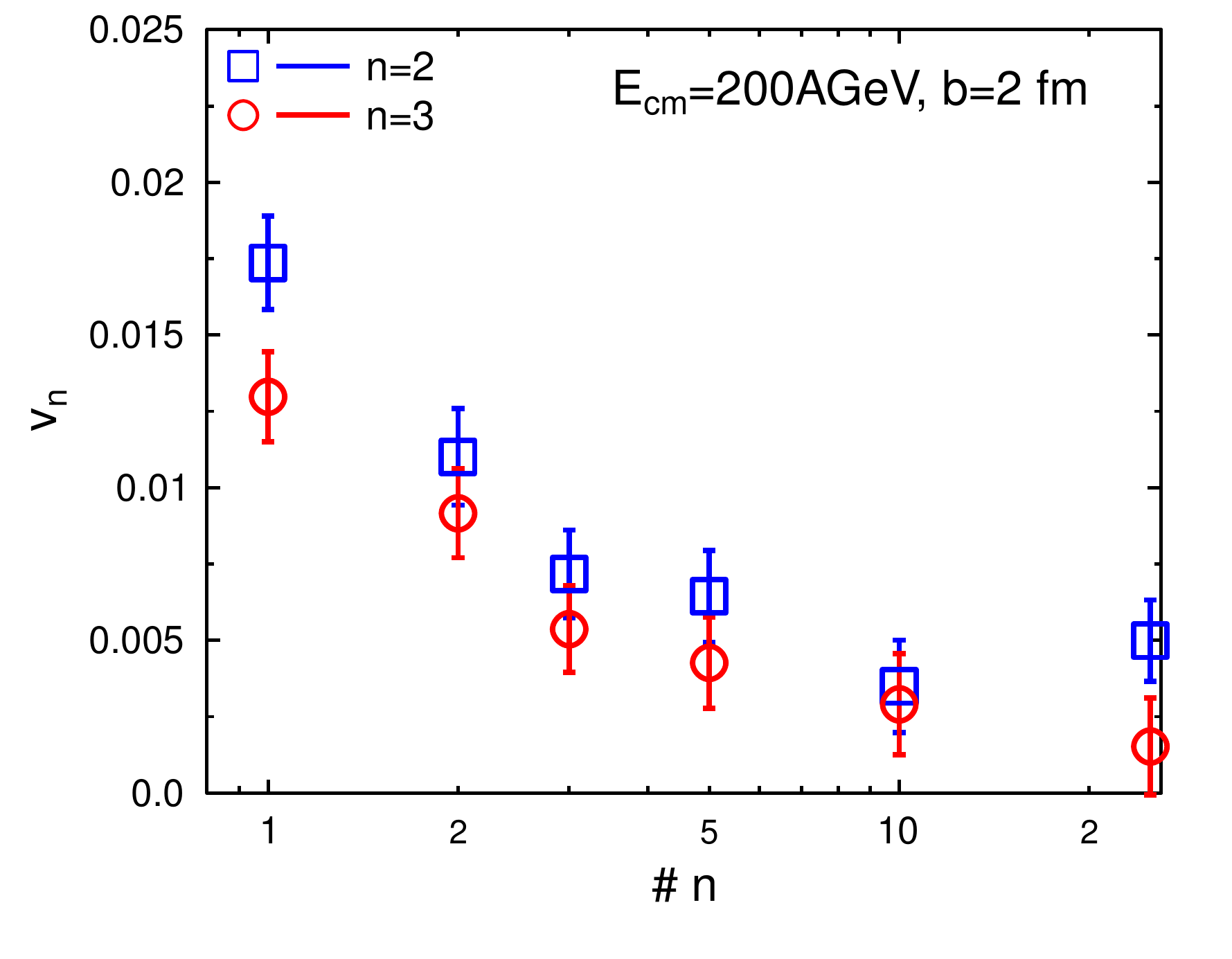}
\includegraphics[width=0.5\textwidth]{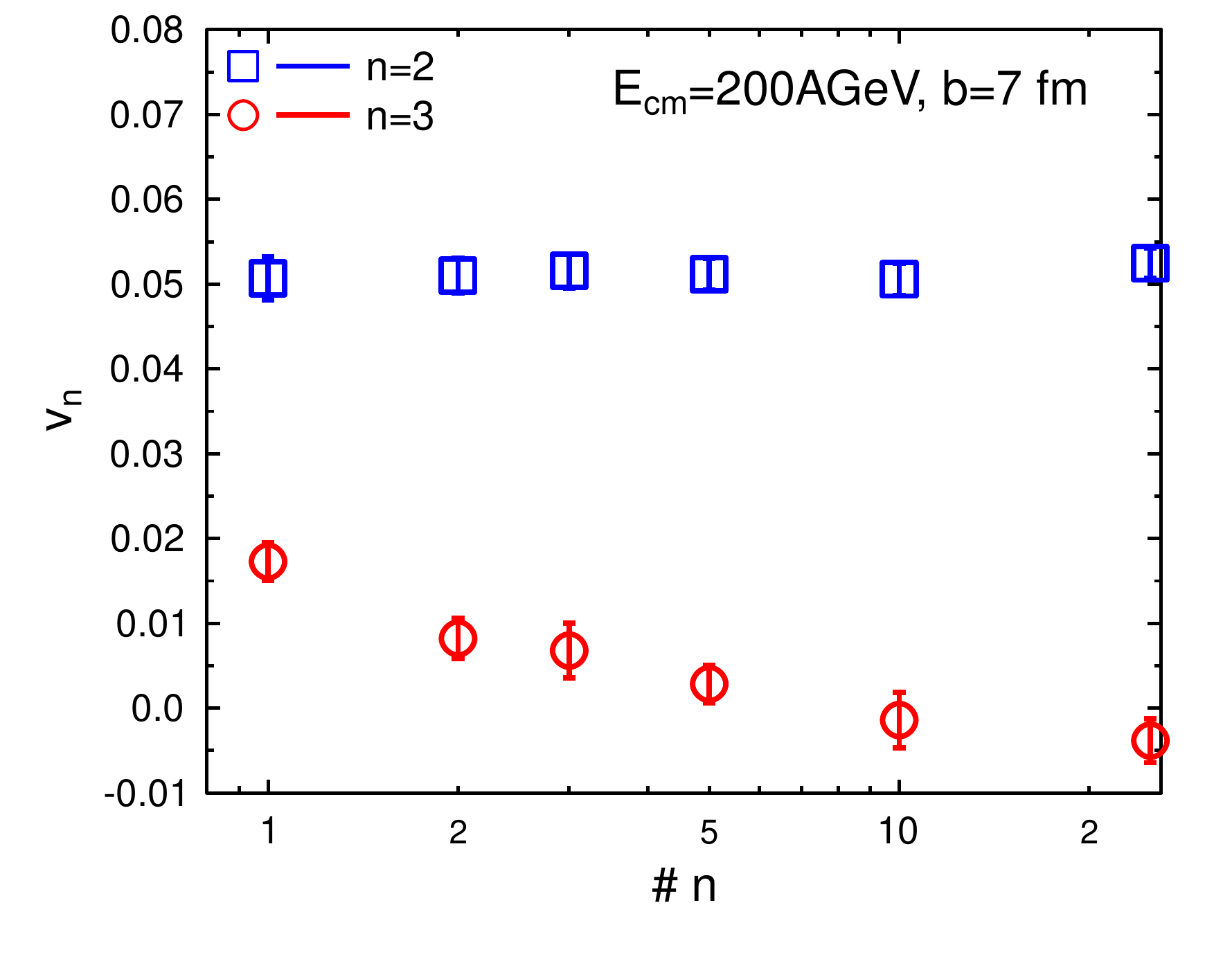}
\caption{Elliptic and triangular flow versus initial state granularity for central (left) and mid-central (right) Au+Au collisions at $\sqrt{s_{\rm NN}}=200$ GeV (taken from \cite{Petersen:2012qc}). }
\label{fig_vn_granularity} 
\end{figure}

To show that odd Fourier coefficients are on the average sensitive to fluctuations in the initial state, a systematic study of the dependence of triangular flow on the granularity of the initial conditions has been carried out in \cite{Petersen:2012qc}. Fig. \ref{fig_vn_granularity} shows for two different impact parameters how $v_2$ and $v_3$ change when initial conditions are smoothed by averaging them over a varying number of events with reaction plane fixed.  Moving to the right in the plots corresponds to an average over more initial profiles, which results in a smoother transverse profile as well as a smaller fluctuation-generated eccentricity.  In central collisions both flow coefficients are very sensitive to fluctuations while in mid-central collisons only triangular flow keeps the sensitivity to fluctuations. 

One of the generic features of any hydrodynamic evolution including local peaks in the initial energy density distribution is that the matter will be squeezed out on both sides of the hot spot. This is easy to understand, since the gradients in the energy density will prevent the matter to flow toward the hotter region and therefore the matter will flow around the obstacle. This effect has been demonstrated in single tube calculations on top of a smooth background profile, where clear two-peak structures in the final state particle distributions are obtained \cite{Andrade:2010xy}. An analytical study in a similar spirit has been carried out in \cite{Staig:2010pn,Staig:2011wj}. 

\begin{figure}
\includegraphics[width=0.5\textwidth]{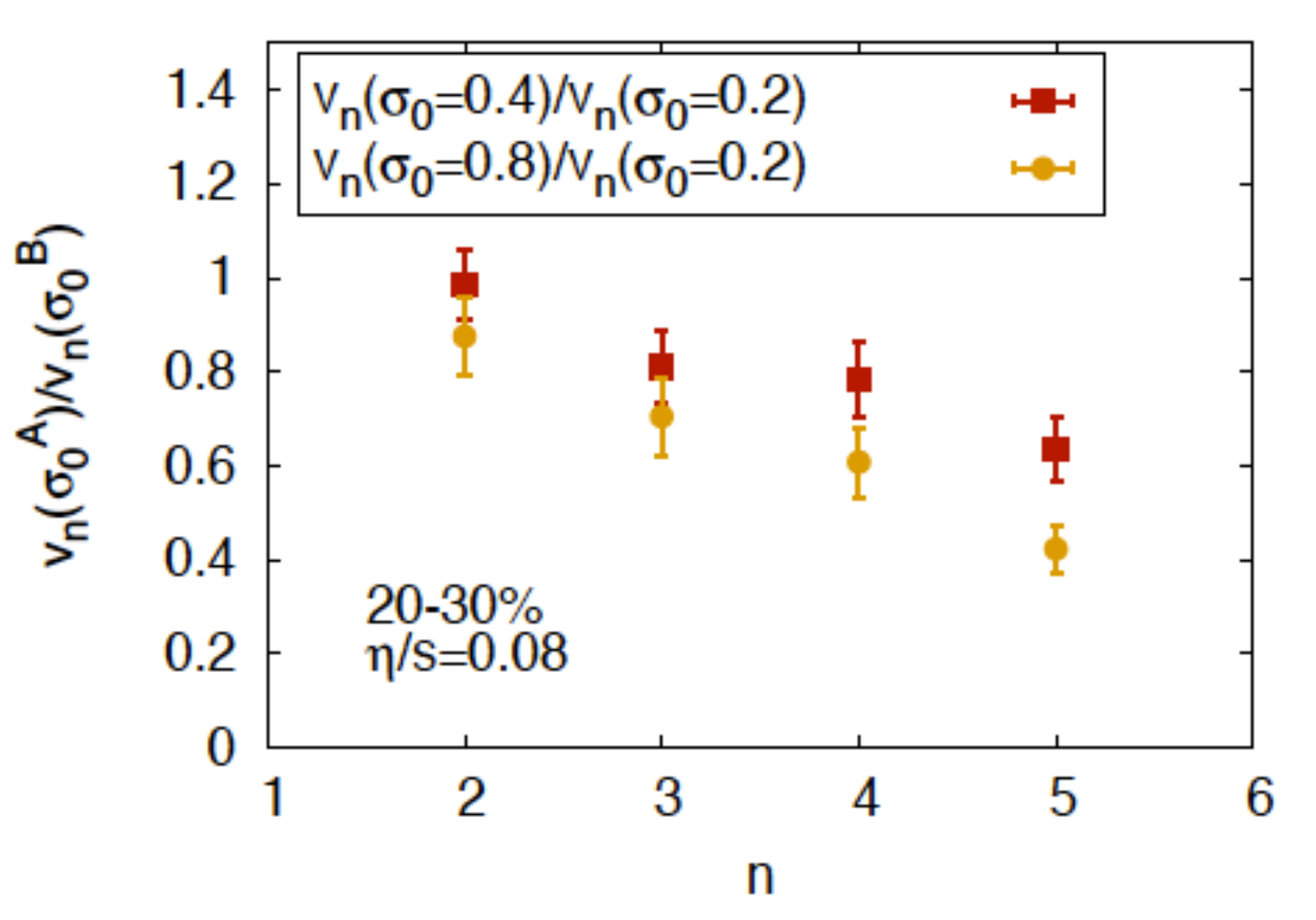}
\includegraphics[width=0.5\textwidth]{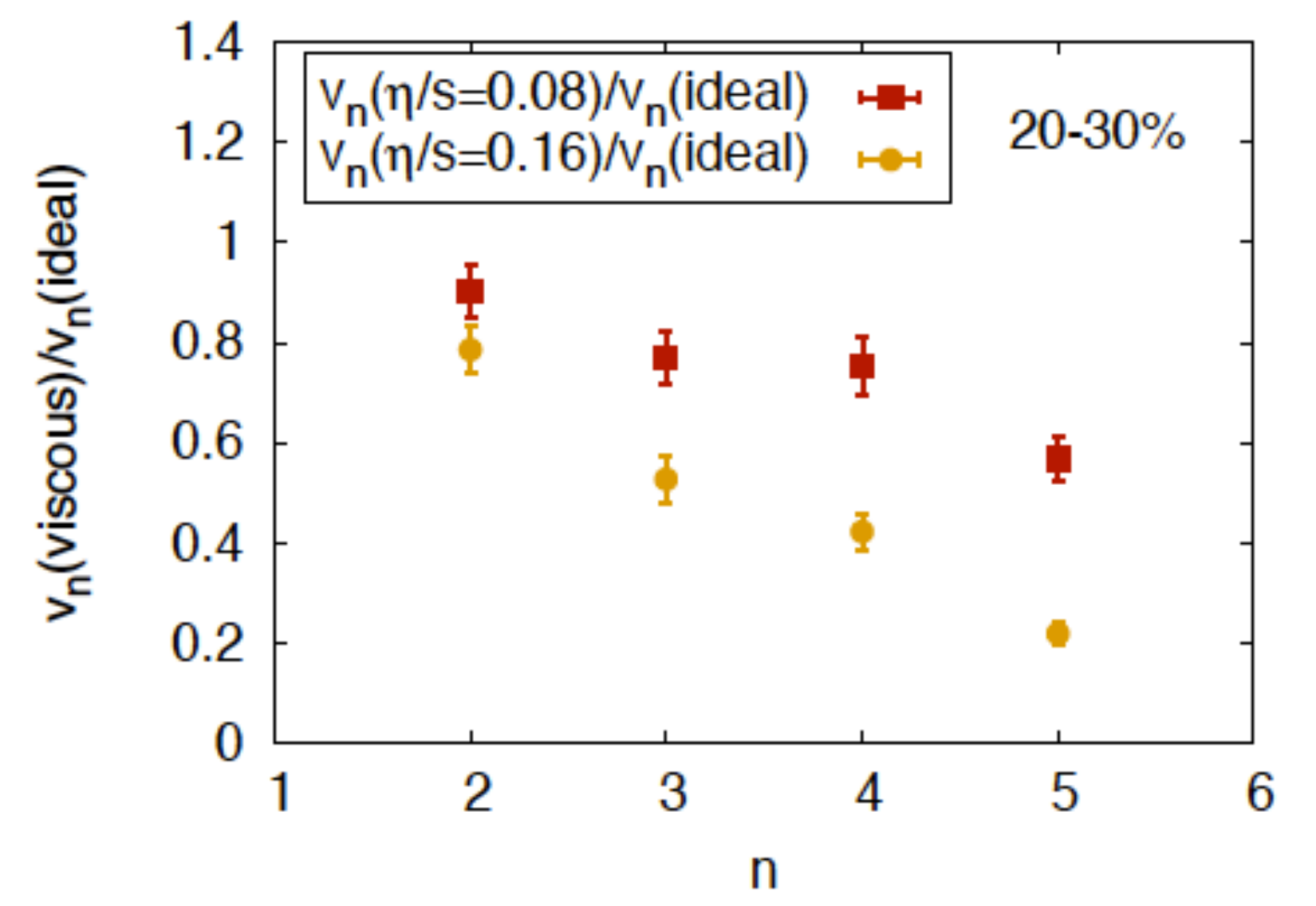}
\caption{$v_n$ coefficients from event-by-event hydrodynamics with different smearing kernel width (left) and different values for the shear viscosity to entropy ration (right) in mid-central Au+Au collisions at $\sqrt{s_{\rm NN}}=200$ GeV (taken from \cite{Schenke:2011bn}). }
\label{fig_vn_sigma_visc} 
\end{figure}

A finite shear viscosity during the fluid dynamic evolution has a very similar effect on the $v_n$ coefficients as varying the amount of fluctuations in the initial state. In \cite{Schenke:2011bn} (Fig. \ref{fig_vn_sigma_visc}) it has been demonstrated that increasing the smearing kernel width for the initial fluctuations results in the same reduction of flow coefficients as an increased viscosity during the evolution. Therefore, one can mimic viscous effects by starting with smoother initial conditions. More detailed comparisons to various flow coefficients and their distributions needs to be pursued to constrain the shear viscosity and the initial state separately. 

A more systematic study of the hydrodynamic response can be accomplished with a systematic characterization of the initial transverse density,
which can be represented by a set of basis functions and corresponding coefficients.
The most studied is the moment or cumulant expansion of the two-dimensional Fourier transform of the transverse energy or entropy density~\cite{Teaney:2010vd}.
In the moment expansion, the Fourier transformed density is expanded in a Fourier series in azimuthal angle and a Taylor series 
\begin{align}
\label{fouriertransform}
\frac {\rho(\bf{k})} {\rho(\bf{0})} &= \{e^{i \bf{k}\cdot \bf{x}}\} \equiv \frac{ \int d^2x \rho(\bf{x}) e^{i \bf{k}\cdot \bf{x}}} {\int d^2x \rho(\bf{x})}\\
&= \sum_{n=-\infty}^{\infty} \sum_{m=0}^{\infty}  \rho_{m,n} k^m e^{-i n\phi_k },
\end{align}
where $\phi_k$ is the azimuthal angle of the momentum vector ${\bf k}$ and the moments $\rho_{m,n}$ are only non-zero when $m>n$ and $(m-n)$ is even, in which case
\begin{equation}
\rho_{m,n} = \frac {i^m}{2^m(\frac{m+n}{2})!(\frac{m-n}{2})!}  \{ r^m e^{in\phi}\} .
\end{equation}
The curly brackets are an average over the transverse density as defined implicitly in Eq.~\eqref{fouriertransform}.
The cumulant expansion instead expands the log of the 2D Fourier transformed density
\begin{equation}
W({\bf k}) \equiv \log(\rho({\bf{k}}))
 = \sum_{m,n} W_{m,n} k^m e^{-i n\phi_k } .
\end{equation}
These have the advantage that all cumulants but the first $W_{1,1}$ are translationally invariant and do not depend on the location
of the origin of the coordinate system, sharing this property with the particle distribution that one would like to map, Eq.~\eqref{vn}.  If one chooses a coordinate system 
so that $\rho_{1,1}$ vanishes, the first few moments and cumulants coincide, so the difference between these bases is often unimportant. So any arbitrary density distribution is uniquely represented by a set of cumulants $W_{m,n}$, and any final observable is then a function of the cumulants with appropriate symmetries.  

One can immediately recognize that the eccentricities of Eqs.~\eqref{stdeps2} and \eqref{eps2pp} 
correspond to the magnitude of the lowest cumulant involving the 
second harmonic, made dimensionless $\varepsilon_2 = |W_{2,2}|/2W_{2,0}$,
and similarly for the generalized eccentricities $\varepsilon_n$ and planes$\Phi_n$~\eqref{eqn_epsilon} with $r^n$ weights.

\begin{figure}
\includegraphics[width=0.5\textwidth]{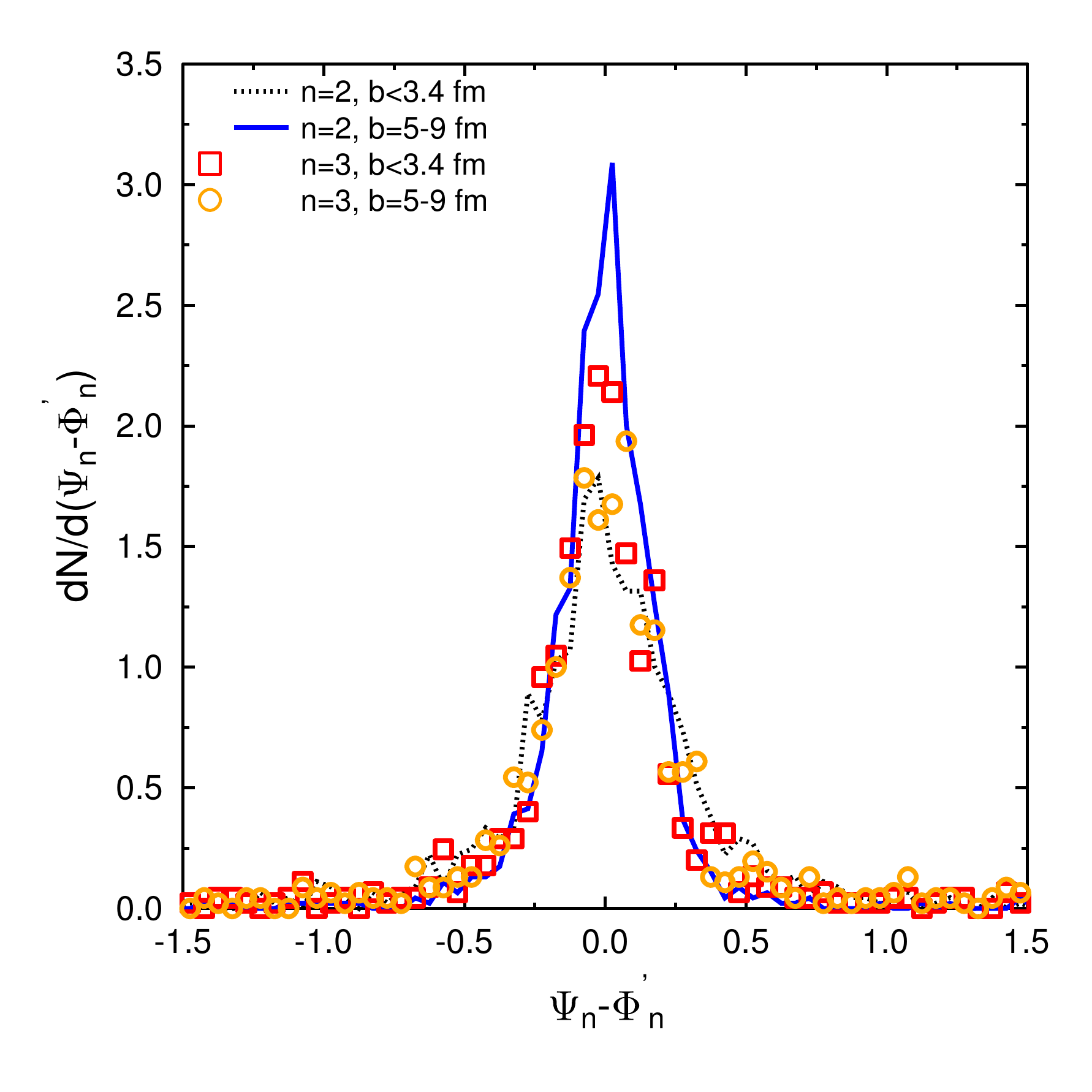}
\includegraphics[width=0.5\textwidth]{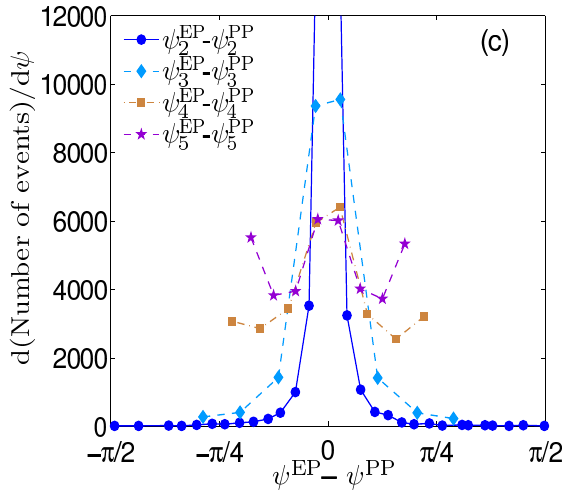}
\caption{There is a strong event-by-event correlation between event planes $\Psi_n$ and participant planes $\Phi_n$ for $n\leq 3$ (left~\cite{Petersen:2010cw}), but not for $n\geq 4$ (right~\cite{Qiu:2011iv}), which has a non-linear contribution~\cite{Gardim:2011xv}.}
\label{fig_phipsi} 
\end{figure}

\begin{figure}
\includegraphics[width=\textwidth]{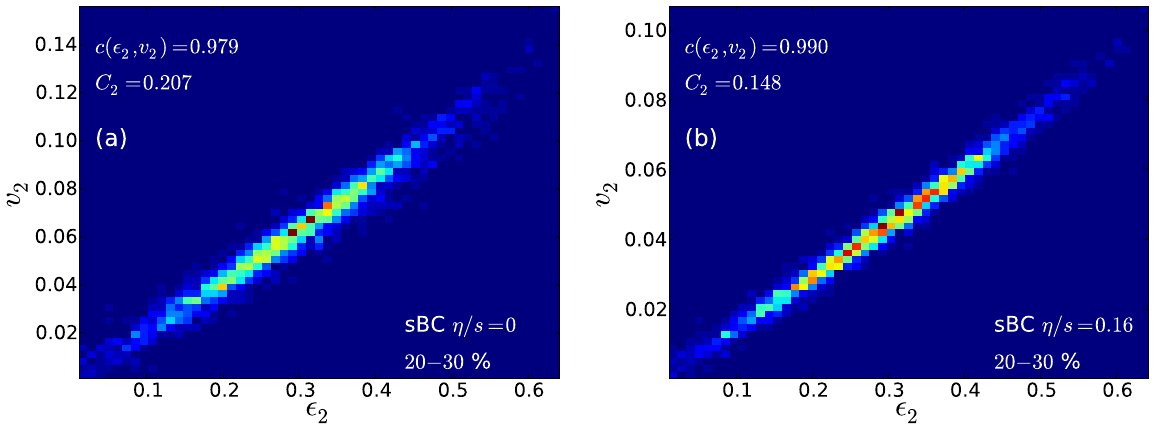}
\caption{2D histogram of event-by-event values of $v_2$ and $\varepsilon_2$, showing strong correlation between the two quantities, which increases with viscosity (taken from \cite{Niemi:2012aj}).}
\label{fig_vnepsn} 
\end{figure}

\begin{figure}
\includegraphics[width=0.5\textwidth]{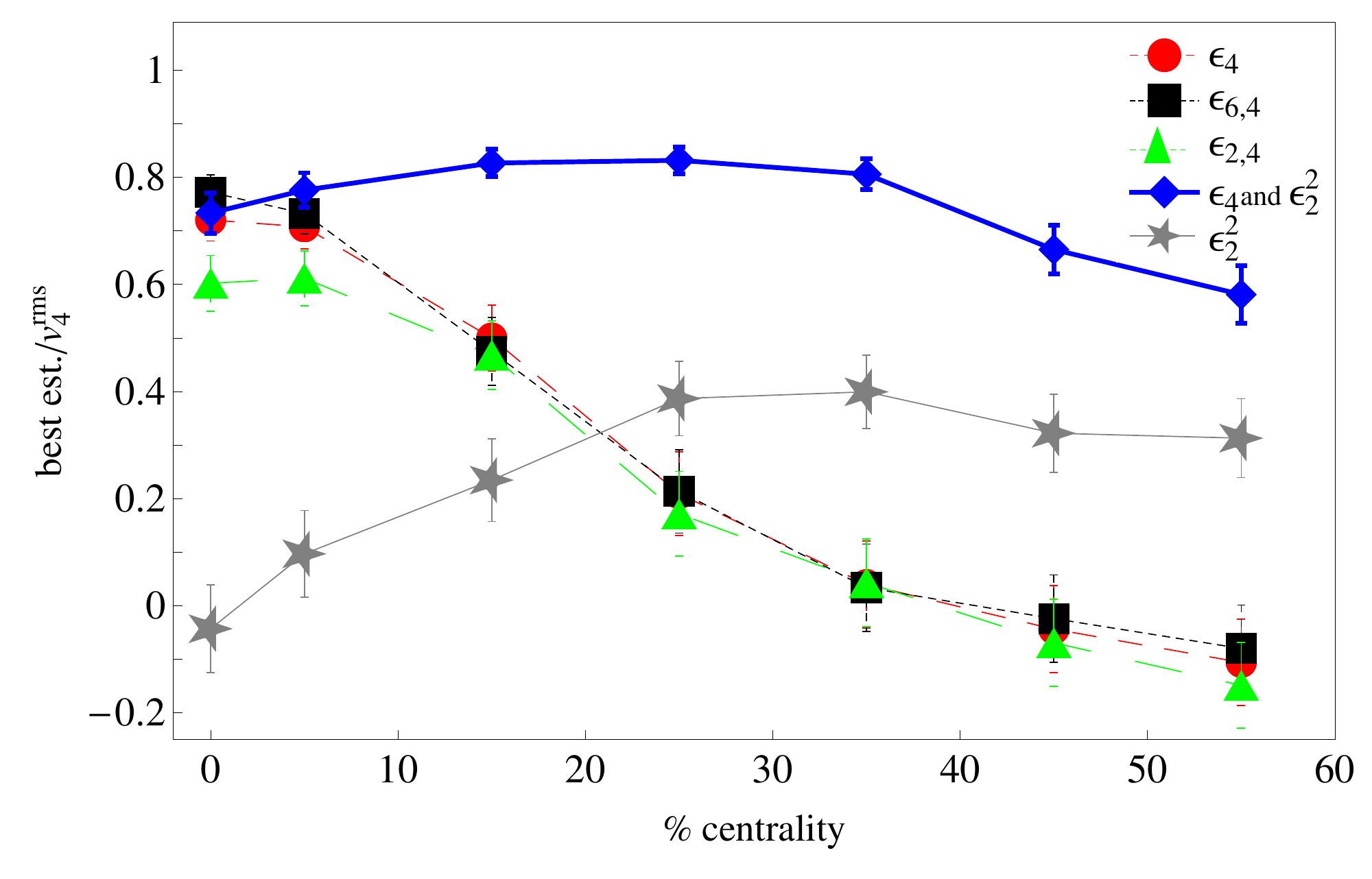}
\includegraphics[width=0.5\textwidth]{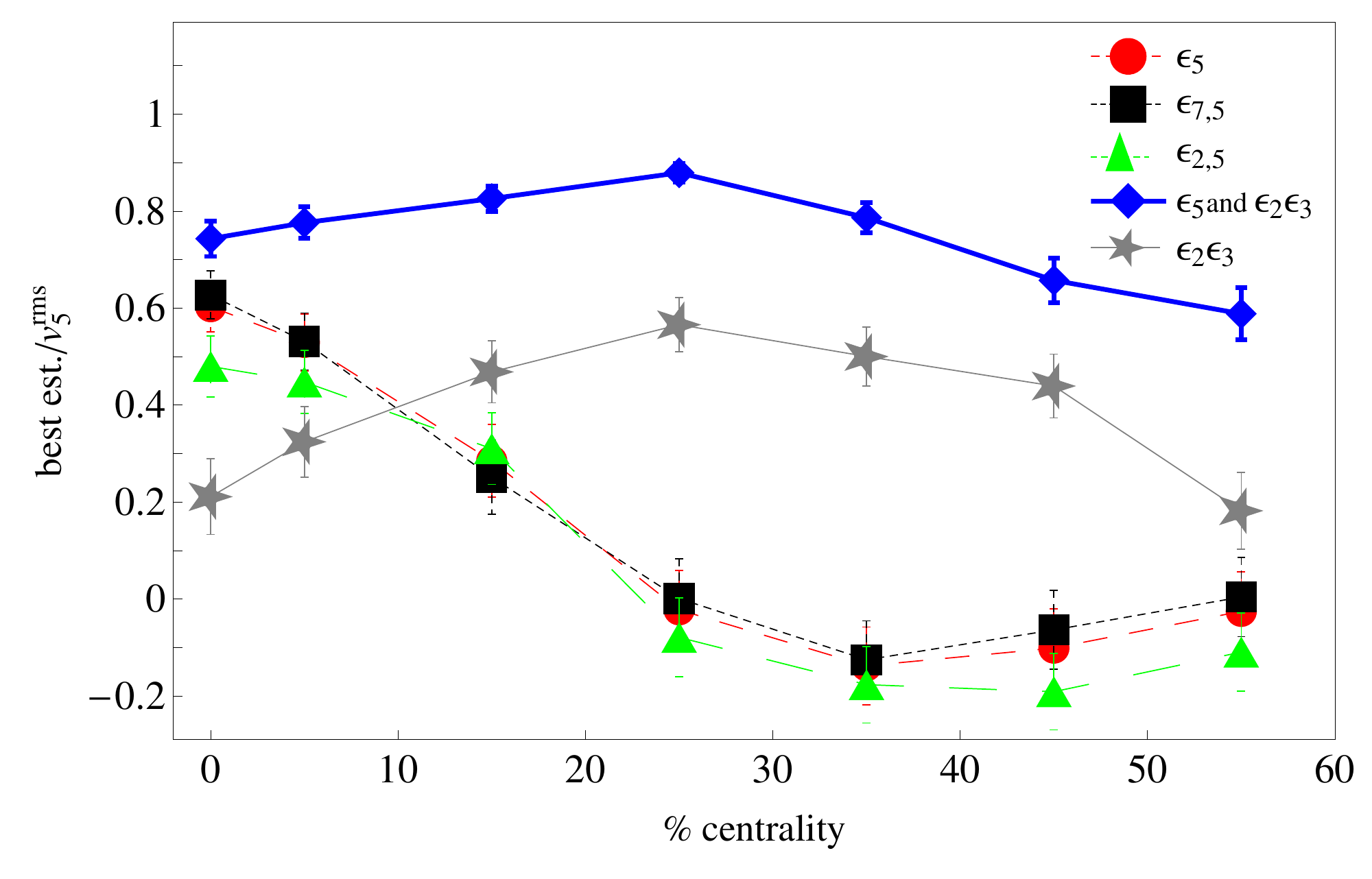}
\caption{Event-by-event correlation between the vector defined by $v_n$ and $\Psi_n$, and various estimators.  A value of 1 indicates perfect event-by-event correlation, 0 indicates no correlation, and a negative values indicate anticorrelation.  See Ref.~\cite{Gardim:2011xv} for details.  The red circles and grey stars represent the first and second term of Eqs.~(\ref{v4eps4},\ref{v5eps5}), respectively, while the blue diamonds represent the sum, showing that both contributions are important except for very central collisions.}
\label{fig_v4v5} 
\end{figure}

It has been seen that for harmonics $n\leq 3$, on an event-by-event basis, the participant plane $\Phi_n$
correlates well with the final event plane $\Psi_n$ (see Fig.~\ref{fig_phipsi}), while the magnitude
$v_n$ is proportional to $\varepsilon_n$ (see Fig.~\ref{fig_vnepsn}) \cite{Petersen:2010cw, Gardim:2011qn}.  This can 
be interpreted as hydrodynamic evolution being most sensitive to the large scale structure represented by the 
lowest momentum mode of the initial transverse density with the correct symmetry, and insensitive to the 
small scale structure represented by higher cumulants.  This interpretation is supported by the fact that the correlation between $v_n$ and $\epsilon_n$ gets even stronger with increasing viscosity (see Fig.~\ref{fig_vnepsn}) ~\cite{Niemi:2012aj} . 

These are not exact relations.  One can, for example, engineer an initial condition for which the lowest cumulant vanishes
and which still has a sizable flow coefficient~\cite{Qian:2013nba}.  However, they are quite accurate for the majority of events
that have been studied.

For $n>3$, one no longer has such a simple relation between $v_n$ and $\varepsilon_n$ (see Fig.~\ref{fig_phipsi})~\cite{Qiu:2011iv}, but
this is not entirely surprising, since lower-order cumulants can contribute, albeit nonlinearly.  It turns out that one can accurately predict the 
event-by-event flow vector (i.e., the magnitude $v_n$ and orientation $\Psi_n$) with the following equations (see, e.g., Fig.~\ref{fig_vnepsn})~\cite{Gardim:2011qn,Gardim:2011xv}:
\begin{align}
\label{v1eps1}
v_1 e^{i\Psi_1} &=C_1\ \varepsilon_2 e^{i\Phi_1}\\
\label{v2eps2}
v_2 e^{i2\Psi_2} &=C_2\ \varepsilon_2 e^{i2\Phi_2}\\
\label{v3eps3}
v_3 e^{i3\Psi_2} &=C_3\ \varepsilon_3 e^{i3\Phi_3}\\
\label{v4eps4}
v_4 e^{i4\Psi_4} &=C_{4,4}\ \varepsilon_4 e^{i4\Phi_4} + C_{4,22}\ \varepsilon_2^2 e^{i4\Phi_2} \\
\label{v5eps5}
v_5 e^{i5\Psi_5} &=C_{5,5}\ \varepsilon_5 e^{i5\Phi_5} + C_{5,23}\ \varepsilon_2 \varepsilon_3 e^{i(2\Phi_2+3\Phi_3)}.
\end{align}
These relations work equally well when defined in terms of the energy density or entropy density, but do not work quite 
as well when $\varepsilon_n$ for $n\neq 2$ is defined with $r^2$ weights~\cite{Gardim:2011xv}.  For example, compare
the green circles and the green triangles in Fig.~\ref{fig_vnepsn}.

Thus, by determining just a handful of (real) coefficients $C$ from a few of hydrodynamic calculations~\cite{Teaney:2012ke},
which contain all useful information about the medium response, 
one can reproduce even complicated multiplane correlations~\cite{Teaney:2013dta}.
These relations can also be used to work backward from data to put constraints on initial conditions~\cite{Retinskaya:2012ky}. 

One can use other bases to characterize the initial transverse density instead of cumulants, which have various advantages 
for theoretical study~\cite{Gubser:2010ui, ColemanSmith:2012ka, Floerchinger:2013rya}. Further progress has been made by the development of analytic solutions to the hydrodynamic equations~\cite{Gubser:2010ze,Gubser:2010ui}, which can give insight into
various properties of the hydrodynamic response~\cite{Staig:2011as}.

\subsection{Confronting experimental data: pair correlations}

Early on, measurements of elliptic flow (e.g., $v_2\{2\}(p_T)$ and $v_2\{{\rm EP}\}(p_T)$) 
showed large, long-range azimuthal asymmetries that could apparently only be 
reproduced by ideal hydrodynamics calculations. However, many models
for initial conditions could be made to fit data by adjusting parameters --- especially when viscosity is introduced~\cite{Song:2007ux, Luzum:2008cw} --- and so neither viscosity nor the initial conditions could be well constrained by data.

\begin{figure}
\begin{center}
\includegraphics[width=0.8\textwidth]{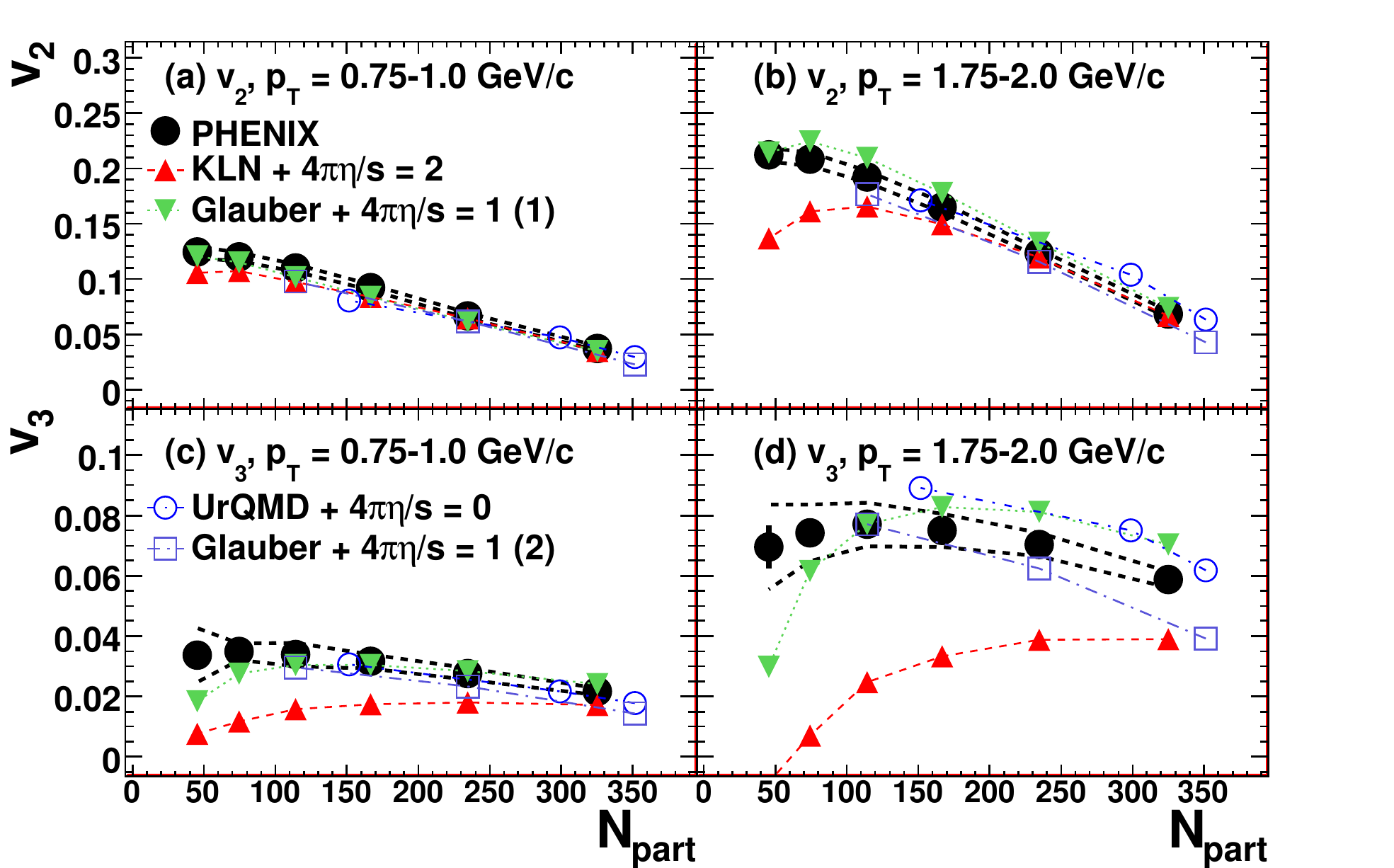}
\end{center}
\caption{Elliptic and triangular flow versus centrality (taken from \cite{Adare:2011tg}), showing that calculations with MC-KLN initial conditions cannot be made to fit both
$v_2$ and $v_3$ measurements.}
\label{fig_PHENIXv3} 
\end{figure}

With the new appreciation for the importance of flow fluctuations, more Fourier harmonics were then measured,    
placing significant additional constraints on models.  A first measurement of triangular flow
demonstrated that the MC-KLN model for initial conditions could not be made to fit both $v_2$ and $v_3$ --- a smaller viscosity 
would result in a $v_2$ that is too large, while a larger viscosity makes $v_3$ too small~\cite{Alver:2010dn,Adare:2011tg} 
(see Fig.~\ref{fig_PHENIXv3}). The MC-KLN is only a particular simple implementation of a model, and  can of course be modified and improved~\cite{ALbacete:2010ad, Dumitru:2012yr},  but the fact that non-trivial constraints can now
be put on such models is one of the most significant developments in the field.

\begin{figure}
\includegraphics[width=0.5\textwidth]{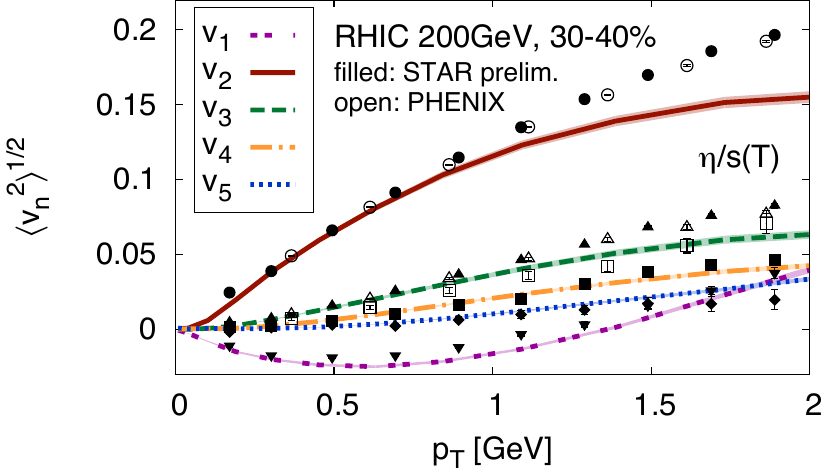}
\includegraphics[width=0.5\textwidth]{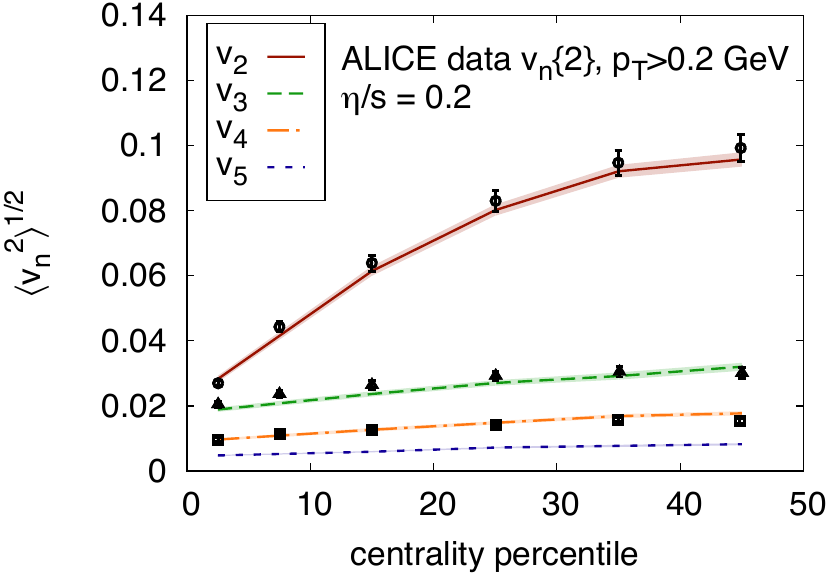}
\caption{$v_1$ through $v_5$ as a function of $p_T$ and centrality, at RHIC and LHC,  from event-by-event viscous hydrodynamics using IP-Glasma initial conditions~\cite{Schenke:2012hg}, compared to experimental data from PHENIX~\cite{Adare:2011tg}, STAR~\cite{Pandit:2012hp}, and ALICE~\cite{ALICE:2011ab} (taken from \cite{Gale:2012rq}).}
\label{fig_Bjoernvn} 
\end{figure}

\begin{figure}
\includegraphics[width=\textwidth]{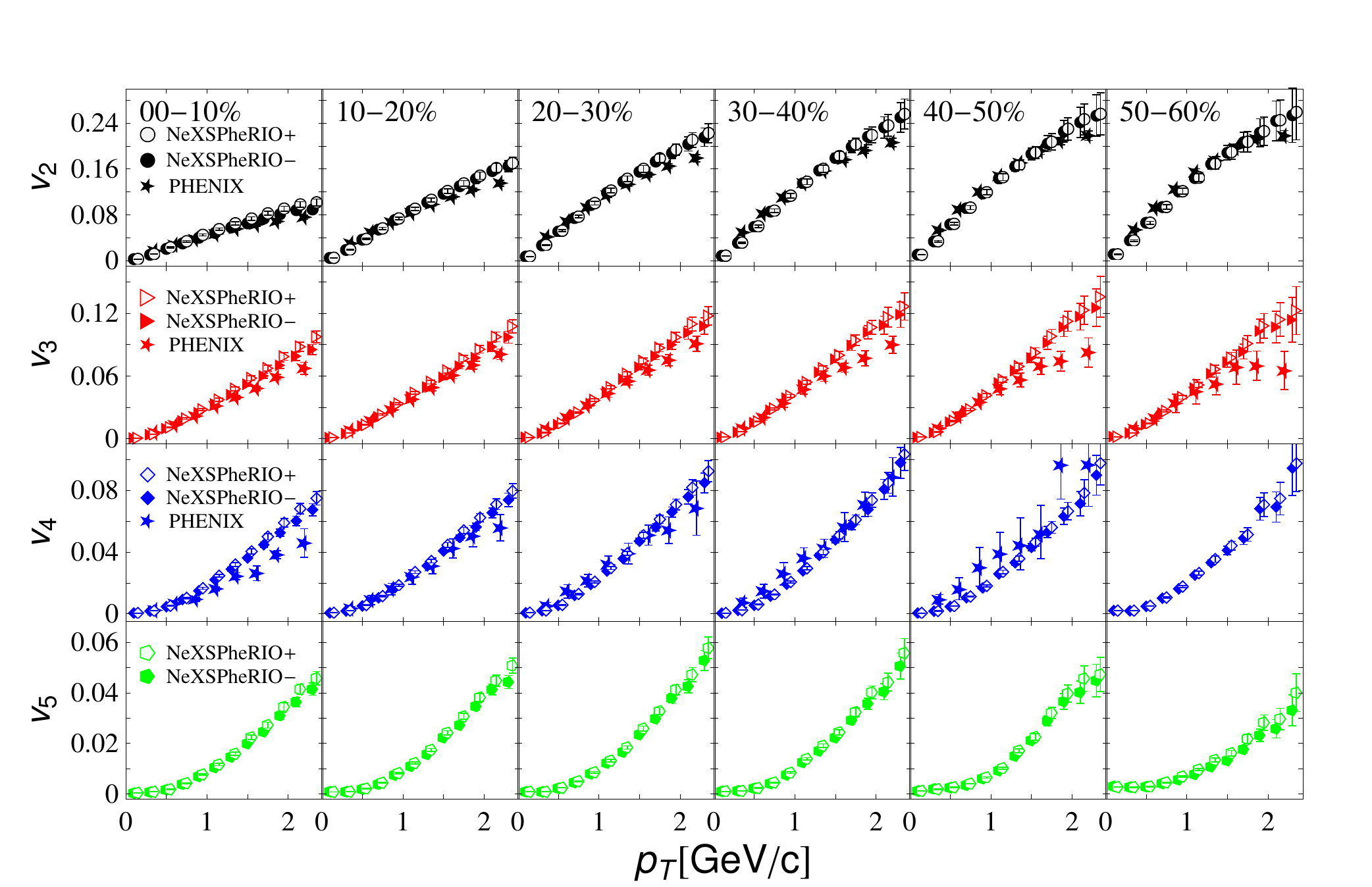}
\caption{$v_n$ versus $p_T$ and centrality at RHIC, from event-by-event ideal hydrodynamic calculations using NeXus initial conditions~\cite{Gardim:2012yp}.  Upper and lower results correspond to the low- and high-resolution limit of the event-plane measurement to which the calculation is compared~\cite{Adare:2011tg}, Eqs.~\eqref{lowres} and~\eqref{highres}, respectively}
\label{fig_Brazilvn} 
\end{figure}

The full set of measured Fourier harmonics places even more constraints on theoretical models, 
and a large number of calculations have now been done using a variety of 
initial conditions, and compared to these data. Fig. \ref{fig_Bjoernvn} shows a calculation within a (3+1)d viscous hydrodynamic calculation starting from IP-Glasma initial conditions fitting all the $v_n$ measurements at RHIC and LHC simultaneously. A similar impressive agreement to differential $v_2-v_5$ measurements from PHENIX can be obtained in NEXshpeRIO ideal hydrodynamic calculations based on NEXUS event generator initial profiles (see Fig. \ref{fig_Brazilvn}). Both of these calculations do not account for hadronic rescattering in the final stages of the reaction and illustrate that there is still ambiguity between the fluctuating initial conditions that have been employed and the transport properties of the medium, even if a large set of measurements is considered. 

Beyond constraints on initial conditions, one can look for observables that are insensitive to initial conditions, and therefore give more direct
access to medium properties.  For example it has been shown that
the set of flow harmonics in ultra-central collisions have a smaller sensitivity to the model for initial conditions, and can
therefore provide the best current constraints on fluid properties such as shear viscosity~\cite{Luzum:2012wu} (Fig. \ref{fig_vnvsn}.

\subsection{Multiparticle correlations}
Ultimately, the long-range part of the pair correlation --- 
including all Fourier harmonics as a function of transverse momentum centrality --- 
can be fit by more than one model~\cite{Gardim:2012yp, Schenke:2012hg}. 
Therefore, it is advantageous to consider measurements beyond two-particle correlations --- 
both to provide more stringent tests of the hydrodynamic picture as a whole (i.e., the
idea that long-range correlations are dominated by collective effects) and to discriminate
further between theoretical models.  Although two-particle correlations are less limited by statistics
and are more studied, multiparticle correlations open up a much larger space of independent
information, only some of which has been explored.

One can consider combinations of observables that isolate the initial state.  
By noting the linear nature of Eqs.~\eqref{v1eps1}--\eqref{v3eps3}, one can immediately see that
ratios of observables that contain the same powers of each harmonic will remove the hydrodynamic
response coefficients $C_n$, and effectively isolate properties of the initial conditions~\cite{Bhalerao:2011yg}.
As just one example, from Eqs.~\eqref{v22} and ~\eqref{diff4pcum} we can see,
\begin{equation}
\frac {2v_n\{2\}^4-
v_n\{4\}^4 } {v_n\{2\}^4}
= \frac {\langle v_n^{\ 4} \rangle} {\langle v_n^{\ 2} \rangle^2}
\simeq \frac {\langle \varepsilon_n^{\ 4} \rangle} {\langle \varepsilon_n^{\ 2} \rangle^2}.
\end{equation}
So then by only calculating event-by-event moments of eccentricities of the spatial distribution of a model for 
the initial conditions, we can immediately see whether it is consistent with data --- e.g., we can immediately
see that  the original MC-KLN and MC-Glauber do not contain the correct fluctuations~\cite{Bhalerao:2011yg}.

Some of the most studied multiparticle correlations are higher cumulants of elliptic flow.  Measurements of the fourth
cumulant elliptic flow $v_2\{4\}$~\cite{Adler:2002pu} are of the same order as $v_2\{2\}$, but systematically smaller.  
This is consistent with what is expected from flow (see Eq.~\eqref{diff4pcum}), and what is seen 
quantitatively in event-by-event hydrodynamic calculations~\cite{DerradideSouza:2011rp,Qiu:2011iv,Hirano:2012kj}.
Measurements of even higher cumulants ($v_2\{6\}$, $v_2\{8\}$, $v_2\{{\rm LYZ}\}$), are consistent with the fourth
cumulant~\cite{Bilandzic:2011ww,Aamodt:2010pa}.  Although they require many statistics, and have therefore not yet been calculated in 
an event-by-event hydrodynamic framework, this is what is expected when flow fluctuations are approximately 
Gaussian~\cite{Voloshin:2007pc}.  Further, it is difficult, if not impossible, 
to generate such a large anisotropy among correlations of such a large number of particles without an underlying collective motion.

The fourth cumulant triangular flow $v_3\{4\}$ has also been measured.  While it is also roughly consistent with hydro 
expectations~\cite{Bhalerao:2011ry}, there is an interesting small discrepancy between measurements at RHIC and
at LHC.  As predicted~\cite{Bhalerao:2011yg}, the fourth cumulant triangular flow in central collision at RHIC is consistent with Gaussian
flow fluctuations ($v_3\{4\}$ = 0)~\cite{Adamczyk:2013waa}.  However, the measurement at LHC is positive and inconsistent with zero~\cite{ALICE:2011ab,Bhalerao:2011ry}.  Although the discrepancy
is small, it's potentially interesting, and not yet understood.

Multiparticle correlations can also be used to probe correlations among different harmonics.  An early example is the rapidity-odd projection
of directed flow measured with
respect to the elliptic flow event plane $v_1\{\Psi_2\}$.   More recently, but still before the flow fluctuation revolution, quadrangular flow
was measured with respect to the second event plane $v_4\{\Psi_2\}$~\cite{Abelev:2007qg,Adare:2010ux}.  
While they are difficult to understand from smooth and symmetric calculations~\cite{Luzum:2010ae,Luzum:2010zz}, these can be reproduced well in event-by-event  
calculations~\cite{Gardim:2012yp, Konchakovski:2012yg}

Even more recently, there have been a large number of mixed harmonic correlation measurements.  In particular a large set of 
event-plane correlations that were recently measured~\cite{Jia:2012sa}, all of which match surprisingly well to hydrodynamic calculations~\cite{Qiu:2012uy,Teaney:2013dta}.

Finally, unfolded event-by-event distributions of the various $v_n$ were measured  (recall Fig. \ref{fig_unfolded}).  
Due to the linear relationships such as Eq.~\eqref{v2eps2}, the mean value of the distributions of harmonics $n\leq3$ is mainly sensitive to the viscosity while the shape of the distributions is mainly related to the shape of the initial eccentricity distributions \cite{Niemi:2012aj} that vary in different initial state models.  
As was seen before from ratios of cumulants, 
MC-Glauber and MC-KLN models do not have the correct fluctuations~\cite{Niemi:2012aj}.  However, the IP-Glasma model 
matches data quite well~\cite{Gale:2012rq}.  

\subsection{Other multiparticle observables}
Beyond correlations of particles in the azimuthal angle $\phi$, there are many other event-by-event observables that are likely sensitive to initial state fluctuations. In \cite{Gavin:2011gr}, the final transverse momentum correlations are related to the density correlation function in the initial profile. The transverse momentum fluctuations have also been studied in \cite{Bozek:2012fw} in an event-by-event viscous hydrodynamic approach and the relation to structures in the initial state profile has been confirmed. Calculating event-by-event observables is computationally very demanding, but has promising potential to constrain initial state fluctuations independently. 

\section{Open Issues}
\label{sec_issues}
\subsection{Longitudinal Fluctuations}

If there are fluctuations in the transverse plane, there should likewise be fluctuations in the longitudinal direction as well. Most models for the initial dynamics rely on some type of flux tube or string picture that motivates that a boost-invariant treatment is reasonable. Experimentally, non-trivial dependencies of triangular flow on $\Delta \eta$ have been observed by STAR \cite{Adamczyk:2013waa}, but not yet confirmed by other experiments. 

Theoretical calculations by Bozek \textit{et al.} \cite{Bozek:2010vz}, Petersen \textit{et al.} \cite{Petersen:2011fp} and Pang \textit{et al.} \cite{Pang:2012he} demonstrate that longitudinal fluctuations may result in non-trival pseudo-rapidity dependencies of the triangular flow event plane and two-particle correlations. 
In some calculations, one can even find that the $v_3$ flow vector is anticorrelated between the far forward and backward regions \cite{Xiao:2012uw}.  Any such non-trivial pseudorapidity dependence can give both a non-trivial $\eta$ and $\Delta\eta$ dependence to measurements.
Before drawing explicit conclusions, the experimental data has to be clarified by adjusting the kinematic cuts and acceptance windows across collaborations to separate these effects from non-trivial longitudinal structures. In addition, 3+1 d viscous hybrid event-by-event calculations including longitudinal fluctuations are needed to identify sensitivities. 

\subsection{Hard and/or Electromagnetic Probes}
Another way to study the properties of the medium created in a heavy-ion collision is by studying particles that are not themselves part of the medium.  If the initial scattering creates a particle with very large transverse momentum, it can escape the collision region before sustaining enough rescatterings so as to thermalize.  Likewise, electromagnetic radiation can be emitted, but interacts so weakly that it will also escape without equilibrating with the rest of the system.

Indeed,  an early motivation to study di-hadron correlations as a function of $\Delta \phi$ and $\Delta \eta$ came from the desire to understand energy loss mechanisms in a more detailed way than can be done with only the global yield (e.g., in the form of the nuclear modification factor $R_{\rm AA}$). 
If a hard process in the initial collision generates jets, there will be a non-trivial modification of correlations due to jet-medium interactions, and this should be visible in correlations involving large transverse momentum trigger particles. 
At more moderate values of transverse momentum, there can be also a significant contribution from thermal particles emitted from the medium.
To fully address the question of how much of the measured correlations results from jet-medium interactions versus bulk response to initial state fluctuations in various ranges of transverse momentum, one needs an integrated approach that treats a hydrodynamic evolution coupled to the propagation of hard particles combined including the back-reaction of the modified fluid on the energy loss mechanism. Since such an approach has not been fully established yet, we are pointing out two studies that assess qualitatively the relation of hard processes and fluctuations in the bulk evolution. 

The Jysv\"{a}skyl\"{a} group has performed a calculation incorporating event-by-event fluctuations in the hydrodynamic evolution to investigate the influence on elastic and radiative parton energy loss \cite{Renk:2011qi}.  It has been found that in central collisions the angular dependence of the nuclear modification factor is very similar in the fluctuating and in the smooth scenario. This can be attributed to the cancellation of two opposing effects, which leads to some sensitivity in the energy loss pattern in non-central collisions. 

Another approach towards a consistent description of hard processes and bulk evolution has been published by Werner et al \cite{Werner:2012xh}. In this framework, EPOS is combined with a 3+1D ideal hydrodynamic evolution, and it is shown that the hard processes influence the transverse momentum spectra event at lower $p_T$ values. Therefore, it is not excluded that hard processes have also some influence on the measured flow coefficients. 

Photons and dileptons are produced at all stages of a heavy ion collision and since they only interact via the electromagnetic interaction they will reach the detector once they are produced, with little modification. Since these electromagnetic probes are produced during the entire collision, they can carry complimentary information compared to hadronic observables, which are largely emitted at later times.  See Ref.~\cite{Gale:2012xq} for a review.

In particular, one can measure correlations between electromagnetic probes and the rest of the system -- i.e., flow.  Recently the elliptic flow and triangular flow of direct photons have been measured, with values similar to charged hadrons \cite{Adare:2011zr,Lohner:2012ct}, and similarly for dileptons \cite{XiangliCuifortheSTAR:2012mya}.

Unsurprisingly, fluctuations are important for these observables as well.  Event-by-event hydrodynamic simulations have been made to study the production of direct photons \cite{Dion:2011pp,Chatterjee:2011dw,Chatterjee:2012dn,Chatterjee:2013naa,Shen:2013cca}, while the effect of fluctuations on leptonic observables is still being studied \cite{Vujanovic:2013jpa}.

\subsection{Beam energy dependence}
In 2010/11 the first low beam energy scan program at RHIC has been carried out where gold nuclei were collided at beam energies from $\sqrt{s_{\rm NN}}=7.7-39$ GeV. The main goal of this enterprise was to search for the critical endpoint of the phase tranisition line in the QCD phase diagram. Refined measurements of observables that showed interesting structures at the SPS were taken in a collider setup that allows for similar acceptance over a large range of beam energies. 

The main results include the disappearance of some signatures for quark gluon plasma formation at high beam energies, e.g. the nuclear modification factor shows an enhancement instead of a suppression and the constituent quark scaling breaks between particles and antiparticles. The structures of the fluctuations in the initial state on the nucleon level are expected to be rather weakly dependent on the beam energy, whereas gluon saturation should vanish as a function of energy. Furthermore, it has to be investigated how well the description in terms of hybrid hydrodynamics and transport approaches that was successful at higher beam energies can be applied at lower beam energies where the non-equilibrium effects gain importance. 

Measurements of higher flow harmonics and their transverse momentum and centrality dependence are expected to be sensitive probes of the increasing viscosity during the evolution and can possibly be used to disentangle fluctuation from critical phenomena associated with the phase transition and initial state geometry fluctuations. 

\begin{figure}
\includegraphics[width=0.5\textwidth]{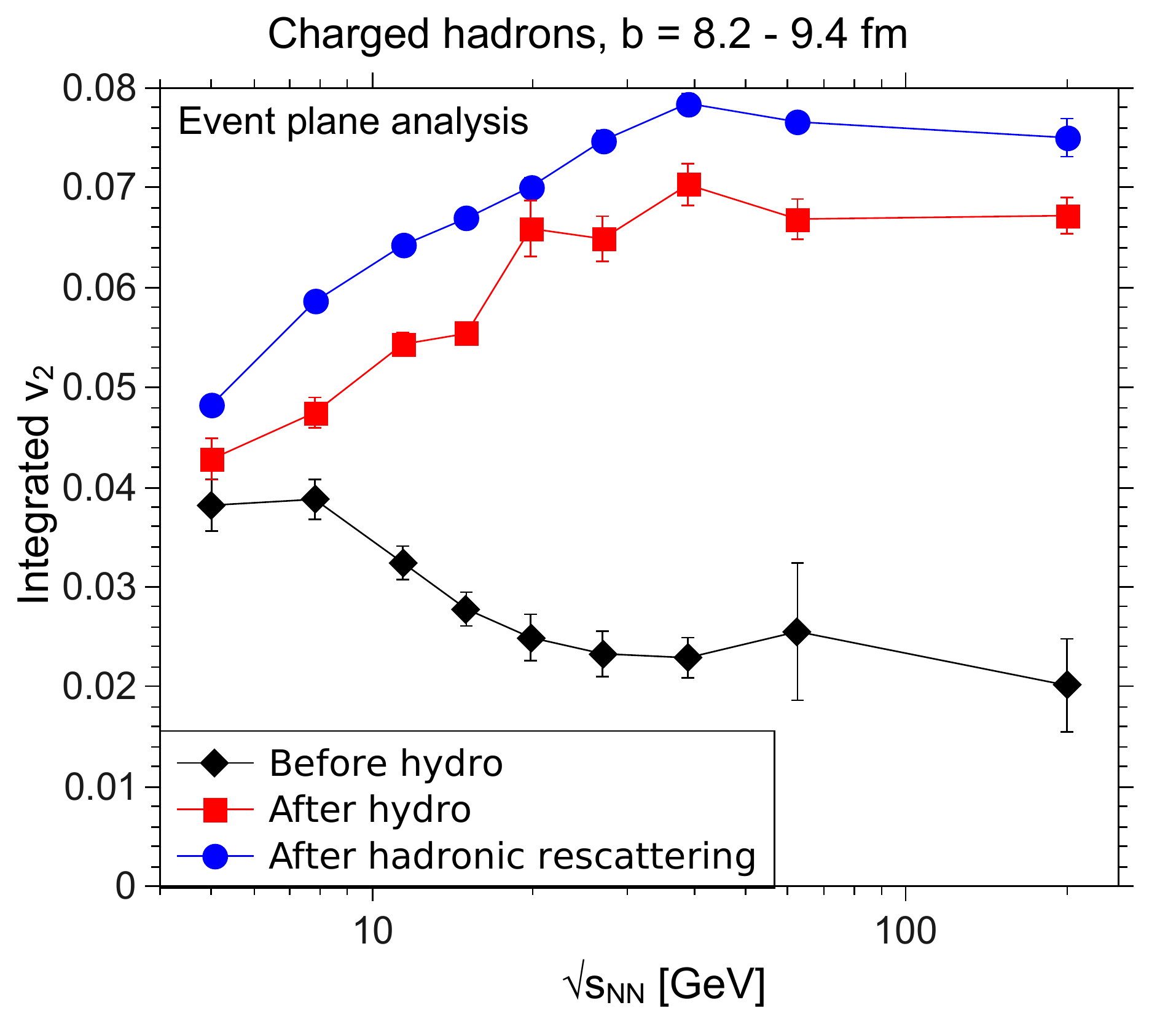}
\includegraphics[width=0.5\textwidth]{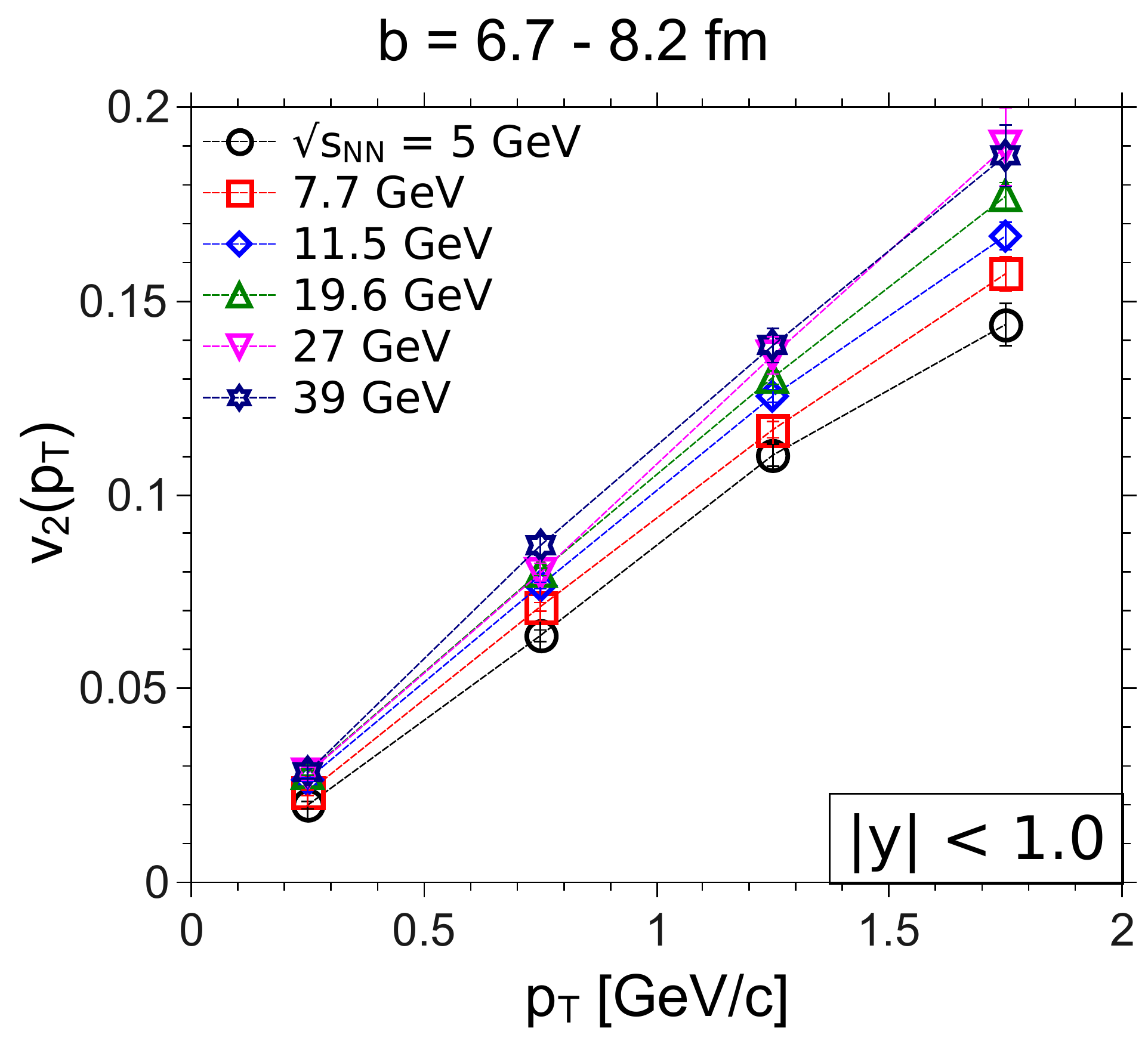}
\caption{Elliptic flow as a function of beam energy in the UrQMD hybrid model. The black diamonds depict the contribution from the early non-equilibrium evolution, the red squares show the result after the hydrodynamic evolution and the blue circles depict the full result including hadronic rescattering (left). Elliptic flow as a function of transverse momentum in the calculated in the UrQMD hybrid model (right) (taken from \cite{Auvinen:2013sba}).}
\label{fig_besv2} 
\end{figure}

One surprising result that has been obtained in the low beam energy scan at RHIC is the fact that the differential elliptic flow as a function of transverse momentum does not change significantly over a large range in beam energies. Qualitative calculations in an event-by-event hybrid approach suggest the conclusion, that the mechanism to build up elliptic flow smoothly changes from an almost ideal fluid to a hot and dense hadron gas \cite{Auvinen:2013qia,Auvinen:2013sba}. In Fig. \ref{fig_besv2} the elliptic flow that is generated during different stages of the dynamic evolution of heavy ion reactions at different beam energies is shown and the resulting $v_2(p_T)$ is roughly constant. 

To fully explore the potential of measurements at lower beam energies, more theory development is needed to make clear predictions for the fluctuations associated with a first order phase transition or the critical endpoint. Since many of the so far suggested event-by-event fluctuation observables are very sensitive to kinematic cuts and require large statistics, a more realistic treatment of non-equilibrium phase transitions in a finite system will be crucial \cite{Herold:2013qda}. 
 
\subsection{Other Systems}
Besides measuring flow observables as a function of centrality in symmetric collisions of round heavy nuclei such as lead and gold, it is of interest to study particle correlations in other systems as well. Especially since the measurements of elliptic flow in Cu+Cu collisions at RHIC played an important role in realizing the importance of event-by-event fluctuations for flow observables. 

Recently, a run of U+U collisions has been carried out at RHIC and first results are about to appear. The motivation to collide uranium nuclei is based on the fact, that it is known from nuclear structure calculations that uranium has a prolate shape like a cigar. Therefore, there are a lot of interesting possibilities: If the two nuclei collide head-on with their round sides, there is going to be high multiplicity and very small elliptic flow, but if they collide with their elongated edges, a highly anisotropic initial state is generated and large elliptic flow and a very high multiplicity is expected. In addition, there is no magnetic field present, since there is only a small number of spectators. Therefore, these special events provide the opportunity to cross-check the predictions for the chiral magnetic effect that are ambiguous in Au+Au calculations where alternative explanations based on elliptic flow and balance functions are also able to explain the charge correlation measurements.    

Collisions of asymmetric systems are also of interest, since one would expect finite odd flow components due to the initial geometry of e.g. an Cu+Au collision. Again data has been recently taken at RHIC and it will be exciting to see the results for triangular flow measurements that are not solely generated by fluctuations. 

\begin{figure}
\includegraphics[width=0.5\textwidth]{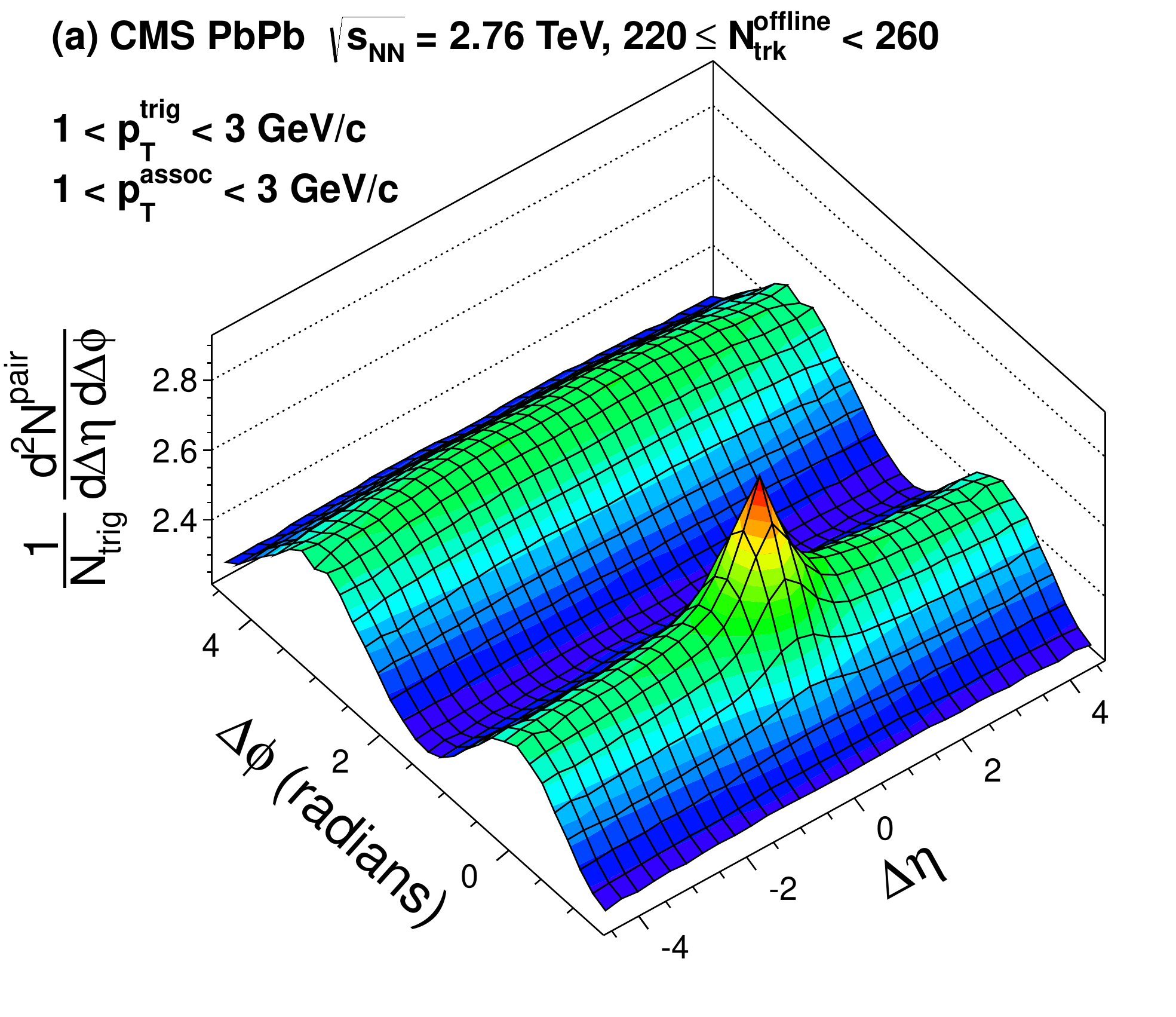}
\includegraphics[width=0.5\textwidth]{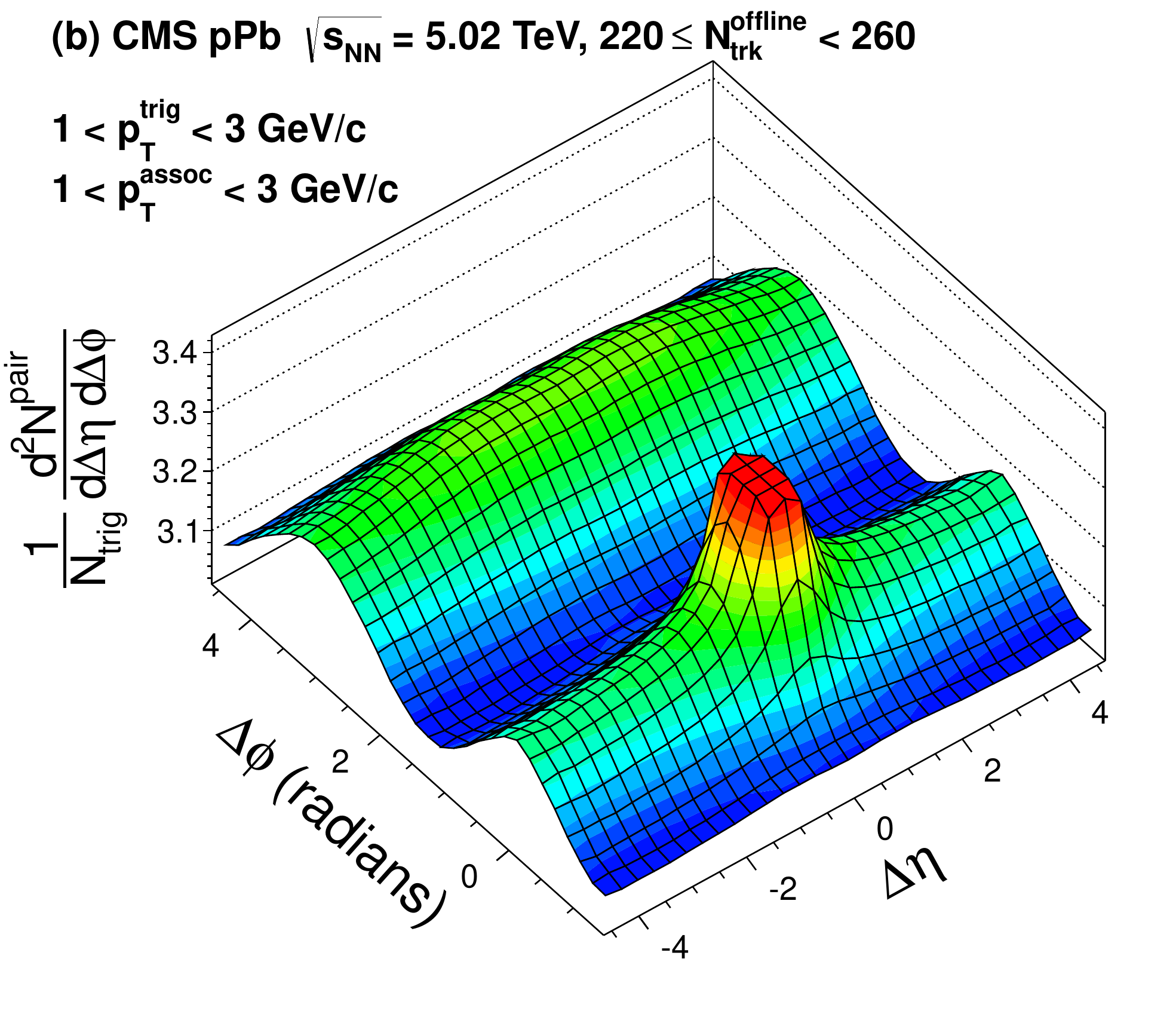}
\caption{2D correlation function for pairs of charged hadrons in both PbPb collisions (Left) and pPb collisions (Right) with the same observed (uncorrected) multiplicity and using the same kinematic cuts, showing striking similarity (taken from \cite{Chatrchyan:2013nka}).}
\label{fig_cmscorrelation} 
\end{figure}

Last but not least, it is of high interest to look at bulk observables in elementary collisions. If the assumption is that anisotropic flow is a clear signal for hydrodynamic behaviour and the formation of a quark gluon plasma, one would expect these collective effects to be reduced as the system becomes smaller.   Measurements have now been performed for p-p, p-Pb and d-Au collisions. 
At the LHC, the collision energies are so high that in some elementary collisions, the multiplicity can be as high as in peripheral Pb+Pb collisions. Within the last year, the LHC experiments have performed correlation measurements and as in the example shown in Fig. \ref{fig_cmscorrelation} by the CMS collaboration correlations the two-particle correlations show surprisingly similar correlation structures in peripheral heavy ion and elementary p-Pb collisions. At the moment, there are two main explanations, either the correlations are generated directly in the initial state, e.g., by early gluon dynamics in a saturation picture \cite{Dusling:2012wy,Dusling:2013oia}, or the correlations are generated by collective expansion, e.g., hydrodynamics \cite{Bozek:2012gr,Bozek:2013uha,Werner:2013ipa,Qin:2013bha,Bzdak:2013zma}. Since p-Pb and p-p collisions are used as a reference for many hard observables it is important to understand them in more detail.

\section[Summary]{Summary and Conclusion}
\label{sec_sum}
To summarize there is a lot of experimental and theoretical effort ongoing in the heavy ion community to understand and determine initial state features by looking at final state correlations. We introduced the flow correlation observables that are investigated, and described the state of the art for dynamical modeling of heavy ion reactions in terms of viscous hydrodynamics coupled to hadron transport. The current plethora of initial state models has been presented in detail. Since there is no full first principle QCD calculation for the early time dynamics yet, the way to progress is to apply a well-constrained bulk evolution model to constrain features of the initial state profiles. These features are connected to properties of the hot and dense state of strongly interacting matter that is initially formed in heavy ion reactions. 

The generic hydrodynamic response to initial state structures can be summarized as follows: For $n\leq 3$ the assumption of a linear response between $\varepsilon_n$ and $v_n$ is justified, while for higher moments non-linear contributions are important. Overall, the full set of experimental multi-particle observables needs to be exploited to derive quantitative values for transport coefficients of the quark gluon plasma and to constrain the initial dynamics. There are many open issues that are important to consider on the road towards precision studies of the properties of hot and dense strongly interacting matter. The longitudinal fluctuations, initial velocity profiles, initial shear tensor components and the effect of hadronic rescattering on higher flow coefficients are beyond the ones that will be studied in the near future. Here, also the study of different system sizes and beam energies will provide important input. 

Open issues that require more fundamental development are the combination of hard processes and jet-medium interactions with the bulk medium in a fully consistent way including the backreactions and to disentangle initial state fluctuations from the fluctuation probes that signal a first order phase transition of a critical endpoint. For both of these a whole new generation of dynamical transport approaches needs to be developed, before quantitative conclusions can be drawn. 
%
%
%
\ack
H.P. acknowledges funding of a Helmholtz Young Investigator Group 
VH-NG-822 from the Helmholtz Association. This
work was supported by the Helmholtz International Center for the Facility for Antiproton
and Ion Research (HIC for FAIR) within the framework of the Landes-Offensive zur Entwick-
lung Wissenschaftlich-\"okonomischer Exzellenz (LOEWE) program launched by the State of
Hesse.
M.L.~was supported by the Natural Sciences and Engineering Research Council of Canada and by the Office of Nuclear Physics in the US
Department of Energy's Office of Science under Contract No.~DE-AC02-05CH11231.  
He would like to thank Jean-Yves Ollitrault for helpful discussion. 

\end{document}